\documentclass[aps, prb, twocolumn, superscriptaddress, showpacs, showpacs, preprintnumbers, amsmath, amssymb]{revtex4-2}
\usepackage{amssymb}
\usepackage{amsmath}
\usepackage{bm}
\usepackage[dvips]{graphicx}
\usepackage{dcolumn}
\usepackage{longtable}
\usepackage{color}
\usepackage{bm}
\usepackage[version = 4]{mhchem}
\usepackage[resetlabels,labeled]{multibib}
\newcommand{\TSC}{$T_{\rm c}$}
\newcommand{\dv}{$\bm{d}$}
\begin{document}
\title{Intimate relationship between spin configuration in the triplet pair and superconductivity in UTe$_2$} 
\author{Hiroki Matsumura}
\email{matsumura.hiroki.75r@st.kyoto-u.ac.jp}
\author{Yuki Takahashi}
\author{Riku Matsubayashi}
\author{Katsuki Kinjo}
\thanks{Present adress: Institute of Multidisciplinary Research for Advanced Materials, Tohoku University, Sendai, Miyagi 980-8577, Japan}  
\author{Shunsaku Kitagawa} 
\author{Kenji Ishida}
\email{kishida@scphys.kyoto-u.ac.jp}
\affiliation{Department of Physics, Graduate School of Science, Kyoto University, Kyoto 606-8502, Japan}
\author{Yo Tokunaga}
\author{Hironori Sakai}
\author{Shinsaku Kambe}
\author{Motoi Kimata}
\affiliation{Advanced Science Research Center, Japan Atomic Energy Agency, Tokai, Ibaraki 319-1195, Japan}
\author{Ai Nakamura}
\author{Yusei Shimizu}
\thanks{Institute for Solid State Physics (ISSP), University of Tokyo, Kashiwa, Chiba 277-8581, Japan}
\author{Yoshiya Homma}
\author{Dexin Li}
\author{Fuminori Honda}
\thanks{Central Institute of Radioisotope Science and Safety, Kyushu University, Fukuoka 819-0395, Japan}
\author{Atsushi Miyake}
\affiliation{Institute for Materials Research, Tohoku University, Oarai, Ibaraki 311-1313, Japan}
\author{Dai Aoki}
\affiliation{Institute for Materials Research, Tohoku University, Oarai, Ibaraki 311-1313, Japan}
\affiliation{Universit\'e Grenoble Alpes, CEA, IRIG, PHELIQS, F-38000 Grenoble, France}
\author{Tetsuya Furukawa}
\author{Takahiro Sasaki}
\affiliation{Institute for Materials Research, Tohoku University, Sendai, Miyagi 311-1313, Japan}

\date{\today}
\begin{abstract}
Spin-triplet superconductivity is an intriguing quantum coherent state with both spin and orbital degrees of freedom, which holds significant potential for future applications in quantum technology. 
However, how the spin of the triplet pairs responds to an external magnetic field remains poorly understood. 
This is mainly due to the absence of suitable spin-triplet superconductors.
Here, we report results of Knight-shift and ac-susceptibility measurements on UTe$_2$.
We demonstrate that the spin susceptibility, which slightly decreases compared to the normal-state value below the superconducting (SC) transition temperature $T_{\rm c}$, is rapidly restored and nearly recovers to the normal-state values around 5 T, well below the SC upper critical field $H_{c2}$ when the magnetic field is applied along the $c$ axis ($H \parallel c$).
In addition, we found that $H_{\rm c2}$ of superconductivity becomes larger when the SC spin aligns with the magnetic field.
By considering the results on $H \parallel b$, our results suggest the presence of a close relationship between the spin configuration of the triplet pair and $H_{\rm c2}$, as well as the anisotropic pinning interaction acting on the triplet pairs. 
These phenomena, which have never been observed in spin-singlet superconductors, represent characteristic features unique to spin-triplet superconductors.
We discuss the similarities between superconductivity in UTe$_2$ and superfluid $^3$He, focusing on their spin-triplet pairing states.
\end{abstract}

\maketitle
\section{Introduction}
Superconductivity occurs when a coherent quantum fluid is formed by electron pairs.
For most superconductors, the total spin ($S$) of the pairs is in the singlet state ($S$ = 0), and it is also possible in the triplet state ($S$ = 1).
Such superconductors, called ``spin-triplet superconductors'', are coherent quantum fluids with spin and orbital degrees of freedom.
This superconducting (SC) state involves rich physics, which can be used in future quantum technology\cite{GulianIEEE2003,LeijnsePRL2013}.
However, spin-triplet superconductors are very rare, thus the physical properties on the spin-triplet superconductivity have not been well understood.
After the discovery of ferromagnetic (FM) superconductors\cite{SaxenaNature2000, AokiNature2001, HuyPRL2007}, they have made it possible to study the spin-triplet pairing state in superconductors.
However, the coexistence of ferromagnetism was not suitable for investigating the spin-triplet properties, since a precise measurement of the spin susceptibility in the SC state could not be performed because of the presence of the internal field arising from ferromagnetism, and tiny contribution of the spin susceptibility related to superconductivity.   
In such a situation, UTe$_2$ has been newly discovered\cite{RanScience2019}.
Although UTe$_2$ shares the similarity with FM superconductors, such as strong Ising anisotropy in the spin susceptibility, the magnetic-field ($H$)-enhanced superconductivity for $H$ applied along the $b$ axis, and occurrence of metamagnetic behavior \cite{RanScience2019, aoki2021JPCMrev,AokiJPSJ2019Rev}, it undergoes a superconducting transition in the paramagnetic state, where a detailed investigation of the spin susceptibility in the SC state is possible.
Thus, UTe$_2$ provides a special opportunity to study the physics of spin-triplet superconductivity.

As is well known, physics of the spin-triplet pairing has been developed in superfluid $^3$He\cite{OsheroffPRL1972,LeggettRMP1975, WheatleyRMP1975,VollhardtHe3}, where the system is clean and ideally isotropic.
However, in superconductivity on Uranium-based compounds, the electronic state is strongly correlated and highly anisotropic. 
As the background realized in the spin-triplet pairing is quite different between two systems, it is important to recognize the similarity and difference between two systems. 
As for the similarity, we point out the presence of multiple phases in both systems.
It is well known that superfluid $^3$He possesses two phases, named A and B phases, which are tuned with pressure ($P$) and magnetic field ($H$) \cite{OsheroffPRL1972,WheatleyRMP1975,VollhardtHe3}.
The physical properties on the two phases were intensively studied and are well understood at present.
In UTe$_2$, two new SC states have been found. 
One appears when $\mu_0H$ greater than 16 T is applied along the $b$ axis.
This SC state is called high-field superconducting (HFSC) state\cite{KnebelJPSJ2019}.
The other emerges under applied $P$\cite{BraithwaiteCommPhy2019}, and is called SC2 in the paper.
Quite recently, it was reported that these two SC states are identical\cite{VasinaPRL2025}.
From angle dependent measurement of the superconductivity, the HFSC state is more anisotropic than the low-field superconducting (LFSC) state, implying that the properties of the two SC states are different   \cite{RanNatPhy2019,KnebelJPSJ2019,KinjoPRB2023,SakaiPRL2023,WuPNA2024}.
Therefore, the interest is focused on the determination of the pairing state in these SC states.

To clarify the SC pairing state, we have measured the $^{125}$Te-NMR Knight shifts of UTe$_2$ to investigate spin susceptibility in the SC state just after the discovery of superconductivity in UTe$_2$\cite{NakamineJPSJ2019, NakaminePRB2021, NakamineJPSJ2021, FujibayashiJPSJ2022, KinjoPRB2023, MatsumuraJPSJ2023, KinjoSciAdv2023}.
It is well known that the SC pairing state is expressed with the orbital and spin parts.
The former is related to the SC gap structure, and the latter can be known from the measurement of the spin susceptibility.
Until now, a lot of experimental results have been reported, providing information on the SC gap structure\cite{NakamineJPSJ2019, KittakaPRR2020,IshiharaNatComm2023, AokiJPSJ2024,HayesPRX2025}. 
However, the results about the SC gap structure are now under extensive debate, as discussed later.

We consider that the measurement of the spin susceptibility is crucial, because the difference between spin-singlet and triplet superconductivity appears in the spin susceptibility in the SC state. 
The Knight shift probing the static field at the nuclear site is one of the most reliable methods to measure spin susceptibility in the SC state\cite{YosidaPhyRev1958,DEMacLaughlin1976}.
We have measured the Knight shift in a high-quality single crystalline sample with the SC transition temperature  \TSC \ $\sim 2.1$~K, which is regarded as a disorder-free sample from the relationship between \TSC \ and the residual electronic term in the specific heat\cite{aoki2021JPCMrev, AokiJPSJ2024}. 

We revealed from the Knight-shift measurements that the spin susceptibility along all three crystalline axes decreases in the SC state\cite{MatsumuraJPSJ2023,MatsumuraPRB2025}.
In particular, we found a notable reduction in Knight shift along the $a$ axis $K_a$ ($\Delta K_a \sim 4\%$) in the SC state, although the large upper critical field $\mu_0 H_{\mathrm{c2}} \sim 12$ T was observed.
The large reduction in $K_a$ is more than 20 times larger than that in $K_b$ and $K_c$ but is still 14\% of the normal-state Knight-shift value. 
As discussed in the paper\cite{MatsumuraJPSJ2023}, such a larger reduction in $K_a$ in the SC state should give rise to the Pauli-depairing effect, if UTe$_2$ were a spin-singlet superconductor.
Therefore, we pointed out that the absence of the Pauli-depairing effect and the large upper critical field exclude the spin-singlet SC state quantitatively but are consistent with the spin-triplet SC state with the alignment of the SC spin by magnetic fields.
In addition, we suggested from the decrease in spin susceptibility in the SC state that the \dv-vector has all three-axis components\cite{MatsumuraJPSJ2023, MatsumuraPRB2025}. 
Here, the \dv\ vector corresponding to the direction of zero spin projection \cite{LeggettRMP1975} is the order parameter in the spin triplet pairing.

As UTe$_2$ has an orthorhombic crystal structure, an odd-parity $p$-wave order parameter with $D_{2h}$ point group in zero field is classified into one of the four irreducible representations shown in Table I.
When the higher-order pairing state is considered, the $B_{iu}$ ($i$ = 1, 2, 3) state includes a higher-order term of $k_ak_bk_c\hat{d}_{i}$ ($\hat{d}_i$ = $\hat{c}$, $\hat{b}$, $\hat{a}$).
When a magnetic field ($H$) is applied e.g, for $H \parallel a, b,$ or $c$, the symmetry decreases from $D_{2h}$ to $C_{2h}$ and induces mixing between the zero-field order parameters. 
Under $C_{2h}$, the odd-parity representations are described by $A_u$ and $B_u$ as shown in Table II\cite{IshizukaPRL2019}.
In the intermixing state, the \dv-vector components can be flexibly changed by applied $H$, which is a characteristic feature of multicomponent SC states.

From the viewpoint of the \dv-vector components, it seems that the larger reduction in the spin susceptibility along the magnetic easy $a$ axis ($K_a$) would be consistent with the odd-parity $A_{u}$ state with the three-axis \dv -vector component\cite{IshizukaPRL2019}, which is one of the irreducible presentations of the point group symmetry in the $D_{2h}$ crystal symmetry.
This pairing state is similar to the B phase in superfluid $^3$He.
However, if the pairing state is extended to the higher-order, the other state becomes possible, as discussed later.

An interesting and fundamental question in spin-triplet superconductors is how the spin of a triplet pair responses against the applied fields, and correspondingly whether the SC properties are affected depending on the spin directions.
Although the spin state of triplet superconductors has been investigated\cite{HeffnerPRL1989,MatsunoJPCP2017,TouPRL1996,TouPRL1998,GannonPRB2012,GannonPRB2017,AoyamaJPSJ2019,YangScAd2021}, the relationship between spin state in the SC state and SC properties has not been clarified.
As for the investigation on the SC properties, it is crucially important to use high-quality samples.
After intensive NMR investigation on the early-stage sample with \TSC\ $\sim$ 1.6 K, we realized that the intrinsic SC properties are covered by the magnetism induced with U deficiency\cite{NakamineJPSJ2019,NakaminePRB2021,NakamineJPSJ2021}. 
In this paper, we report the results of the Knight-shift and ac magnetic susceptibility ($\chi_{ac}$) measurements in $H \parallel b$ and $H \parallel c$ in the wide field range up to 24 T on the high-quality UTe$_2$ with \TSC\ $\sim$ 2.1 K .

\begin{table}[htb]
\begin{center}
\caption{\label{t1}Classification of the odd-parity SC order parameters for point groups with $D_{\rm 2h}$ in zero field. Irreducible representation (IR) and its basis functions are listed. Dominant spin component in the SC state is also shown. The node structure in the SC gap is shown. N(FG) means the nodeless full gap, and P($k_c$) means point nodes at the $k_c$ point. When the higher-order pairing state is considered, the $B_{iu}$ ($i$ = 1, 2, 3) state includes the higher-order term of $k_ak_bk_c\hat{\bm{d}}_i$ ($\hat{d}_i$ = $\hat{c}$, $\hat{b}$, $\hat{a}$).\cite{IshizukaPRL2019, aoki2021JPCMrev}}
\vspace{3mm}
  \begin{tabular}{cccc}\hline \hline 
 \multicolumn{1}{c}{$D_{\rm 2h}$ (zero field)} \\ \hline 
IR & Basis functions &  SC spin comp. & Node \\ \hline
$A_{\rm u}$ & $k_a \hat{a}$, $k_b \hat{b}$, $k_c \hat{c}$ & &  N (FG) \\
$B_{\rm 1u}$ & $k_b \hat{a}$, $k_a \hat{b}$  & $c$ &  P ($k_c$)\\
$B_{\rm 2u}$ & $k_a \hat{c}$, $k_c \hat{a}$  & $b$ &  P ($k_b$)\\
$B_{\rm 3u}$ & $k_c \hat{b}$, $k_b \hat{c}$  & $a$ &  P ($k_a$)\\ \hline \hline
  \end{tabular}
  \end{center}
\end{table}
\begin{table}[htb]
\begin{center}
\caption{\label{t2}Classification of odd-parity superconducting phases in UTe$_2$ under a magnetic field along each axis\cite{IshizukaPRL2019}. 
Typical order parameters belonging to each IR are listed in Table \ref{t1}.}
\vspace{3mm}
  \begin{tabular}{cccc}\hline \hline 
 \multicolumn{1}{c}{IR of $C_{\rm 2h}$ (under field)} \\ \hline 
$H$ direction         & $A_{\rm u}$                & &$B_{\rm u}$               \\ \hline
$H \parallel c$ & $A_{\rm u}$ +$B_{\rm 1u}$ &  &$B_{\rm 2u} + B_{\rm 3u}$ \\
$H \parallel b$ & $A_{\rm u}$ +$B_{\rm 2u}$ &  &$B_{\rm 3u} + B_{\rm 1u}$ \\ 
$H \parallel a$ & $A_{\rm u}$ +$B_{\rm 3u}$ &  &$B_{\rm 1u} + B_{\rm 2u}$ \\ \hline \hline
  \end{tabular}
  \end{center}
\end{table}


\section{Experimental Method}
The sample we utilized was a $^{125}$Te-enriched high-quality sample with the size of $3 \times 1 \times 0.5$ mm$^3$, which was prepared with the newly developed molten salt flux method\cite{Sakai2022PRM} with natural uranium U and 99.9~\% $^{125}$Te-enriched metals for starting elements. 
This sample was used in the previous measurements\cite{MatsumuraJPSJ2023, MatsumuraPRB2025}.
Single crystals under an optimized growth condition with excess U exhibit a sharp SC transition at $T_\mathrm{c}$ = 2.06~K, which was determined with measurements of specific heat and ac susceptibility.
The characterization of the sample was described in a supplemental material of the previous paper\cite{MatsumuraJPSJ2023}.
The NMR spectra as a function of frequency were recorded using the Fourier transform of a spin-echo signal observed after a radio-frequency (RF) pulse sequence at a fixed magnetic field. 
The magnetic field was calibrated using a $^{65}$Cu (gyromagnetic ratio $^{65}\gamma /2\pi = 12.089$ MHz/T, $K = 0.2385$~\%)-NMR signal from a Cu coil\cite{GCCarter1976}. 
We observed two $^{125}$Te-NMR signals ($^{125}\gamma/2\pi$ = 13.454 MHz/T, and nuclear spin $I$ = 1/2), when $H$ is applied to the $b$ or $c$ axis, respectively, since there are two crystallographically distinct Te sites in UTe$_2$.
Following the previous papers, we call the $^{125}$Te-NMR signal with the smaller [larger] Knight shift value Te(I) [Te(II)] signal in $H \parallel b$, and this relationship is reversed in $H \parallel c$ \cite{TokunagaJPSJ2019,NakamineJPSJ2019,NakaminePRB2021,FujibayashiJPSJ2023}.
This was confirmed from the angle dependence of the $^{125}$Te-NMR spectrum in the $bc$ plane\cite{NakaminePRB2021}.
Meissner signal was detected with the change in ac magnetic susceptibility defined as $\Delta \chi_{\rm ac} \equiv -\Delta f$ in the resonance frequency of the NMR tank circuit. 
Low-temperature NMR and $\Delta \chi_{\rm ac}$ measurements below 15.5 T were performed using a $^3$He / $^4$He dilution refrigerator with an uniaxial piezoelectric rotator (ANRv51/ULT/RES+, attocube) down to $\sim 55$ mK.
$^{125}$Te NMR measurements above 15.5 T were performed using a 25-T cryogen-free SC magnet in the High-Field Laboratory for Superconducting Materials at the Institute for Materials Research at Tohoku University.
In the High-Field Laboratory, NMR and $\Delta \chi_{\rm ac}$ measurements down to $\sim 0.5$ K were performed using $^3$He cryostat with a double-axis rotator.
The sample was immersed into the $^3$He / $^4$He mixture or $^3$He, and was rotated to apply the magnetic field $H$ precisely parallel to the $b$ or $c$ axis, respectively.
In the SC state, the heat-up effect by the NMR RF pulses was checked, as shown in Appendix I.
The energy of the RF pulses was reduced to ensure that the NMR results were unchanged by the power of the RF pulses.
 
\section{Results}
\begin{figure*}[tbp]
\begin{center}
\includegraphics[width=15cm]{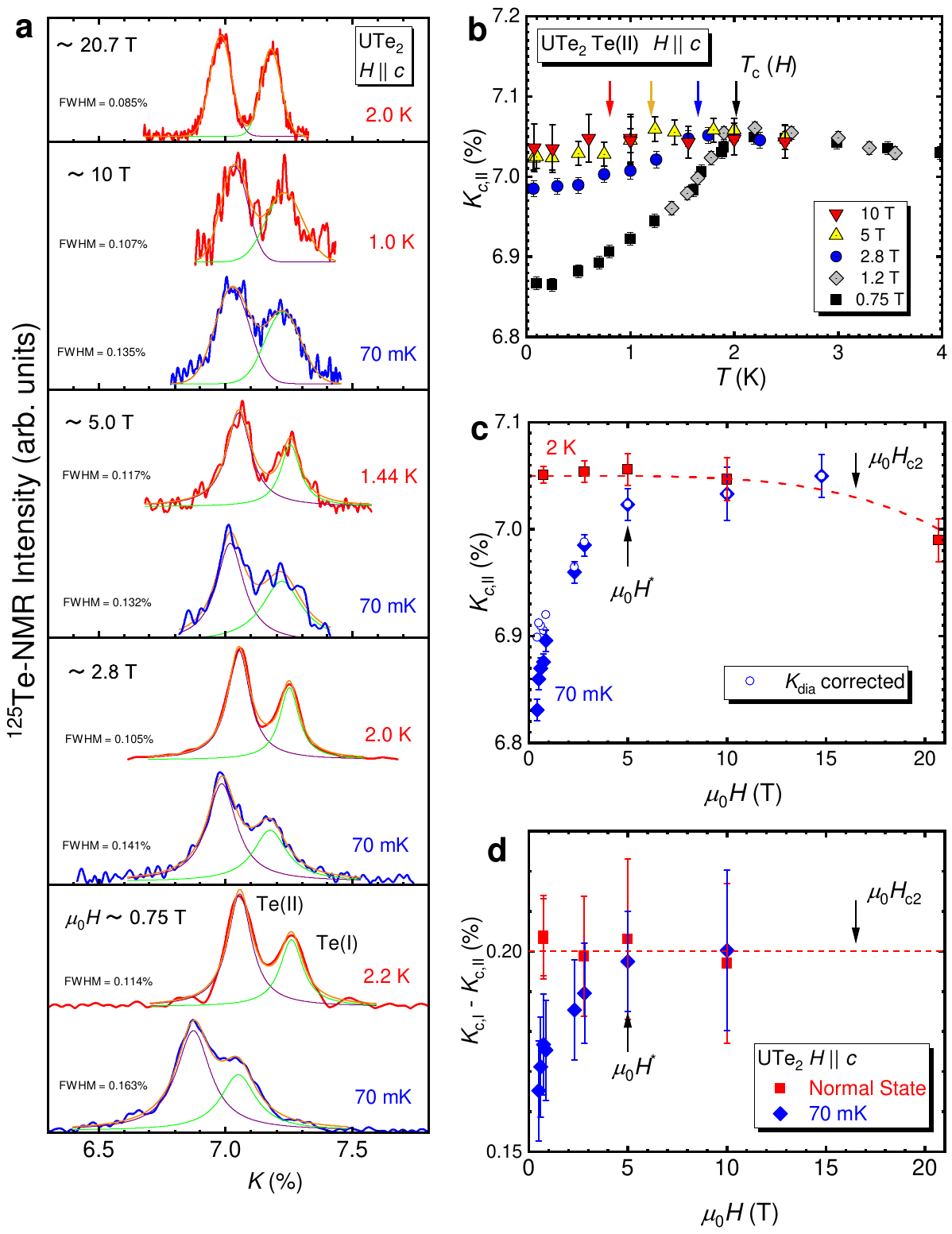}
\end{center}
\caption{(Color online) \textbf{a} Te(II) and Te(I) NMR spectra measured under various $H$ at 70 mK and the normal state are shown against $K$. To determine the Te(II) and Te(I) peak positions, the double-peak spectrum was fitted with the two Lorentian functions and $K_{c,{\rm II}}$ (purple peak) and $K_{c,{\rm I}}$ (green peak) were evaluated. Full Width at Half Maximum (FWHM) of the Te(II) peak estimated with the fitting is shown. \textbf{b} Temperature dependence of the Knight shift measured at the Te(II)-NMR peak $K_{c,{\rm II}}$ under various magnetic fields in $H \parallel c$. The arrows shows the SC transition temperature under $H$, $T_c$($H$). \textbf{c} $H$ dependence of $K_{c,{\rm II}}$ measured at 70 mK and 2 K. The open circles indicates $K_{c}^{\rm dia}$ corrected $K_{c,{\rm II}}$ at 70 mK. The dotted curve shows the eye-guide of the $H$ dependence of $K_{c,{\rm II}}$ at 2 K. The SC upper-critical field $H_{\rm c2}$ is shown by arrow. The characteristic field $H^*$ determined in Fig.~\ref{f3} is shown.     \textbf{d} $H$ dependence of the difference between $K_{c,{\rm II}}$ and $K_{c,{\rm I}}$ measured at $T \sim 70$ mK and normal state in $H \parallel c$. The dotted line shows the average of $K_{c,{\rm II}} - K_{c,{\rm I}}$ measured at 2 K.  }
\label{f1}
\end{figure*}
\subsection{$H \parallel c$}
We observed two $^{125}$Te-NMR signals when $H$ is applied to the $b$ or $c$ axis, since there are two crystallographically distinct Te sites in UTe$_2$. 
Figure \ref{f1} \textbf{a} shows the Te(II) and Te(I) NMR spectra measured under various $H$ in $H \parallel c$ at 70 mK and the normal state, which are plotted against $K$. 
To determine the peak positions of Te (II) and Te (I), the double-peak spectrum was fitted with two Lorentzian or Gaussian functions, and $K_{c,{\rm II}}$ (purple peak) and $K_{c,{\rm I}}$ (green peak) were evaluated. 

Figure \ref{f1} \textbf{b} shows the temperature dependence of $K_{c,{\rm II}}$ measured at various magnetic fields.   
In the normal state, $K_{c,{\rm II}}$ gradually increases with decreasing $T$, but is almost constant below 3 K to \TSC. 
The value of $K_{c,{\rm II}}$ at 2 K is almost constant with $\sim$ 7.05\% up to 10 T.   
In the SC state, $K_{c,{\rm II}}$ in 0.75 T decreases appreciably below \TSC, but the decrease in $K_{c,{\rm II}}$ $\sim$ 0.2\% is much smaller than the normal-state value.   
The decrease in the SC state is significantly suppressed in the small-field region with increasing applied field, as seen in Fig.~\ref{f1} \textbf{b}.
This is consistent with our previous result obtained in the 1.6 K sample\cite{NakaminePRB2021}.
It is noted that the decrease in $K_{c,{\rm II}}$ cannot be interpreted solely by the SC diamagnetic effect, as discussed in the previous paper\cite{MatsumuraPRB2025}, but indicates the decrease in spin susceptibility in the SC state.  
To check the $H$ dependence of the $K_{c,{\rm II}}$ decrease in the SC state, $K_{c,{\rm II}}$ at 70 mK and 2 K is plotted against $H$ in Fig.~\ref{f1} \textbf{c}. 
It was revealed that the $K_{c,{\rm II}}$ decrease is rapidly suppressed by applying $H$ in the low field range, and is almost recovered already at around 5 T, far below $H_{\rm c2}$.
The recovery of $K_{c,{\rm II}}$ against $H$ is directly compared with that of the residual Sommerfeld coefficient $\gamma_0(H)$ reported by Lee {\it et al.}\cite{LeePRR2025} in Appendix II.
The recovery of the Knight shift is significantly faster than that of $\gamma_0(H)$.
Figure \ref{f1} \textbf{d} shows the $H$ dependence of the Knight-shift difference between at the Te(I) and Te(II) sites in the normal state and 70 mK. 

Here, we analyze the experimental results following our previous paper\cite{MatsumuraPRB2025}.
The total Knight shift along the $\alpha$ axis ($\alpha$ = $b$ or $c$ in the paper) at the Te($i$) ($i$ = I or II) site $ K_{\alpha, i}$ in the SC state is given by  
\[
K_{\alpha, i}(T, H) = K_{\alpha, i}^{\rm spin}(T, H) + K_{\alpha,i}^{\rm orb} + K_{\alpha}^{\rm dia}(T, H). 
\]
$K_{\alpha, i}^{\rm spin}(T)$ is the spin part of the Knight shift, which is coupled to spin susceptibility $\chi^{\rm spin}$ with the relation of $K_{\alpha, i}^{\rm spin}(T, H) = A_{\alpha, i} \chi^{\rm spin}(T, H)$.
Here, $A_{\alpha, i}$ is the hyperfine coupling constant along the $\alpha$ axis at the Te($i$) site.
$K_{\alpha, i}^{\rm orb}$ is so-called ``orbital shift", which results from the field-induced orbital electron magnetism and is considered to be independent of $T$ and $H$ at low temperatures. 
$K_{\alpha}^{\rm dia}(T, H)$ is the shift originating from bulk SC diamagnetism, which results from the Meissner shielding effect and is reasonably assumed to work at both sites in the same manner, and thus site independent.
This gives the fractional change in the average internal field in the sample.

In the following, $K_{\alpha}^{\rm dia}(T, H)$ is considered in two ways.
First, by taking the difference between $K_{c, {\rm I}}$ and $K_{c, {\rm II}}$, $K_{c}^{\rm dia}(T, H)$ can be experimentally canceled out as follows,
\begin{equation*}
K_{c, \rm{I}} - K_{c, \rm{II}}  =  (A_{c,\rm{I}} - A_{c, \rm{II}})\chi^{\rm spin}(T, H) +const. .
\end{equation*} 
Figure~\ref{f1} \textbf{d} shows the $H$ dependence of $K_{c, \rm{I}} - K_{c, \rm{II}}$ measured at the same temperature ($T \sim 70$ mK and normal state).
Since $(A_{c,\rm{I}} - A_{c, \rm{II}})$ is $H$ independent in general, the figure indicates that the decrease in spin susceptibility in the SC state rapidly recovers with increasing $H$ and becomes almost the normal-state values already around 5 T far below $H_{c2}$.
This qualitative trend cannot be altered by any quantitative analysis.

Next, $K_{\alpha}^{\rm dia}(T, H)$ is estimated with the theoretical equation\cite{deGennes} with the SC parameters reported experimentally\cite{IshiharaNatComm2023,IshiharaPRR2023}.
The Knight shift due to the SC diamagnetism along the $\alpha$ ($\alpha$ = $b$, or $c$) axis is expressed as\cite{deGennes}, 
\begin{eqnarray*} \label{eq1}
K_{\alpha}^{\rm dia} &=& - \frac{H_{{\rm c1}, \alpha}}{H} \frac{\ln(\frac{\beta \lambda_d}{\sqrt{2.7} \xi})}{\ln{\kappa_{\alpha}}} \\
 & =&- \frac{H_{{\rm c2}, \alpha}}{H} \frac{1}{2\kappa_{\alpha}^2} \ln\left[0.62\left(\frac{H_{{\rm c2}, \alpha}}{H}\right)^{0.5}\right].
\end{eqnarray*}
Here, $H_{c1}$ and $\xi$ are the lower SC critical field and Ginzburg-Landau (GL) coherence length. 
$\beta$ is a factor that depends on the vortex structure and is 0.38 for the triangular vortex lattice, $\lambda_d$ is the distance between the vortices and is calculated using the relation $\phi_0 = \frac{\sqrt{3}}{2} \lambda_d^2(\mu_0H_{\rm ext})$, and $\kappa_{\alpha}$ is the GL parameter along the $\alpha$ direction.
Using the definition of $H_{{\rm c1},\alpha}$, $H_{{\rm c2}, \alpha}$ and $\kappa_{\alpha}$, $K_{\alpha}^{\rm dia}(T, H)$ is expressed with the parameter $\kappa_{\alpha}$, as expressed in the second line of the above equation, since $\mu_0 H_{{\rm c2}, \alpha}$ was determined from the measurements ($\mu_0 H_{{\rm c2}, b}$ of the low-field SC phase $\sim $ 25 T, $\mu_0 H_{{\rm c2}, c} \sim $ 16.5 T). 
The open circles in Fig.~\ref{f1} \textbf{c} indicate the $K_{c}^{\rm dia}$ corrected Knight-shift result at 70 mK, where the GL parameter along the $c$ axis $\kappa_{c}$ is adopted to be 200 so as to give a smooth variation of the corrected Knight-shift result in $H \parallel c$.
Since the normal-state $K_{c,{\rm II}}$ at 2 K is almost independent of $H$ up to 10 T as seen in Fig.~\ref{f1} \textbf{b} and \ref{f1} \textbf{c}, the $H$ dependence of the difference between $K_{c}^{\rm dia}$-corrected (subtracted) $K_{c,{\rm II}}$ at 70 mK and $K_{c, {\rm II}}$ at 2 K corresponds to $H$ variation of the decrease in the $c$-axis spin-part of Knight shift $\Delta K_c^{\rm spin}$ or spin susceptibility $\Delta \chi^{\rm spin}_c$ at 70 mK.
The rapid recovery in $\Delta K_c^{\rm spin}$ in the small $H$ region is consistent with the above $K_{c, \rm{I}}(H) - K_{c, \rm{II}}(H)$ result in Fig. \ref{f1} \textbf{d}.

\begin{figure}[tbp]
\begin{center}
\includegraphics[width=8cm]{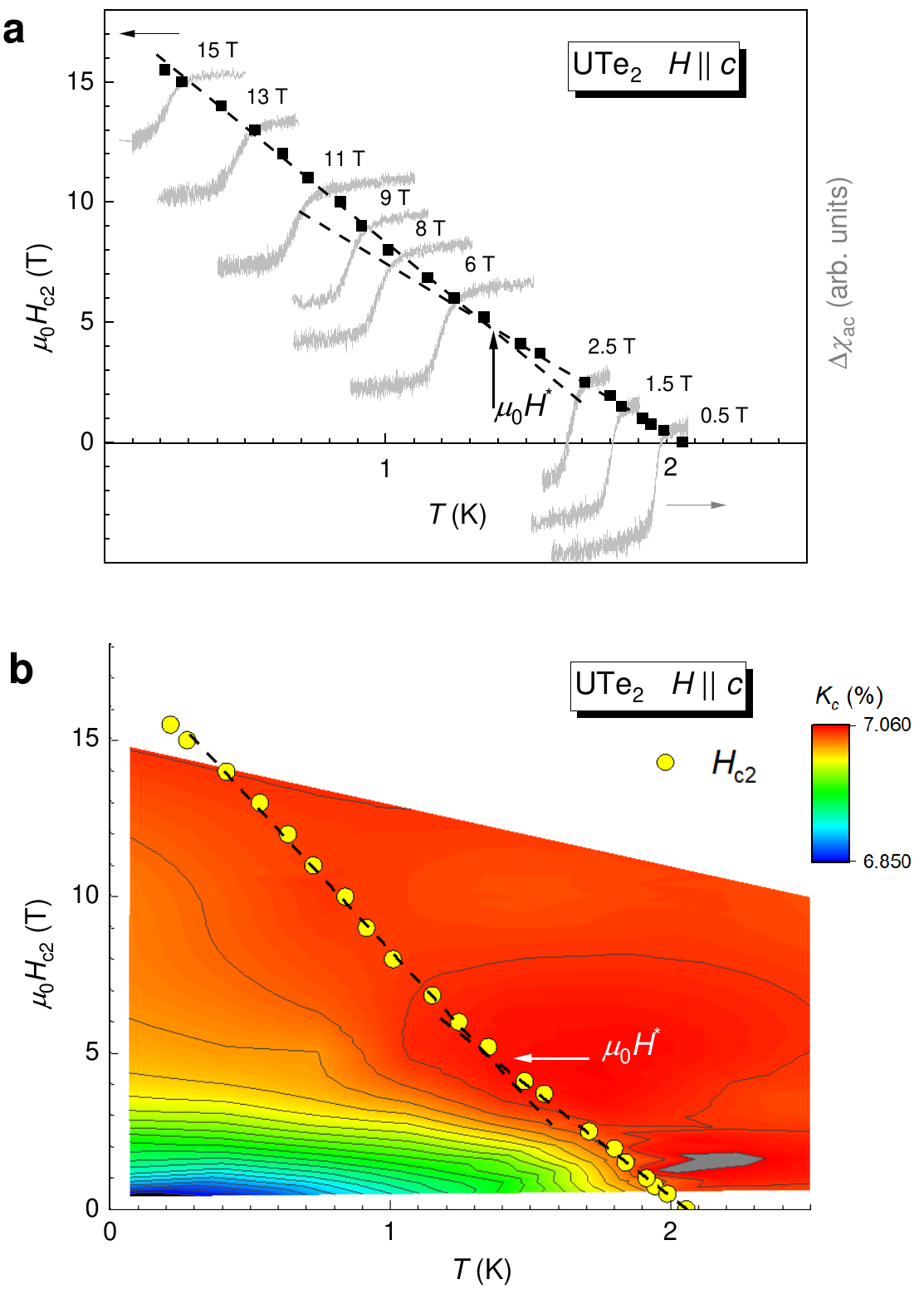}
\end{center}
\caption{(Color online) \textbf{a} Temperature dependence of the SC upper critical field in $H \parallel c$, $H_{c2}^{H \parallel c}$ is plotted to the left axis. In the figure, $\Delta \chi_{\rm ac}$ corresponding to the Meissner signals measured at various fields are also shown. The magnetic field, where the $T$ dependence of $H_{c2}$ is changed, is denoted as $H^*$. The kink becomes clearer when two lines are drawn. The slope change around 13 T is a conventional suppression of superconductivity near $H_{\rm c2}$ seen in the WHH model\cite{WHH}, thus the high-field region is not included in the linear fitting.  \textbf{b} the relationship between the Knight shift and $H_{c2}$ along the $c$ axis. The temperature dependence of $\mu_0H_{c2}(T)$ is overlaid on a color contour map of the $K_c(H,T)$ in the $\mu_0H$–$T$ plane.
  }
\label{f2}
\end{figure}
Figure \ref{f2} \textbf{a} shows the $H$ - $T$ phase diagram of the superconductivity for $H \parallel c$ determined with the ac magnetic susceptibility $\Delta \chi_{\rm ac}$ measurements.
The magnitude of $\Delta \chi_{\rm ac}$ in the SC state is almost unchanged up to the maximum field of measurement (15.5 T), indicating bulk superconductivity up to $H_{\rm c2}$.
$H_{c2}$ linearly increases with decreasing $T$, but if we carefully check the behavior of $H_{c2}$, we notice that the slope of $H_{c2}$ becomes steeper around $\mu_0 H^* \sim 5$ T. 
This trend was also observed in the results in the 2.1 K sample by other groups\cite{TokiwaPRL2025, LeePRR2025}, but was not observed in the 1.6 K sample, in which a small amount of U deficiency was reported\cite{Haga2022JPCM}.
Figure \ref{f2} \textbf{b} shows the relationship between the Knight shift and $H_{c2}$ along the $c$ axis by contour plot. 
The figure represents the correspondence between the recovery of the spin susceptibility and the enhancement of superconductivity.

\subsection{$H \parallel b$}
\begin{figure}
\begin{center}
\includegraphics[width=8cm]{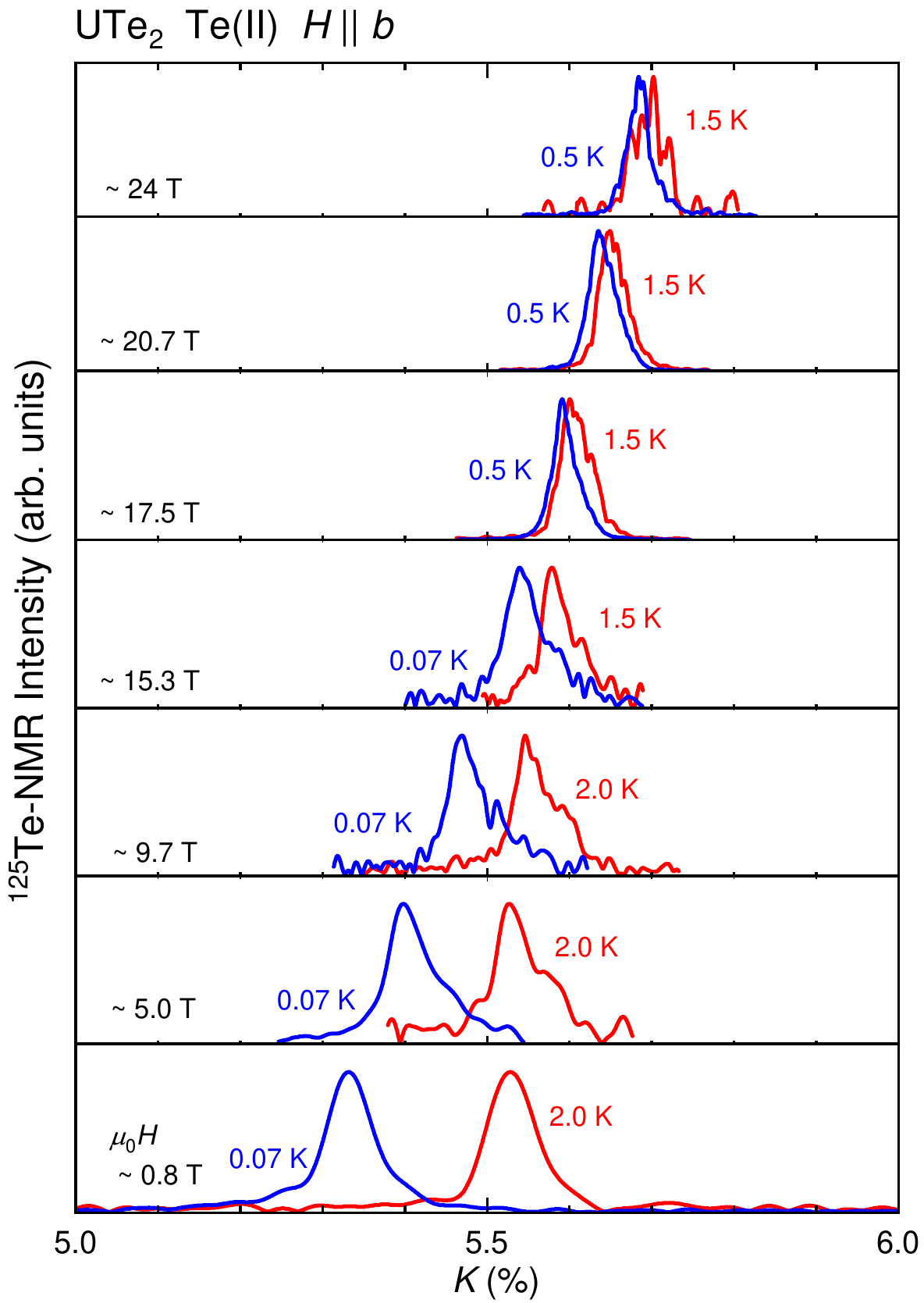}
\end{center}
\caption{(Color online) Te(II) NMR spectra measured under various $H$ in the SC state (0.07 K or 0.5 K) and the normal state (2.0 or 1.5 K). The NMR spectra above 15 T were measured at the High-Field Laboratory for Superconducting Materials at the Institute for Materials Research at Tohoku University.} 
\label{f3}
\end{figure}
\begin{figure}
\begin{center}
\includegraphics[width=8cm]{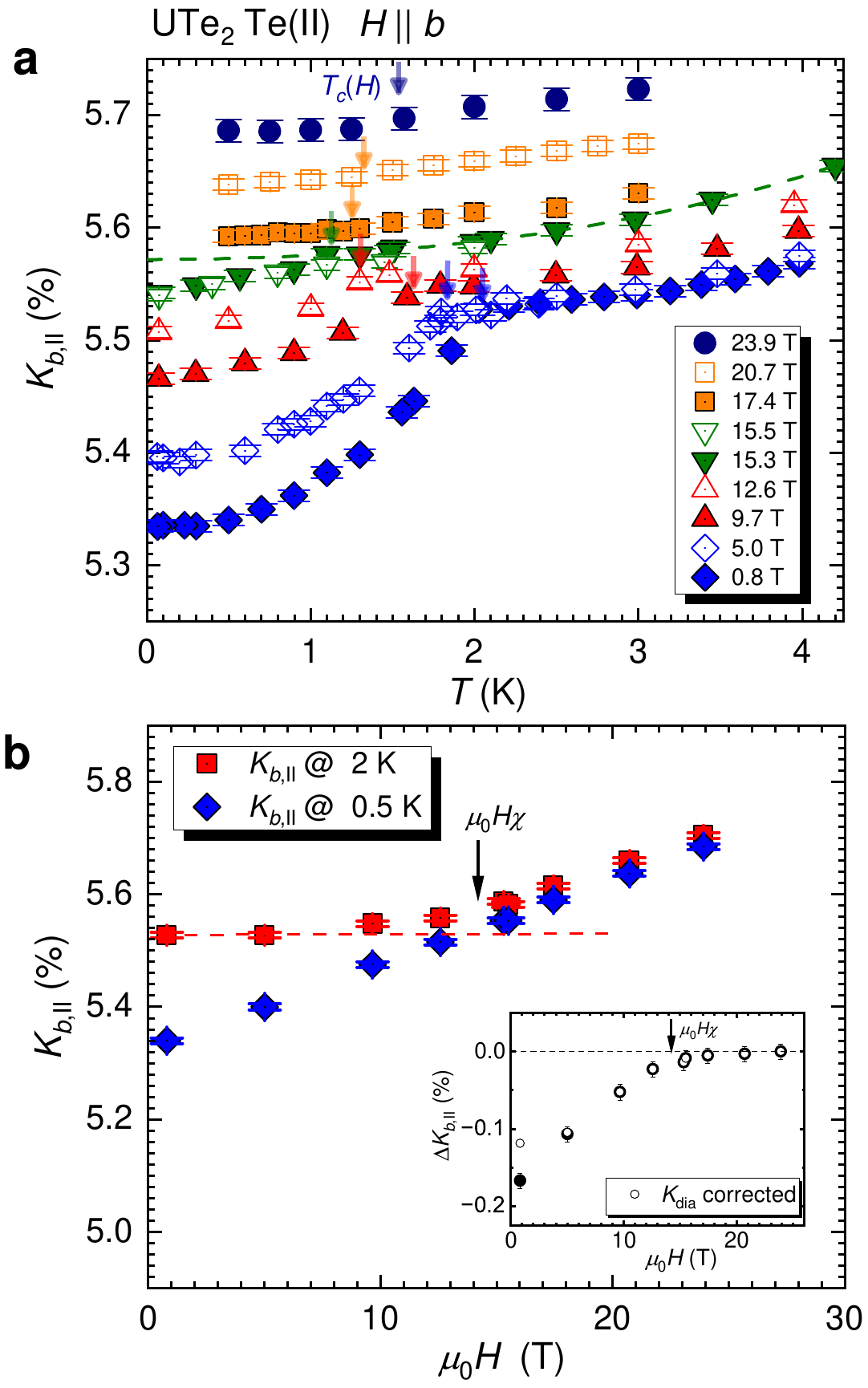}
\end{center}
\caption{(Color online) \textbf{a} Temperature dependence of the Knight shift measured at the Te(II)-NMR peak under various fields in $H \parallel b$, $K_{b,{\rm II}}$. The arrows indicates $T_c(H)$ determined from the $\Delta \chi_{\rm ac}$ measurements. 
The dotted curve shows the fitting of the temperature dependence of $K_{b,{\rm II}}$ in the normal-state data at 15.3 T with the quartic function, which is extrapolate to $T = 0$ so as to estimate the $K_{b,{\rm II}}$  decrease in the SC state. 
\textbf{b} Field dependence of $K_{b,{\rm II}}$ measured at 0.5 K and 2 K. The horizontal dotted line shows the value of the normal-state $K_{b,{\rm II}}$ in low-$H$ region, where $K_{b,{\rm II}}$ is almost constant with $H$. The magnetic field of $\mu_0H_{\chi}$ determined in Fig.~\ref{f5} \textbf{b} is shown by an arrow. The inset shows the decrease in $K_{b,{\rm II}}$ ascribed to the SC transition $\Delta K_{b,{\rm II}}$. The open circles indicates $K_{b}^{\rm dia}$ corrected $K_{b,{\rm II}}$, which corresponds to the decrease in the spin-part Knight shift in the SC state at 0.5 K.  }
\label{f4}
\end{figure}
We move on to the results on $H \parallel b$.
Figure \ref{f3} shows Te(II) NMR spectra measured in the SC state (0.07 K or 0.5 K) and the normal state (2.0 or 1.5 K) under various $H$ up to 24 T.
Figure \ref{f4} \textbf{a} shows the $T$ dependence of the Knight-shift of Te(II) ($K_{b,{\rm II}}$), which is determined at the peak of the spectrum shown in Fig.~\ref{f3}.   
In all magnetic fields, the normal state $K_{b,{\rm II}}$ gradually decreases with decreasing $T$ below 4 K. 
As a representative example and a visual guide, a dotted curve shows the fitting of the $T$ dependence of $K_{b,{\rm II}}$ in the normal-state data at 15.3 T with the quartic function, which is extrapolate to $T = 0$, in Fig.~\ref{f4} \textbf{a}.
The decrease is related to the maxmum in magnetic susceptibility observed at around $T_{\chi{\rm max}} \sim 40$ K, below which the magnetic susceptibility and Knight shift gradually decrease.
The normal-state $K_{b,{\rm II}}$ is not only $T$ but also $H$ dependent.
The $H$ dependence is noticeable above 5 T, and becomes remarkable above 15 T with preservation almost the same $T$ dependence up to 24 T. 

In the SC state, $K_{b,{\rm II}}$ decreases sharply below $T_{\rm c}$ in 0.8 T, as observed in $K_{c,{\rm II}}$ in 0.75 T.
However, the recovery of the decrease with increasing applied field in $H \parallel b$ is weaker than that observed in $H \parallel c$, and $K_{b,{\rm II}}$ decrease in the SC state was observed even in $\mu_0 H = 15.5$ T, which is recognized with the deviation from the dotted curve below \TSC. 
To investigate the variation of $K_{b,{\rm II}}$ against $H$, $K_{b,{\rm II}}$ at 0.5 K and 2 K is plotted in Fig.~\ref{f4} \textbf{b}.
It should be noted that similar increase in the normal state was reported in the electronic term $\gamma_n$ in the specific heat\cite{MiyakeJPSJ2019}.
The result indicates that the normal-state density of states (DOS) increases with applied fields in $H \parallel b$.
Thus, the $K_{b,{\rm II}}$ decrease in the SC state at 0.5 K ($\Delta K_{b, \rm{II}}$) is evaluated with the deviation from the extrapolated normal-state $K_{b,{\rm II}}$, which is plotted by solid circles in the inset of Fig.~\ref{f4} \textbf{b}.
In the same inset, the $K_{b}^{\rm dia}$-corrected (subtracted) $\Delta K_{b, {\rm II}}$ is shown by open circles, which corresponds to the decrease in the $b$-axis spin-susceptibility in the SC state, $\Delta \chi^{\rm spin}_b$. 
Here, $\kappa_b \sim 200$ is adopted as in the case of $ H\parallel c$. 
$\Delta \chi^{\rm spin}_b$ decreases linearly with increasing $H$ above 5 T, and almost zero at $\sim$ 16 T. 
To investigate the relationship between $\Delta \chi^{\rm spin}_b$ and superconductivity, we measured $\Delta \chi_{\rm ac}$ in $H \parallel b$.
 
\begin{figure}
\begin{center}
\includegraphics[width=8.5cm]{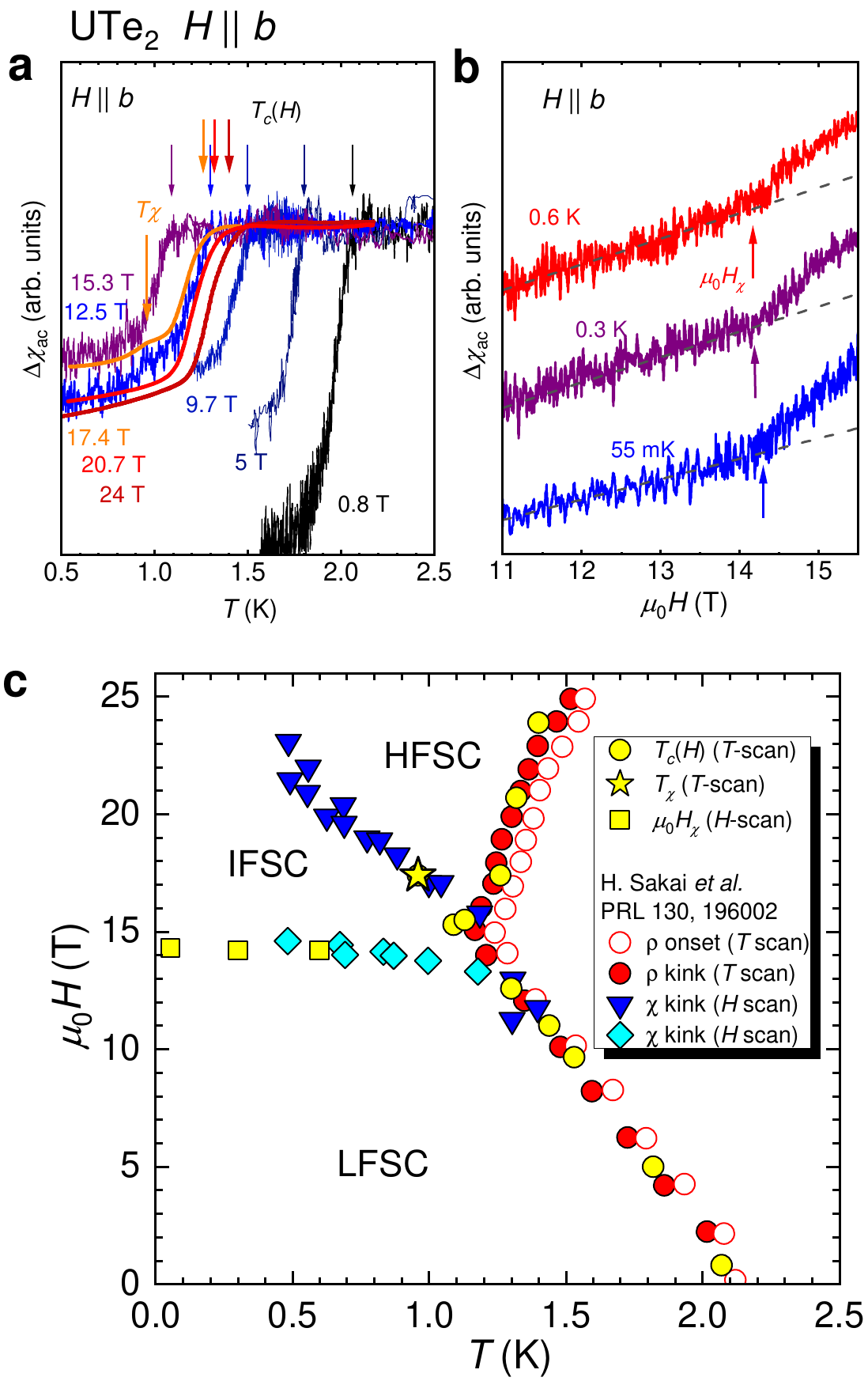}
\end{center}
\caption{(Color online) \textbf{a} Temperature dependence of $\Delta \chi_{\rm ac}$ corresponding to the Meissner signals measured at various $H$ is shown. Arrows show $T_c(H)$ determined with $T$ dependence of $\Delta \chi_{\rm ac}$. The low-$T$ arrow in 17.4 T shows the kink in $\Delta \chi_{\rm ac}$ corresponding to the boundary between HFSC and LFSC. The results above 16 T was obtained at the High-Field Laboratory for Superconducting Materials at the Institute for Materials Research at Tohoku University. 
\textbf{b} $H$ dependence of $\Delta \chi_{\rm ac}$ measured at 0.6, 0.3, and 0.055 K. The arrows show the change in the $H$ dependence of $\Delta \chi_{\rm ac}$. 
\textbf{c} $H$-$T$ phase diagram of UTe$_2$ determined by $\Delta \chi_{\rm ac}$ measurements. The previous results obtained on the same-quality UTe$_2$ with the resistivity and $\Delta \chi_{\rm ac}$ measurements are also plotted\cite{SakaiPRL2023}. LFSC, IFSC, and HFSC indicate the low-field, intermediate-field and high-field superconducting state, respectively.  }
\label{f5}
\end{figure}
Figure \ref{f5} \textbf{a} shows $T$ dependence of $\Delta \chi_{\rm ac}$ corresponding to the Meissner signals measured at various $H$. 
Arrows show $T_c(H)$ determined with $T$ dependence of $\Delta \chi_{\rm ac}$. 
In $\Delta \chi_{\rm ac}$ at 17.4 T, the additional kink was observed below \TSC($H$).
The temperature is denoted as $T_{\chi}$, which corresponds to the SC-SC transition reported previously\cite{RosuelPRX2023,SakaiPRL2023}.
We point out that the similar anomaly in $\Delta \chi_{\rm ac}$ was reported in the SC state under $P$\cite{WuPRL2025}.
This might also suggest the similar pairing state in the HFSC and SC2 states.   
Figure \ref{f5} \textbf{b} shows the $H$ dependence of $\Delta \chi_{\rm ac}$ at low temperatures. 
The arrows show the field where the $H$ dependence of $\Delta \chi_{\rm ac}$ changes.
This $H$ is denoted as $H_{\chi}$, which are plotted on the $H$ - $T$ phase diagram for $H \parallel b$ (Fig.~\ref{f5} \textbf{c}) developed by Sakai {\it et al.}\cite{SakaiPRL2023}.
The data obtained by the two measurements are quite consistent, as the same-quality samples were used.
It was revealed that the anomaly at around 14 T is almost $T$ independent down to 55 mK. 
At present, the origin of this anomaly was not clear, but the anomaly was detected not only by $\Delta \chi_{\rm ac}$ but also by the flux-flow resistance measurements\cite{SakaiPRL2023, TokiwaPRB2023A}.
It is noted that the anomaly field almost coincides with the inflection of $H_{\rm c2}$, above which the HFSC phase appears.
It seems that the LFSC and HFSC phases coexist in the intermediate-field (IF) region. 
We denote the superconductivity observed in the IF region as IFSC.
We discuss the phase diagram and possible SC pairing state in the next section.

\section{Discussions}
\subsection{phase diagram}
\begin{figure}
\begin{center}
\includegraphics[width=8.5cm]{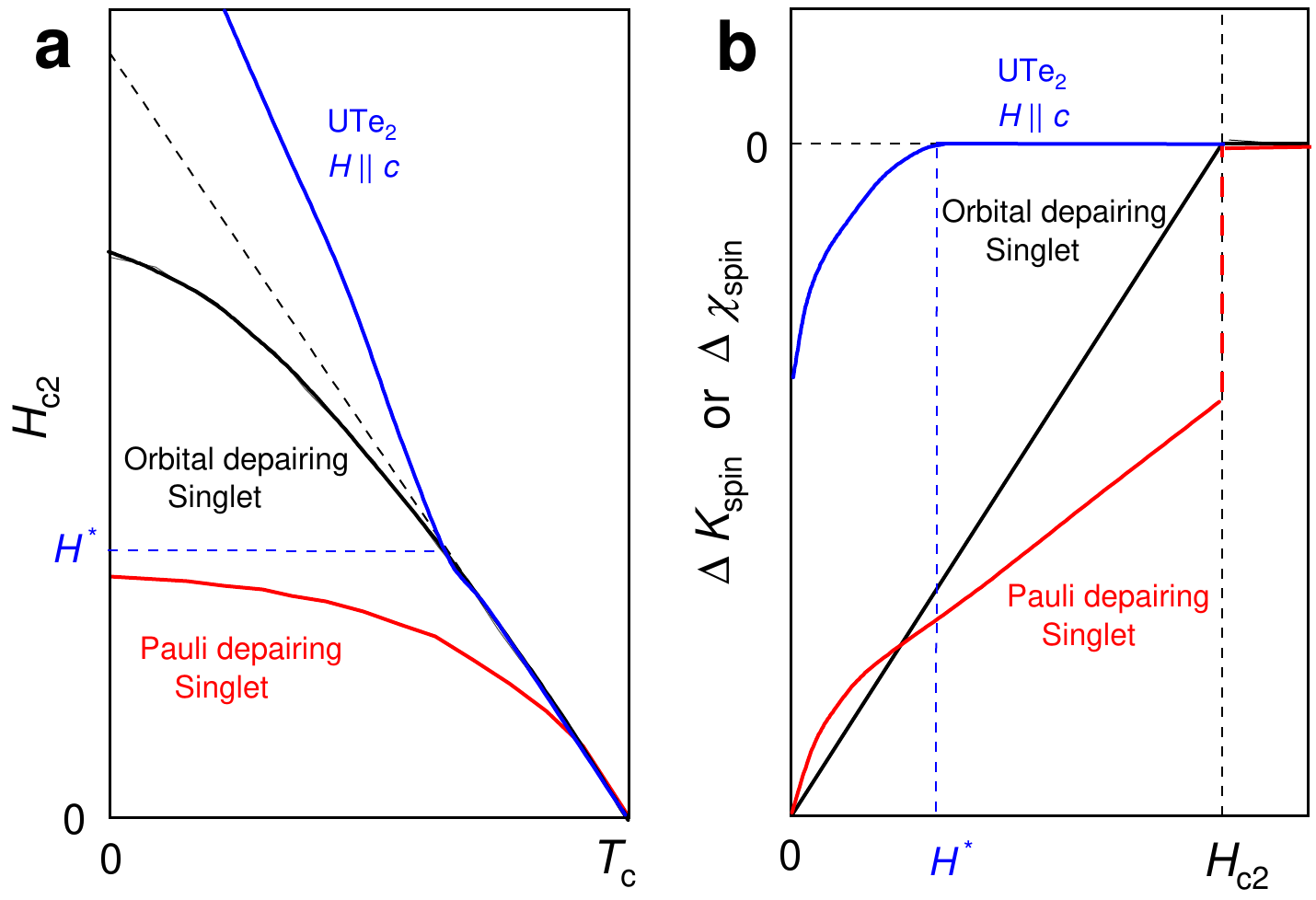}
\end{center}
\caption{(Color online) Sketches illustrating between \textbf{a} superconducting critical fields and \textbf{b} the decrease in spin susceptibility ($\chi_{\rm spin}$) or spin-part of the Knight shift ($K_{\rm spin}$) in spin-singlet superconductors. In the \textbf{a} $H_{\rm c2} - T$ phase diagram, the initial slope at \TSC [$d(\mu_0 H_{\rm c2})/dT$ at $T$ = \TSC] is fixed in all superconductors. In superconductors with orbital depairing, the suppression of superconductivity is second order, and $\Delta K_{\rm spin}$ ($\Delta \chi_{\rm spin}$) is linearly decreased to $H_{\rm c2}$.
Here, $\Delta K_{\rm spin}$ ($\Delta \chi_{\rm spin}$) is defined as the decrease from the normal state value $K_{\rm N}$ ($\chi_{\rm N}$) in the SC state, $\Delta K_{\rm spin} = K_{\rm spin} - K_{\rm N}$ ($\Delta \chi_{\rm spin} = \chi_{\rm spin} - \chi_{\rm N}$).   
On the other hand, in superconductors with Pauli depairing, the suppression of superconductivity is, in most cases, first order and thus the abrupt change is observed at $H_{\rm c2}$.
The behaviors observed in UTe$_2$ ($H \parallel c$) are qualitatively different from those in spin-singlet superconductors.}
\label{f6}
\end{figure}

First, we compare $H_{c2}$ and $\Delta K_{\rm spin}$ observed in UTe$_2$ along the $c$-axis with those in typical spin-singlet superconductors, as shown in Fig.~\ref{f6}.
As for spin singlet superconductors, in the case of the orbital depairing, $\lvert \Delta K_{\rm spin} \lvert$ proportional to $\vert \Delta \chi_{\rm spin} \vert$ decreases linearly toward $H_{\rm c2}$ and smoothly goes to zero at $H_{\rm c2}$\cite{LiPRB2022,KinjoJPSJ2019}, whereas in the case of the Pauli depairing, $\lvert \Delta K_{\rm spin} \lvert$ shows a sudden change at $H_{\rm c2}$\cite{KoutroulakisPRL2008,ChronisterPNAS2021}.
A large response in spin susceptibility in the SC state is not expected by applying much smaller field than $H_{c2}$ in the spin-singlet pairing, and thus the $H$ variation in spin susceptibility in UTe$_2$ would be a characteristic feature of a spin-triplet superconductor.
In addition, when the SC spin component is almost aligned above $\mu_0H^* \sim 5$ T, the slope of $H_{c2}$ in Fig.~\ref{f2} \textbf{b} changes at $H^*$ and becomes steeper, resulting in the robustness of superconductivity against $H$ above $H^*$.
Because $\Delta \chi_{\rm spin} \sim 0$ under $H$ means the destruction of superconductivity in the spin-singlet case, the further enhancement of $H_{\rm c2}$ above $H^*$ cannot be explained within a spin-singlet scenario but is most naturally attributed to spin-triplet superconductivity.
The robustness of the superconductivity when $\Delta \chi_{\rm spin} \sim 0$, indicating that the SC spin (the \dv\ vector) is oriented parallel (perpendicular) to the applied magnetic field, can be understood in terms of the Zeeman energy in a spin-triplet superconducting state. 
In an equal-spin triplet pairing state, the configuration with the \dv\ vector perpendicular to $H$ minimizes the Zeeman energy and is therefore the most stable against magnetic fields.
In particular, the large response in $H \parallel c$ seems to be consistent with the quasi-2-D fermi surface along the $c$ axis\cite{aokiJPSJ2022,EatonNatComm2024}, because the $c$-axis component in \dv -vector would be small due to the small dispersion along $k_c$.  

Next, we discuss the $H$ - $T$ phase diagram in $H \parallel b$. 
The $\Delta K_{b, {\rm II}}$ is almost zero at around 16 T, above which the HFSC phase was observed, as shown in Fig.~\ref{f5} \textbf{c}.
It seems that the field shown with the horizontal line at $\mu_0H_{\chi} \sim 14$ T simply stands for the characteristic field above which two SC states coexist, but this field would also be a characteristic field for the SC pairs, where the spin of the pairs is almost aligned to the applied field direction as seen in the inset of Fig.~\ref{f4} \textbf{b}.
We point out that $H_{\chi}$ is quantitatively consistent with the Pauli-limiting field $H_{\rm P}$, estimated from the relation of 
\begin{equation*}
H_{\rm P} = \frac{H_c}{\sqrt{\Delta \chi_{\rm spin}}} = H_c \sqrt{\frac{A_{b, \rm II }}{\Delta K_{b, {\rm II}}}},
\end{equation*}
where $\Delta \chi_{\rm spin}$ is related to the \dv -vector component along the $b$ axis, 
$\chi_{\rm N} \langle d_{b}^2 / \lvert \bm{d} \rvert^2 \rangle$, 
and $H_c$ and $A_{b, \rm II}$ are SC critical field and hyperfine coupling constant at the Te(II) site along the $b$ axis.
In $H \parallel b$, $\mu_0 H_{\rm P}$ in UTe$_2$ is estimated as $\sim 14.6$ T with $H_c$ = 76.8 mT\cite{AokiJPSJ2024, MatsumuraJPSJ2023, IshiharaPRR2023}, $\Delta K_{b, {\rm II}}$ = 0.11\% and $A_{b, \rm II}$ = 5.18 T / $\mu_{\rm B}$\cite{TokunagaJPSJ2019,AmbikaPRB2022}. 
This field is regarded as the \dv -vector pinning field, above which $\langle d_b^2 \rangle$ = 0  ($\Delta K_b$ = 0) i.e., the SC spin (\dv -vector) is parallel (perpendicular) to the applied $H$. 
It should be noted that an enhancement of superconductivity is observed when the spin of the SC pair is aligned to the magnetic field direction in both $H \parallel b$ and $H \parallel c$.
These results suggest the existence of an intimate relationship between the alignment of the triplet-pair spin and the pairing interactions. 

\subsection{possible SC pairing state}
Next, we discuss possible SC pairing states based on the field dependence of the \dv-vector components.  
Until now, various experiments have provided information about the SC gap structure.
Some experiments suggest a nodeless full gap\cite{MatsumuraJPSJ2023,SuetsuguSciAdv2024}, but most of the results suggest the presence of nodes in the SC gap\cite{AokiJPSJ2024,IshiharaNatComm2023, HayesPRX2025}.
However, according to Table \ref{t1}, the presence of nodes is inconsistent with our Knight-shift result within the $p$-wave state.
To reconcile this discrepancy, one might consider a higher-order pairing state (e.g., $f$-wave), in which odd-parity states belonging to $B_{\rm {iu}}$ representations can, in principle, generate finite \dv -vector components along all three crystallographic axes. 
However, such higher-order contributions are expected to be subleading and would most naturally appear as small additional components. 
Therefore, while alternative odd-parity states with higher-order terms cannot be strictly excluded, the Knight shift results, which show a strongly anisotropic \dv-vector with a dominant $a$-axis component, most straightforwardly favor the $A_{\rm u}$ scenario, which we focus on in the following discussion.
In this case, when $H$ applies along the $b$ or $c$ axis, the $A_{\rm u}$ state can mix with zero-field $B_{\rm 2u}$ or $B_{\rm 1u}$, which has the SC spin component along the applied $H$ direction. 
The gradual decrease in $\Delta K_{\rm spin}$ in the SC state with increasing $H$ can be understood as the decrease in the corresponding \dv\ -vector component, and the boundary between LFSC and IFSC in $H \parallel b$ and $H^*$ in $H \parallel c$ can be a crossover field where the corresponding \dv\ -vector component is almost zero.
On the other hand, in $H \parallel b$, the clear specific-heat anomaly between IFSC and HFSC indicates that the phase transition occurs between the two states.
Although spin-singlet pairing was suggested in HFSC\cite{RosuelPRX2023}, the unchanged Knight shift ($\Delta K_{\rm spin} = 0$) in HFSC seems to be incompatible with spin-singlet pairing, but the $B_{\rm u}^{H \parallel b} = B_{\rm 1u} + B_{\rm 3u}$ state shown in Table \ref{t2} is possible from the point group theory.
In this case, the constraint that the $\hat{b}$ component of the \dv\ vector is almost zero is needed from $\Delta K_{\rm spin} = 0$.

As described in Introduction part, the HFSC state is connected to the $P$-induced superconducting state (SC2). 
Theoretical studies have suggested that the SC state in SC2 belongs to the $B_{\rm 1u}$ representation, driven by the development of antiferromagnetic fluctuations with (0, $\pi$, 0)\cite{TeiPRB2024, HakunoPRB2024}.
Based on this connection, a $B_{\rm 1u}$ pairing state in the HFSC phase for $H \parallel b$ is theoretically plausible. 
A key issue, however, is whether similar antiferromagnetic fluctuations actually develop in the HFSC phase under magnetic fields above 16 T, as they do under applied pressure.
Our NMR relaxation measurements for $H \parallel b$ above 15 T reveal enhanced magnetic fluctuations, but these are predominantly longitudinal fluctuations along the $b$ axis \cite{TokunagaPRL2023}, which appear to be distinct from the antiferromagnetic fluctuations proposed for stabilizing the SC2 phase.
In addition, recent measurements of magnetization suggested that the metamagnetic transition at 34~T is associated with $\bm{Q}=0$ magnetic fluctuations\cite{TokiwaPRL2025}.
Therefore, the realization of a $B_{\rm 1u}$ pairing state in the HFSC phase is not conclusive, although both the SC2 and HFSC states are stabilized by magnetic fluctuations. 
Clarifying the precise character of the magnetic fluctuations responsible for the SC2 phase remains an important experimental issue.

Another important issue concerns the nature of the transitions within the SC multiphase. 
The transition between IFSC and HFSC appears to be second order, as indicated by a clear $\lambda$-type anomaly in the specific heat\cite{RosuelPRX2023}. 
In contrast, the boundary between LFSC and IFSC, which is delineated by ac susceptibility and resistivity\cite{SakaiPRL2023, TokiwaPRB2024}, shows no pronounced thermodynamic signature and is therefore consistent with a crossover. 
The nature of this boundary remains under debate. 
Clarifying the magnetic-fluctuation spectrum in each phase and establishing the order of the boundaries using precision thermodynamic probes (for example, magnetocaloric measurements) are crucial next steps.

Furthermore, we comment on $H$-induced spin susceptibility observed in $H \parallel b$.
As shown in Fig.~\ref{f4} \textbf{c}, the normal-state spin susceptibility along the $b$ axis gradually increases with increasing $H$, suggesting an enhancement of DOS in $H \parallel b$.
Such an increase in DOS in the normal state was not observed in $H \parallel c$ (Fig.~\ref{f1} \textbf{c}) and in $H \parallel a$.
The specific-heat jump at $T_{\rm c}$ in $\mu_0 H \sim$ 16 T just above the emergence of the HFSC phase is much smaller than that observed in LFSC\cite{RosuelPRX2023}.
In addition, it seems that $H_{\rm c2}$ dependence in LFSC is smoothly connected to the boundary between IFSC and HFSC above 16 T. 
These suggest that HFSC would be related to the $H$-induced DOS and be caused by magnetic fluctuations along the $b$ axis, which were detected with our nuclear spin-spin relaxation measurements\cite{TokunagaPRL2023}. 
Since it is considered that LFSC is caused by the intrachain fluctuations along the magnetic easy axis ($a$ axis)\cite{AokiJPSJ2019Rev,IshizukaPRB2021,TeiPRB2024}, the SC character would be different between LFSC and HFSC due to the difference of the magnetic fluctuations that induce superconductivity.  
We point out the possibility that HFSC might occur in the $H$-induced DOS and the magnetic fluctuations along the $b$ axis, and that the clear anomaly between HFSC and IFSC might be due to the difference in the Fermi surfaces and magnetic fluctuations that are related to superconductivity.   

Finally, it is interesting to note an analogy between the SC multiphase in UTe$_2$ and the superfluid multiphase in $^3$He, although many points remain unclear. 
As discussed above, we show that the Knight shift behavior in the LFSC/HFSC phases of UTe$_2$ reflects changes in the spin state of the Cooper pairs, which bears a resemblance to the behavior observed across the B and A phases of superfluid $^3$He, although the nature of the transition between two phases is different: the transition between B and A phase in $^3$He is first order, whereas the transition between the IFSC and HFSC appears to be second order. 
Furthermore, the NMR relaxation measurements reveal that the HFSC phase in UTe$_2$ is stabilized in the presence of strong magnetic fluctuations, drawing an analogy to the A phase of superfluid $^3$He, which is often discussed in terms of paramagnon-mediated pairing interactions\cite{AndersonPRL1973}. 
We believe that this comparison highlights a nontrivial similarity between the two systems\cite{TokunagaPRL2023}, which provides a useful framework for interpreting the field-induced evolution of the pairing state in the complex superconducting phases of UTe$_2$.


In conclusion, we found that the spin of the SC pairing responds immediately to the applied magnetic field and becomes nearly aligned with the field direction at $\mu_0 H \sim$ 5 T well below $H_{\rm c2}$ when $H \parallel c$. 
In contrast, this alignment is much weaker when $H \parallel b$.
This suggests the presence of anisotropic pinning interactions for the SC spins in the triplet pairs. 
We also observed that superconductivity is enhanced when the SC spin is parallel to the applied field, demonstrating a strong relationship between the spin configuration of the triplet pairs and the upper critical field $H_{\rm c2}$.
Such behavior has not been observed in spin-singlet superconductors and likely represents an intrinsic characteristic of spin-triplet superconductivity.
Our findings provide strong evidence for the spin-triplet superconductivity realized in UTe$_2$, and new insight for a spin-triplet superconductor.
We also point out the similarities between superconductivity in UTe$_2$ and superfluid $^3$He, since both systems have multiphase characterized by the different spin states, and the HFSC phase in UTe$_2$ is stabilized in the presence of strong magnetic fluctuations, drawing an analogy to the A phase of superfluid $^3$He.

\section*{Acknowledgments}
The authors would like to thank J-P. Brison, J. Flouquet, S. Fujimoto, J. Ishizuka, W. Knafo, G. Knebel,  K. Machida, Y. Maeno, Y. Matsuda, V. P. Mineev,  M. Shimizu, Y. Yanase, and S. Yonezawa for their valuable inputs in our discussions. 
This work was supported by Grants-in-Aid for Scientific Research (KAKENHI Grant No. JP20KK0061, No. JP20H00130, No. JP21K18600, No. JP22H04933, No. JP22H01168, No. JP23H01124, No. JP23K19022, No. JP23K22439 No. JP23K25821, No. 24K00590, and 25H00609) from the Japan Society for the Promotion of Science, by JST SPRING(Grant No. JPMJSP2110) from Japan Science and Technology Agency, by research support funding from The Kyoto University Foundation, by ISHIZUE 2024 of Kyoto University Research Development Program, by Murata Science and Education Foundation, and by the JGC-S Scholarship Foundation.
Work at IMR Tohoku University was supported under the IMR-GIMRT Program (Proposal Number 202212-HMKPB-0007, 202212-HMKPB-0008, 202212-IRKAC-0041, 202312-HMKPB-0019, 202312-HMKPB-0022, 202312-IRKAC-0007).
In addition, liquid helium is supplied from the Low Temperature and Materials Sciences Division, Agency for Health, Safety and Environment, Kyoto University. 

\vspace{10mm}

\appendix
\section*{Appendix I: Heat-up effect by NMR RF pulses}
\begin{figure}
\vspace{5mm}
\begin{center}
\includegraphics[width=8.5cm]{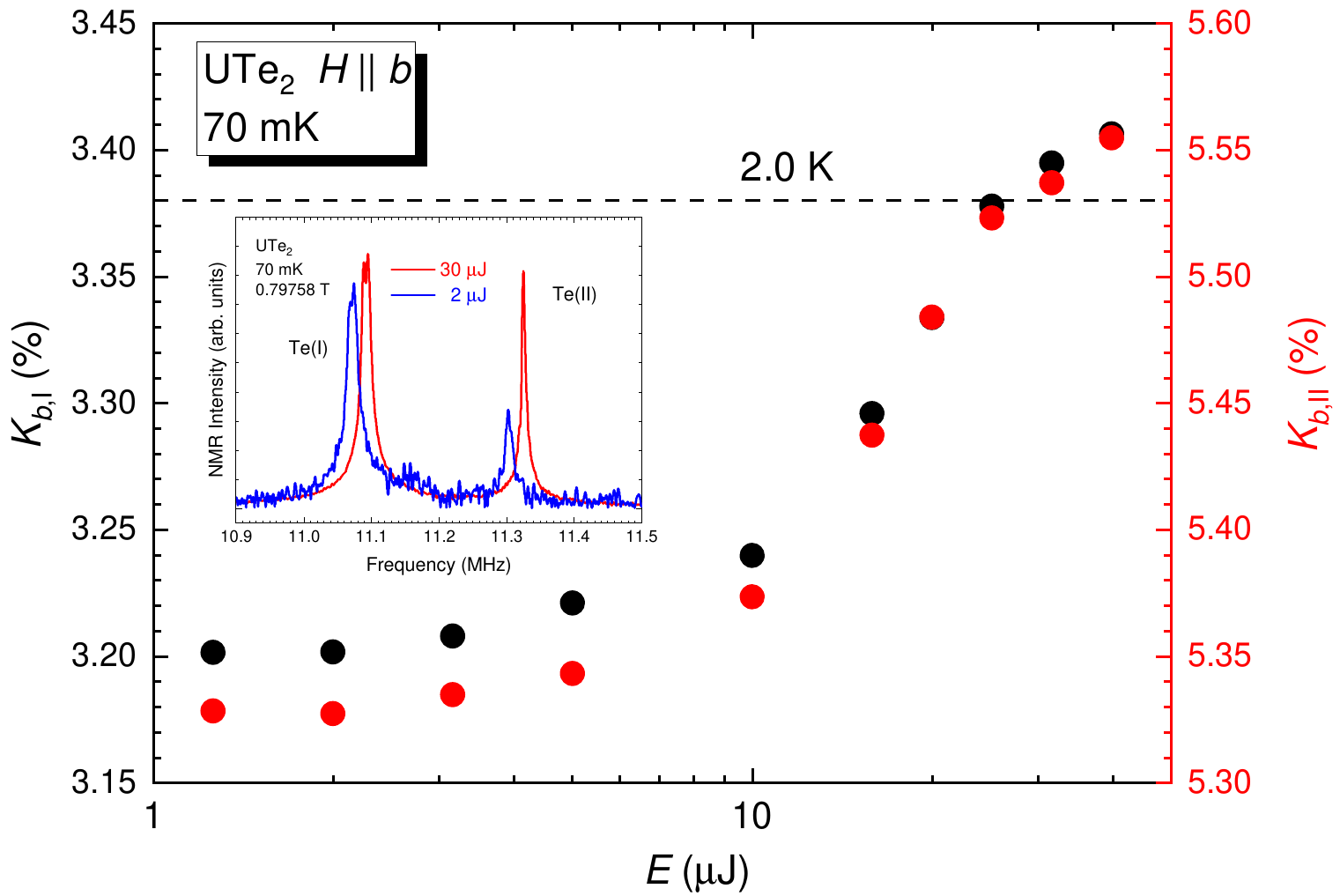}
\end{center}
\caption{(Color online) RF-pulse energy dependence of $K_{b, \rm I}$ and $K_{b, \rm II}$ measured at 70 mK. The inset shows the NMR Free-Induction Decay (FID) spectrum of Te(I) and Te(II) measured with a RF-pulse energy of 30 and 2 $\mu$J. The $K_{b, \rm I}$ and $K_{b, \rm II}$ were determined with a peak of each spectrum. 
The dotted line shows $K_{b, \rm I}$ and $K_{b, \rm II}$ at 2.0 K. }
\label{f7}
\end{figure}

In the present measurement, an NMR spectrometer with a 100W (at 0 dB input) power amplifier (Thamway Product: N146-5049A) was used. 
In general, it is difficult to estimate how much Joule heating occurs after applying an NMR RF pulse.
This is because various factors such as an impedance of the NMR tank circuit and eddy currents induced by the RF pulses should be taken into consideration; thus in the heat-up test, the energy of the RF pulses related to the Joule heating is expressed by the product of nominal output value of the NMR power amplifier and the pulse duration.
Figure \ref{f7} shows the RF-pulse energy dependence of $K_{b, {\rm I}}$ and $K_{b, {\rm II}}$ measured at 70 mK, which was determined from the NMR Free-Induction Decay (FID) spectrum of Te(I) and Te(II), shown in the inset.
The FID spectrum was measured by changing the RF-pulse energies, where pulse duration (10 $\mu$s) was fixed and the output of the NMR power amplifier was reduced.
When applying a RF-pulse energy of 25 $\mu$J, $K_{b, {\rm I}}$ and $K_{b,{\rm II}}$ show the Knight-shift value at 2 K, indicating that superconductivity is destroyed by the applied RF pulse. 
The NMR measurements at low temperatures was performed with the reduced RF-pulse energy smaller than 3 $\mu$J or comparable, in which the Knight shift is unchanged. 
The occurrence of superconductivity was also checked from the linewidth broadening or the shift of the NMR spectrum, which are shown in Fig.~\ref{f1} \textbf{a}, and Fig.~\ref{f3}.

\section*{Appendix II: Comparison between $H$ dependence of $K_{\rm spin}$ and the residual Sommerfeld coefficient $\gamma_0 (H)$ \cite{LeePRR2025} }
\begin{figure}
\vspace{5mm}
\begin{center}
\includegraphics[width=8.5cm]{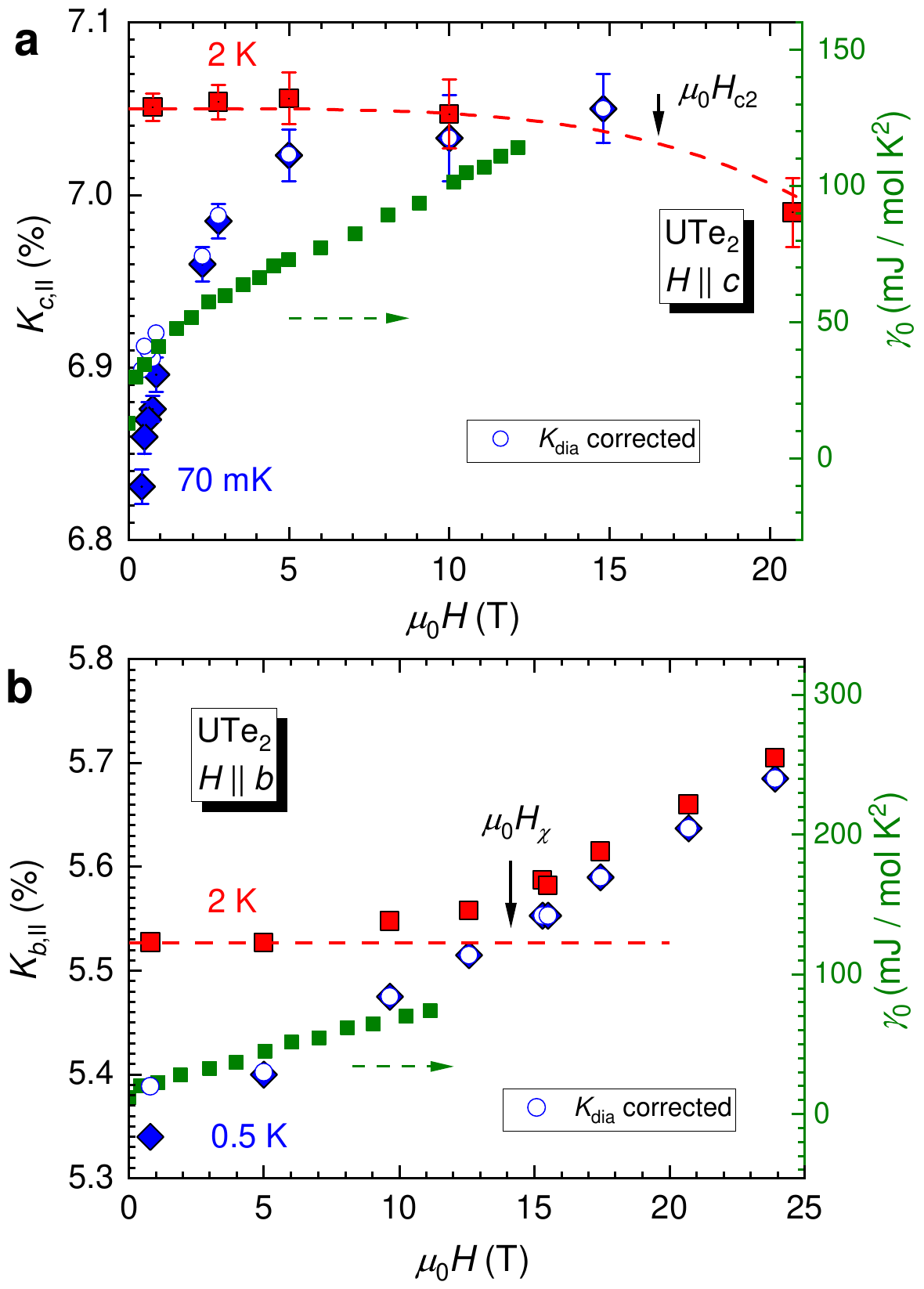}
\end{center}
\caption{(Color online) Comparison between the field dependence of the Knight shift, $K_{c, {\rm II}}$ ($K_{b, {\rm II}}$) and that of the residual Sommerfeld coefficient $\gamma_0(H)$ reported for UTe$_2$ in \textbf{a} $H \parallel c$ (\textbf{b} $H \parallel b$.) }
\label{f8}
\end{figure}

Recently, the zero-temperature extrapolation of $C / T$ (Sommerfeld coefficient) as a function of $H$ $\gamma_0(H)$ in UTe$_2$ was reported by Lee {\it et al} \cite{LeePRR2025}. 
Thus, we directly compare the $H$ dependence of the spin part of the Knight shift, $K_{\rm spin}$, with $\gamma_0$($H$). 
Figure \ref{f8} shows the comparison between the $H$ dependence of $K_{\rm spin}$ and that of $\gamma_0(H)$ reported for UTe$_2$ in \textbf{a} $H \parallel c$ and \textbf{b} $H \parallel b$.
As shown in Fig.~\ref{f8} \textbf{a}, while $\gamma_0(H)$ increases gradually with $H$, $K_{\rm spin}$ along the c axis already recovers to about 90\% of its normal-state value at 5 T. 
In both directions, the recovery of the Knight shift is therefore significantly faster than that of $\gamma_0$, and its field dependence is qualitatively different. 
This clear difference demonstrates that the observed Knight-shift changes cannot be accounted for solely by field-induced quasiparticles, in contrast to the behavior reported for even-parity superconductivity\cite{ChronisterPNAS2021,LiPRB2022}. 
In addition, the field dependence of $T_c$ becomes steeper above 5 T, indicating an enhancement of superconductivity with increasing magnetic field. 
Such behavior is not generally expected for spin-singlet pairing, but is consistent with spin-triplet pairing.

\clearpage


\begin{thebibliography}{69}%
\makeatletter
\providecommand \@ifxundefined [1]{%
 \@ifx{#1\undefined}
}%
\providecommand \@ifnum [1]{%
 \ifnum #1\expandafter \@firstoftwo
 \else \expandafter \@secondoftwo
 \fi
}%
\providecommand \@ifx [1]{%
 \ifx #1\expandafter \@firstoftwo
 \else \expandafter \@secondoftwo
 \fi
}%
\providecommand \natexlab [1]{#1}%
\providecommand \enquote  [1]{``#1''}%
\providecommand \bibnamefont  [1]{#1}%
\providecommand \bibfnamefont [1]{#1}%
\providecommand \citenamefont [1]{#1}%
\providecommand \href@noop [0]{\@secondoftwo}%
\providecommand \href [0]{\begingroup \@sanitize@url \@href}%
\providecommand \@href[1]{\@@startlink{#1}\@@href}%
\providecommand \@@href[1]{\endgroup#1\@@endlink}%
\providecommand \@sanitize@url [0]{\catcode `\\12\catcode `\$12\catcode `\&12\catcode `\#12\catcode `\^12\catcode `\_12\catcode `\%12\relax}%
\providecommand \@@startlink[1]{}%
\providecommand \@@endlink[0]{}%
\providecommand \url  [0]{\begingroup\@sanitize@url \@url }%
\providecommand \@url [1]{\endgroup\@href {#1}{\urlprefix }}%
\providecommand \urlprefix  [0]{URL }%
\providecommand \Eprint [0]{\href }%
\providecommand \doibase [0]{https://doi.org/}%
\providecommand \selectlanguage [0]{\@gobble}%
\providecommand \bibinfo  [0]{\@secondoftwo}%
\providecommand \bibfield  [0]{\@secondoftwo}%
\providecommand \translation [1]{[#1]}%
\providecommand \BibitemOpen [0]{}%
\providecommand \bibitemStop [0]{}%
\providecommand \bibitemNoStop [0]{.\EOS\space}%
\providecommand \EOS [0]{\spacefactor3000\relax}%
\providecommand \BibitemShut  [1]{\csname bibitem#1\endcsname}%
\let\auto@bib@innerbib\@empty
\bibitem [{\citenamefont {Gulian}\ and\ \citenamefont {Wood}(2003)}]{GulianIEEE2003}%
  \BibitemOpen
  \bibfield  {author} {\bibinfo {author} {\bibfnamefont {A.}~\bibnamefont {Gulian}}\ and\ \bibinfo {author} {\bibfnamefont {K.}~\bibnamefont {Wood}},\ }\bibfield  {title} {\bibinfo {title} {\text{T}riplet superconductors from the viewpoint of basic elements for quantum computers},\ }\href {https://doi.org/10.1109/tasc.2003.814156} {\bibfield  {journal} {\bibinfo  {journal} {{IEEE} Trans. Appl. Supercond.}\ }\textbf {\bibinfo {volume} {13}},\ \bibinfo {pages} {944} (\bibinfo {year} {2003})}\BibitemShut {NoStop}%
\bibitem [{\citenamefont {Leijnse}\ and\ \citenamefont {Flensberg}(2013)}]{LeijnsePRL2013}%
  \BibitemOpen
  \bibfield  {author} {\bibinfo {author} {\bibfnamefont {M.}~\bibnamefont {Leijnse}}\ and\ \bibinfo {author} {\bibfnamefont {K.}~\bibnamefont {Flensberg}},\ }\bibfield  {title} {\bibinfo {title} {\text{C}oupling \text{S}pin \text{Q}ubits via \text{S}uperconductors},\ }\href@noop {} {\bibfield  {journal} {\bibinfo  {journal} {Phys. Rev. Lett.}\ }\textbf {\bibinfo {volume} {111}} (\bibinfo {year} {2013})}\BibitemShut {NoStop}%
\bibitem [{\citenamefont {Saxena}\ \emph {et~al.}(2000)\citenamefont {Saxena}, \citenamefont {Agarwal}, \citenamefont {Ahilan}, \citenamefont {Grosche}, \citenamefont {Haselwimmer}, \citenamefont {Steiner}, \citenamefont {Pugh}, \citenamefont {Walker}, \citenamefont {Julian}, \citenamefont {Monthoux}, \citenamefont {Lonzarich}, \citenamefont {Huxley}, \citenamefont {Sheikin}, \citenamefont {Braithwaite},\ and\ \citenamefont {Flouquet}}]{SaxenaNature2000}%
  \BibitemOpen
  \bibfield  {author} {\bibinfo {author} {\bibfnamefont {S.~S.}\ \bibnamefont {Saxena}}, \bibinfo {author} {\bibfnamefont {P.}~\bibnamefont {Agarwal}}, \bibinfo {author} {\bibfnamefont {K.}~\bibnamefont {Ahilan}}, \bibinfo {author} {\bibfnamefont {F.~M.}\ \bibnamefont {Grosche}}, \bibinfo {author} {\bibfnamefont {R.~K.~W.}\ \bibnamefont {Haselwimmer}}, \bibinfo {author} {\bibfnamefont {M.~J.}\ \bibnamefont {Steiner}}, \bibinfo {author} {\bibfnamefont {E.}~\bibnamefont {Pugh}}, \bibinfo {author} {\bibfnamefont {I.~R.}\ \bibnamefont {Walker}}, \bibinfo {author} {\bibfnamefont {S.~R.}\ \bibnamefont {Julian}}, \bibinfo {author} {\bibfnamefont {P.}~\bibnamefont {Monthoux}}, \bibinfo {author} {\bibfnamefont {G.~G.}\ \bibnamefont {Lonzarich}}, \bibinfo {author} {\bibfnamefont {A.}~\bibnamefont {Huxley}}, \bibinfo {author} {\bibfnamefont {I.}~\bibnamefont {Sheikin}}, \bibinfo {author} {\bibfnamefont {D.}~\bibnamefont {Braithwaite}},\ and\ \bibinfo {author} {\bibfnamefont {J.}~\bibnamefont {Flouquet}},\ }\bibfield
  {title} {\bibinfo {title} {Superconductivity on the border of itinerant-electron ferromagnetism in $\text{UGe}_2$},\ }\href@noop {} {\bibfield  {journal} {\bibinfo  {journal} {Nature}\ }\textbf {\bibinfo {volume} {406}},\ \bibinfo {pages} {587} (\bibinfo {year} {2000})}\BibitemShut {NoStop}%
\bibitem [{\citenamefont {Aoki}\ \emph {et~al.}(2001)\citenamefont {Aoki}, \citenamefont {Huxley}, \citenamefont {Ressouche}, \citenamefont {Braithwaite}, \citenamefont {Flouquet}, \citenamefont {Brison}, \citenamefont {Lhotel},\ and\ \citenamefont {Paulsen}}]{AokiNature2001}%
  \BibitemOpen
  \bibfield  {author} {\bibinfo {author} {\bibfnamefont {D.}~\bibnamefont {Aoki}}, \bibinfo {author} {\bibfnamefont {A.}~\bibnamefont {Huxley}}, \bibinfo {author} {\bibfnamefont {E.}~\bibnamefont {Ressouche}}, \bibinfo {author} {\bibfnamefont {D.}~\bibnamefont {Braithwaite}}, \bibinfo {author} {\bibfnamefont {J.}~\bibnamefont {Flouquet}}, \bibinfo {author} {\bibfnamefont {J.~P.}\ \bibnamefont {Brison}}, \bibinfo {author} {\bibfnamefont {E.}~\bibnamefont {Lhotel}},\ and\ \bibinfo {author} {\bibfnamefont {C.}~\bibnamefont {Paulsen}},\ }\bibfield  {title} {\bibinfo {title} {Coexistence of superconductivity and ferromagnetism in \text{URhGe}},\ }\href@noop {} {\bibfield  {journal} {\bibinfo  {journal} {Nature}\ }\textbf {\bibinfo {volume} {413}},\ \bibinfo {pages} {613} (\bibinfo {year} {2001})}\BibitemShut {NoStop}%
\bibitem [{\citenamefont {Huy}\ \emph {et~al.}(2007)\citenamefont {Huy}, \citenamefont {Gasparini}, \citenamefont {de~Nijs}, \citenamefont {Huang}, \citenamefont {Klaasse}, \citenamefont {Gortenmulder}, \citenamefont {de~Visser}, \citenamefont {Hamann}, \citenamefont {G{\"{o}}rlach},\ and\ \citenamefont {v.~L{\"{o}}hneysen}}]{HuyPRL2007}%
  \BibitemOpen
  \bibfield  {author} {\bibinfo {author} {\bibfnamefont {N.~T.}\ \bibnamefont {Huy}}, \bibinfo {author} {\bibfnamefont {A.}~\bibnamefont {Gasparini}}, \bibinfo {author} {\bibfnamefont {D.~E.}\ \bibnamefont {de~Nijs}}, \bibinfo {author} {\bibfnamefont {Y.}~\bibnamefont {Huang}}, \bibinfo {author} {\bibfnamefont {J.~C.~P.}\ \bibnamefont {Klaasse}}, \bibinfo {author} {\bibfnamefont {T.}~\bibnamefont {Gortenmulder}}, \bibinfo {author} {\bibfnamefont {A.}~\bibnamefont {de~Visser}}, \bibinfo {author} {\bibfnamefont {A.}~\bibnamefont {Hamann}}, \bibinfo {author} {\bibfnamefont {T.}~\bibnamefont {G{\"{o}}rlach}},\ and\ \bibinfo {author} {\bibfnamefont {H.}~\bibnamefont {v.~L{\"{o}}hneysen}},\ }\bibfield  {title} {\bibinfo {title} {Superconductivity on the \text{B}order of \text{W}eak \text{I}tinerant \text{F}erromagnetism in $\text{UCoGe}$},\ }\href@noop {} {\bibfield  {journal} {\bibinfo  {journal} {Phys. Rev. Lett.}\ }\textbf {\bibinfo {volume} {99}},\ \bibinfo {pages} {067006} (\bibinfo {year} {2007})}\BibitemShut
  {NoStop}%
\bibitem [{\citenamefont {Ran}\ \emph {et~al.}(2019{\natexlab{a}})\citenamefont {Ran}, \citenamefont {Eckberg}, \citenamefont {Ding}, \citenamefont {Furukawa}, \citenamefont {Metz}, \citenamefont {Saha}, \citenamefont {Liu}, \citenamefont {Zic}, \citenamefont {Kim}, \citenamefont {Paglione},\ and\ \citenamefont {Butch}}]{RanScience2019}%
  \BibitemOpen
  \bibfield  {author} {\bibinfo {author} {\bibfnamefont {S.}~\bibnamefont {Ran}}, \bibinfo {author} {\bibfnamefont {C.}~\bibnamefont {Eckberg}}, \bibinfo {author} {\bibfnamefont {Q.-P.}\ \bibnamefont {Ding}}, \bibinfo {author} {\bibfnamefont {Y.}~\bibnamefont {Furukawa}}, \bibinfo {author} {\bibfnamefont {T.}~\bibnamefont {Metz}}, \bibinfo {author} {\bibfnamefont {S.~R.}\ \bibnamefont {Saha}}, \bibinfo {author} {\bibfnamefont {I.-L.}\ \bibnamefont {Liu}}, \bibinfo {author} {\bibfnamefont {M.}~\bibnamefont {Zic}}, \bibinfo {author} {\bibfnamefont {H.}~\bibnamefont {Kim}}, \bibinfo {author} {\bibfnamefont {J.}~\bibnamefont {Paglione}},\ and\ \bibinfo {author} {\bibfnamefont {N.~P.}\ \bibnamefont {Butch}},\ }\bibfield  {title} {\bibinfo {title} {Nearly ferromagnetic spin-triplet superconductivity},\ }\href {https://doi.org/10.1126/science.aav8645} {\bibfield  {journal} {\bibinfo  {journal} {Science}\ }\textbf {\bibinfo {volume} {365}},\ \bibinfo {pages} {684} (\bibinfo {year} {2019}{\natexlab{a}})}\BibitemShut
  {NoStop}%
\bibitem [{\citenamefont {Aoki}\ \emph {et~al.}(2022{\natexlab{a}})\citenamefont {Aoki}, \citenamefont {Brison}, \citenamefont {Flouquet}, \citenamefont {Ishida}, \citenamefont {Knebel}, \citenamefont {Tokunaga},\ and\ \citenamefont {Yanase}}]{aoki2021JPCMrev}
  \BibitemOpen
  \bibfield  {author} {\bibinfo {author} {\bibfnamefont {D.}~\bibnamefont {Aoki}}, \bibinfo {author} {\bibfnamefont {J.-P.}\ \bibnamefont {Brison}}, \bibinfo {author} {\bibfnamefont {J.}~\bibnamefont {Flouquet}}, \bibinfo {author} {\bibfnamefont {K.}~\bibnamefont {Ishida}}, \bibinfo {author} {\bibfnamefont {G.}~\bibnamefont {Knebel}}, \bibinfo {author} {\bibfnamefont {Y.}~\bibnamefont {Tokunaga}},\ and\ \bibinfo {author} {\bibfnamefont {Y.}~\bibnamefont {Yanase}},\ }\bibfield  {title} {\bibinfo {title} {Unconventional superconductivity in $\text{UTe$_2$}$},\ }\href {https://doi.org/10.1088/1361-648x/ac5863} {\bibfield  {journal} {\bibinfo  {journal} {J. Phys.: Condens. Matter}\ }\textbf {\bibinfo {volume} {34}},\ \bibinfo {pages} {243002} (\bibinfo {year} {2022}{\natexlab{a}})}\BibitemShut {NoStop}%
\bibitem [{\citenamefont {Aoki}\ \emph {et~al.}(2019)\citenamefont {Aoki}, \citenamefont {Ishida},\ and\ \citenamefont {Flouquet}}]{AokiJPSJ2019Rev}%
  \BibitemOpen
  \bibfield  {author} {\bibinfo {author} {\bibfnamefont {D.}~\bibnamefont {Aoki}}, \bibinfo {author} {\bibfnamefont {K.}~\bibnamefont {Ishida}},\ and\ \bibinfo {author} {\bibfnamefont {J.}~\bibnamefont {Flouquet}},\ }\bibfield  {title} {\bibinfo {title} {Review of u-based ferromagnetic superconductors: Comparison between $\text{UGe}_2$, \text{URhGe}, and \text{UCoGe}},\ }\href {https://doi.org/10.7566/JPSJ.88.022001} {\bibfield  {journal} {\bibinfo  {journal} {J. Phys. Soc. Jpn.}\ }\textbf {\bibinfo {volume} {88}},\ \bibinfo {pages} {022001} (\bibinfo {year} {2019})}\BibitemShut {NoStop}%
\bibitem [{\citenamefont {Osheroff}\ \emph {et~al.}(1972)\citenamefont {Osheroff}, \citenamefont {Richardson},\ and\ \citenamefont {Lee}}]{OsheroffPRL1972}%
  \BibitemOpen
  \bibfield  {author} {\bibinfo {author} {\bibfnamefont {D.~D.}\ \bibnamefont {Osheroff}}, \bibinfo {author} {\bibfnamefont {R.~C.}\ \bibnamefont {Richardson}},\ and\ \bibinfo {author} {\bibfnamefont {D.~M.}\ \bibnamefont {Lee}},\ }\bibfield  {title} {\bibinfo {title} {Evidence for a new phase of solid ${\mathrm{he}}^{3}$},\ }\href {https://doi.org/10.1103/PhysRevLett.28.885} {\bibfield  {journal} {\bibinfo  {journal} {Phys. Rev. Lett.}\ }\textbf {\bibinfo {volume} {28}},\ \bibinfo {pages} {885} (\bibinfo {year} {1972})}\BibitemShut {NoStop}%
\bibitem [{\citenamefont {Leggett}(1975)}]{LeggettRMP1975}%
  \BibitemOpen
  \bibfield  {author} {\bibinfo {author} {\bibfnamefont {A.~J.}\ \bibnamefont {Leggett}},\ }\bibfield  {title} {\bibinfo {title} {A theoretical description of the new phases of liquid $^{3}\mathrm{He}$},\ }\href {https://doi.org/10.1103/RevModPhys.47.331} {\bibfield  {journal} {\bibinfo  {journal} {Rev. Mod. Phys.}\ }\textbf {\bibinfo {volume} {47}},\ \bibinfo {pages} {331} (\bibinfo {year} {1975})}\BibitemShut {NoStop}%
\bibitem [{\citenamefont {Wheatley}(1975)}]{WheatleyRMP1975}%
  \BibitemOpen
  \bibfield  {author} {\bibinfo {author} {\bibfnamefont {J.~C.}\ \bibnamefont {Wheatley}},\ }\bibfield  {title} {\bibinfo {title} {Experimental properties of superfluid \text{$^3$He}},\ }\href {https://doi.org/10.1103/revmodphys.47.415} {\bibfield  {journal} {\bibinfo  {journal} {Reviews of Modern Physics}\ }\textbf {\bibinfo {volume} {47}},\ \bibinfo {pages} {415–470} (\bibinfo {year} {1975})}\BibitemShut {NoStop}%
\bibitem [{\citenamefont {Vollhardt}\ and\ \citenamefont {W{\"{o}}lfle}(1990)}]{VollhardtHe3}%
  \BibitemOpen
  \bibfield  {author} {\bibinfo {author} {\bibfnamefont {D.}~\bibnamefont {Vollhardt}}\ and\ \bibinfo {author} {\bibfnamefont {P.}~\bibnamefont {W{\"{o}}lfle}},\ }\href@noop {} {\emph {\bibinfo {title} {The Superfluid Phases of Helium 3}}}\ (\bibinfo  {publisher} {Taylor and Francis},\ \bibinfo {year} {1990})\BibitemShut {NoStop}%
\bibitem [{\citenamefont {Knebel}\ \emph {et~al.}(2019)\citenamefont {Knebel}, \citenamefont {Knafo}, \citenamefont {Pourret}, \citenamefont {Niu}, \citenamefont {Vali{\v{s}}ka}, \citenamefont {Braithwaite}, \citenamefont {Lapertot}, \citenamefont {Nardone}, \citenamefont {Zitouni}, \citenamefont {Mishra}, \citenamefont {Sheikin}, \citenamefont {Seyfarth}, \citenamefont {Brison}, \citenamefont {Aoki},\ and\ \citenamefont {Flouquet}}]{KnebelJPSJ2019}%
  \BibitemOpen
  \bibfield  {author} {\bibinfo {author} {\bibfnamefont {G.}~\bibnamefont {Knebel}}, \bibinfo {author} {\bibfnamefont {W.}~\bibnamefont {Knafo}}, \bibinfo {author} {\bibfnamefont {A.}~\bibnamefont {Pourret}}, \bibinfo {author} {\bibfnamefont {Q.}~\bibnamefont {Niu}}, \bibinfo {author} {\bibfnamefont {M.}~\bibnamefont {Vali{\v{s}}ka}}, \bibinfo {author} {\bibfnamefont {D.}~\bibnamefont {Braithwaite}}, \bibinfo {author} {\bibfnamefont {G.}~\bibnamefont {Lapertot}}, \bibinfo {author} {\bibfnamefont {M.}~\bibnamefont {Nardone}}, \bibinfo {author} {\bibfnamefont {A.}~\bibnamefont {Zitouni}}, \bibinfo {author} {\bibfnamefont {S.}~\bibnamefont {Mishra}}, \bibinfo {author} {\bibfnamefont {I.}~\bibnamefont {Sheikin}}, \bibinfo {author} {\bibfnamefont {G.}~\bibnamefont {Seyfarth}}, \bibinfo {author} {\bibfnamefont {J.-P.}\ \bibnamefont {Brison}}, \bibinfo {author} {\bibfnamefont {D.}~\bibnamefont {Aoki}},\ and\ \bibinfo {author} {\bibfnamefont {J.}~\bibnamefont {Flouquet}},\ }\bibfield  {title} {\bibinfo {title}
  {Field-reentrant superconductivity close to a metamagnetic transition in the heavy-fermion superconductor ute2},\ }\href {https://doi.org/10.7566/JPSJ.88.063707} {\bibfield  {journal} {\bibinfo  {journal} {J. Phys. Soc. Jpn.}\ }\textbf {\bibinfo {volume} {88}},\ \bibinfo {pages} {063707} (\bibinfo {year} {2019})},\ \Eprint {https://arxiv.org/abs/https://doi.org/10.7566/JPSJ.88.063707} {https://doi.org/10.7566/JPSJ.88.063707} \BibitemShut {NoStop}%
\bibitem [{\citenamefont {Braithwaite}\ \emph {et~al.}(2019)\citenamefont {Braithwaite}, \citenamefont {Vali{\v{s}}ka}, \citenamefont {Knebel}, \citenamefont {Lapertot}, \citenamefont {Brison}, \citenamefont {Pourret}, \citenamefont {Zhitomirsky}, \citenamefont {Flouquet}, \citenamefont {Honda},\ and\ \citenamefont {Aoki}}]{BraithwaiteCommPhy2019}%
  \BibitemOpen
  \bibfield  {author} {\bibinfo {author} {\bibfnamefont {D.}~\bibnamefont {Braithwaite}}, \bibinfo {author} {\bibfnamefont {M.}~\bibnamefont {Vali{\v{s}}ka}}, \bibinfo {author} {\bibfnamefont {G.}~\bibnamefont {Knebel}}, \bibinfo {author} {\bibfnamefont {G.}~\bibnamefont {Lapertot}}, \bibinfo {author} {\bibfnamefont {J.-P.}\ \bibnamefont {Brison}}, \bibinfo {author} {\bibfnamefont {A.}~\bibnamefont {Pourret}}, \bibinfo {author} {\bibfnamefont {M.~E.}\ \bibnamefont {Zhitomirsky}}, \bibinfo {author} {\bibfnamefont {J.}~\bibnamefont {Flouquet}}, \bibinfo {author} {\bibfnamefont {F.}~\bibnamefont {Honda}},\ and\ \bibinfo {author} {\bibfnamefont {D.}~\bibnamefont {Aoki}},\ }\bibfield  {title} {\bibinfo {title} {Multiple superconducting phases in a nearly ferromagnetic system},\ }\href {https://doi.org/10.1038/s42005-019-0248-z} {\bibfield  {journal} {\bibinfo  {journal} {Commun. Phys.}\ }\textbf {\bibinfo {volume} {2}},\ \bibinfo {pages} {147} (\bibinfo {year} {2019})}\BibitemShut {NoStop}%
\bibitem [{\citenamefont {Vasina}\ \emph {et~al.}(2025)\citenamefont {Vasina}, \citenamefont {Aoki}, \citenamefont {Miyake}, \citenamefont {Seyfarth}, \citenamefont {Pourret}, \citenamefont {Marcenat}, \citenamefont {Amano~Patino}, \citenamefont {Lapertot}, \citenamefont {Flouquet}, \citenamefont {Brison}, \citenamefont {Braithwaite},\ and\ \citenamefont {Knebel}}]{VasinaPRL2025}%
  \BibitemOpen
  \bibfield  {author} {\bibinfo {author} {\bibfnamefont {T.}~\bibnamefont {Vasina}}, \bibinfo {author} {\bibfnamefont {D.}~\bibnamefont {Aoki}}, \bibinfo {author} {\bibfnamefont {A.}~\bibnamefont {Miyake}}, \bibinfo {author} {\bibfnamefont {G.}~\bibnamefont {Seyfarth}}, \bibinfo {author} {\bibfnamefont {A.}~\bibnamefont {Pourret}}, \bibinfo {author} {\bibfnamefont {C.}~\bibnamefont {Marcenat}}, \bibinfo {author} {\bibfnamefont {M.}~\bibnamefont {Amano~Patino}}, \bibinfo {author} {\bibfnamefont {G.}~\bibnamefont {Lapertot}}, \bibinfo {author} {\bibfnamefont {J.}~\bibnamefont {Flouquet}}, \bibinfo {author} {\bibfnamefont {J.-P.}\ \bibnamefont {Brison}}, \bibinfo {author} {\bibfnamefont {D.}~\bibnamefont {Braithwaite}},\ and\ \bibinfo {author} {\bibfnamefont {G.}~\bibnamefont {Knebel}},\ }\bibfield  {title} {\bibinfo {title} {Connecting high-field and high-pressure superconductivity in ${\mathrm{ute}}_{2}$},\ }\href {https://doi.org/10.1103/PhysRevLett.134.096501} {\bibfield  {journal} {\bibinfo  {journal}
  {Phys. Rev. Lett.}\ }\textbf {\bibinfo {volume} {134}},\ \bibinfo {pages} {096501} (\bibinfo {year} {2025})}\BibitemShut {NoStop}%
\bibitem [{\citenamefont {Ran}\ \emph {et~al.}(2019{\natexlab{b}})\citenamefont {Ran}, \citenamefont {Liu}, \citenamefont {Eo}, \citenamefont {Campbell}, \citenamefont {Neves}, \citenamefont {Fuhrman}, \citenamefont {Saha}, \citenamefont {Eckberg}, \citenamefont {Kim}, \citenamefont {Graf}, \citenamefont {Balakirev}, \citenamefont {Singleton}, \citenamefont {Paglione},\ and\ \citenamefont {Butch}}]{RanNatPhy2019}%
  \BibitemOpen
  \bibfield  {author} {\bibinfo {author} {\bibfnamefont {S.}~\bibnamefont {Ran}}, \bibinfo {author} {\bibfnamefont {I.-L.}\ \bibnamefont {Liu}}, \bibinfo {author} {\bibfnamefont {Y.~S.}\ \bibnamefont {Eo}}, \bibinfo {author} {\bibfnamefont {D.~J.}\ \bibnamefont {Campbell}}, \bibinfo {author} {\bibfnamefont {P.~M.}\ \bibnamefont {Neves}}, \bibinfo {author} {\bibfnamefont {W.~T.}\ \bibnamefont {Fuhrman}}, \bibinfo {author} {\bibfnamefont {S.~R.}\ \bibnamefont {Saha}}, \bibinfo {author} {\bibfnamefont {C.}~\bibnamefont {Eckberg}}, \bibinfo {author} {\bibfnamefont {H.}~\bibnamefont {Kim}}, \bibinfo {author} {\bibfnamefont {D.}~\bibnamefont {Graf}}, \bibinfo {author} {\bibfnamefont {F.}~\bibnamefont {Balakirev}}, \bibinfo {author} {\bibfnamefont {J.}~\bibnamefont {Singleton}}, \bibinfo {author} {\bibfnamefont {J.}~\bibnamefont {Paglione}},\ and\ \bibinfo {author} {\bibfnamefont {N.~P.}\ \bibnamefont {Butch}},\ }\bibfield  {title} {\bibinfo {title} {Extreme magnetic field-boosted superconductivity},\ }\href
  {https://doi.org/10.1038/s41567-019-0670-x} {\bibfield  {journal} {\bibinfo  {journal} {Nat. Phys.}\ }\textbf {\bibinfo {volume} {15}},\ \bibinfo {pages} {1250} (\bibinfo {year} {2019}{\natexlab{b}})}\BibitemShut {NoStop}%
\bibitem [{\citenamefont {Kinjo}\ \emph {et~al.}(2023{\natexlab{a}})\citenamefont {Kinjo}, \citenamefont {Fujibayashi}, \citenamefont {Kitagawa}, \citenamefont {Ishida}, \citenamefont {Tokunaga}, \citenamefont {Sakai}, \citenamefont {Kambe}, \citenamefont {Nakamura}, \citenamefont {Shimizu}, \citenamefont {Homma}, \citenamefont {Li}, \citenamefont {Honda}, \citenamefont {Aoki}, \citenamefont {Hiraki}, \citenamefont {Kimata},\ and\ \citenamefont {Sasaki}}]{KinjoPRB2023}%
  \BibitemOpen
  \bibfield  {author} {\bibinfo {author} {\bibfnamefont {K.}~\bibnamefont {Kinjo}}, \bibinfo {author} {\bibfnamefont {H.}~\bibnamefont {Fujibayashi}}, \bibinfo {author} {\bibfnamefont {S.}~\bibnamefont {Kitagawa}}, \bibinfo {author} {\bibfnamefont {K.}~\bibnamefont {Ishida}}, \bibinfo {author} {\bibfnamefont {Y.}~\bibnamefont {Tokunaga}}, \bibinfo {author} {\bibfnamefont {H.}~\bibnamefont {Sakai}}, \bibinfo {author} {\bibfnamefont {S.}~\bibnamefont {Kambe}}, \bibinfo {author} {\bibfnamefont {A.}~\bibnamefont {Nakamura}}, \bibinfo {author} {\bibfnamefont {Y.}~\bibnamefont {Shimizu}}, \bibinfo {author} {\bibfnamefont {Y.}~\bibnamefont {Homma}}, \bibinfo {author} {\bibfnamefont {D.~X.}\ \bibnamefont {Li}}, \bibinfo {author} {\bibfnamefont {F.}~\bibnamefont {Honda}}, \bibinfo {author} {\bibfnamefont {D.}~\bibnamefont {Aoki}}, \bibinfo {author} {\bibfnamefont {K.}~\bibnamefont {Hiraki}}, \bibinfo {author} {\bibfnamefont {M.}~\bibnamefont {Kimata}},\ and\ \bibinfo {author} {\bibfnamefont {T.}~\bibnamefont
  {Sasaki}},\ }\bibfield  {title} {\bibinfo {title} {Change of superconducting character in \text{UTe$_2$} induced by magnetic field},\ }\href {https://doi.org/10.1103/physrevb.107.l060502} {\bibfield  {journal} {\bibinfo  {journal} {Phys. Rev. B}\ }\textbf {\bibinfo {volume} {107}},\ \bibinfo {pages} {L060502} (\bibinfo {year} {2023}{\natexlab{a}})}\BibitemShut {NoStop}%
\bibitem [{\citenamefont {Sakai}\ \emph {et~al.}(2023)\citenamefont {Sakai}, \citenamefont {Tokiwa}, \citenamefont {Opletal}, \citenamefont {Kimata}, \citenamefont {Awaji}, \citenamefont {Sasaki}, \citenamefont {Aoki}, \citenamefont {Kambe}, \citenamefont {Tokunaga},\ and\ \citenamefont {Haga}}]{SakaiPRL2023}%
  \BibitemOpen
  \bibfield  {author} {\bibinfo {author} {\bibfnamefont {H.}~\bibnamefont {Sakai}}, \bibinfo {author} {\bibfnamefont {Y.}~\bibnamefont {Tokiwa}}, \bibinfo {author} {\bibfnamefont {P.}~\bibnamefont {Opletal}}, \bibinfo {author} {\bibfnamefont {M.}~\bibnamefont {Kimata}}, \bibinfo {author} {\bibfnamefont {S.}~\bibnamefont {Awaji}}, \bibinfo {author} {\bibfnamefont {T.}~\bibnamefont {Sasaki}}, \bibinfo {author} {\bibfnamefont {D.}~\bibnamefont {Aoki}}, \bibinfo {author} {\bibfnamefont {S.}~\bibnamefont {Kambe}}, \bibinfo {author} {\bibfnamefont {Y.}~\bibnamefont {Tokunaga}},\ and\ \bibinfo {author} {\bibfnamefont {Y.}~\bibnamefont {Haga}},\ }\bibfield  {title} {\bibinfo {title} {\text{F}ield \text{I}nduced \text{M}ultiple \text{S}uperconducting \text{P}hases in \text{UTe$_2$} along \text{H}ard \text{M}agnetic \text{A}xis},\ }\bibfield  {journal} {\bibinfo  {journal} {Physical Review Letters}\ }\textbf {\bibinfo {volume} {130}},\ \href {https://doi.org/10.1103/physrevlett.130.196002}
  {10.1103/physrevlett.130.196002} (\bibinfo {year} {2023})\BibitemShut {NoStop}%
\bibitem [{\citenamefont {Wu}\ \emph {et~al.}(2024)\citenamefont {Wu}, \citenamefont {Weinberger}, \citenamefont {Chen}, \citenamefont {Cabala}, \citenamefont {Chichinadze}, \citenamefont {Shaffer}, \citenamefont {Pospíšil}, \citenamefont {Prokleška}, \citenamefont {Haidamak}, \citenamefont {Bastien}, \citenamefont {Sechovský}, \citenamefont {Hickey}, \citenamefont {Mancera-Ugarte}, \citenamefont {Benjamin}, \citenamefont {Graf}, \citenamefont {Skourski}, \citenamefont {Lonzarich}, \citenamefont {Vališka}, \citenamefont {Grosche},\ and\ \citenamefont {Eaton}}]{WuPNA2024}%
  \BibitemOpen
  \bibfield  {author} {\bibinfo {author} {\bibfnamefont {Z.}~\bibnamefont {Wu}}, \bibinfo {author} {\bibfnamefont {T.~I.}\ \bibnamefont {Weinberger}}, \bibinfo {author} {\bibfnamefont {J.}~\bibnamefont {Chen}}, \bibinfo {author} {\bibfnamefont {A.}~\bibnamefont {Cabala}}, \bibinfo {author} {\bibfnamefont {D.~V.}\ \bibnamefont {Chichinadze}}, \bibinfo {author} {\bibfnamefont {D.}~\bibnamefont {Shaffer}}, \bibinfo {author} {\bibfnamefont {J.}~\bibnamefont {Pospíšil}}, \bibinfo {author} {\bibfnamefont {J.}~\bibnamefont {Prokleška}}, \bibinfo {author} {\bibfnamefont {T.}~\bibnamefont {Haidamak}}, \bibinfo {author} {\bibfnamefont {G.}~\bibnamefont {Bastien}}, \bibinfo {author} {\bibfnamefont {V.}~\bibnamefont {Sechovský}}, \bibinfo {author} {\bibfnamefont {A.~J.}\ \bibnamefont {Hickey}}, \bibinfo {author} {\bibfnamefont {M.~J.}\ \bibnamefont {Mancera-Ugarte}}, \bibinfo {author} {\bibfnamefont {S.}~\bibnamefont {Benjamin}}, \bibinfo {author} {\bibfnamefont {D.~E.}\ \bibnamefont {Graf}}, \bibinfo {author}
  {\bibfnamefont {Y.}~\bibnamefont {Skourski}}, \bibinfo {author} {\bibfnamefont {G.~G.}\ \bibnamefont {Lonzarich}}, \bibinfo {author} {\bibfnamefont {M.}~\bibnamefont {Vališka}}, \bibinfo {author} {\bibfnamefont {F.~M.}\ \bibnamefont {Grosche}},\ and\ \bibinfo {author} {\bibfnamefont {A.~G.}\ \bibnamefont {Eaton}},\ }\bibfield  {title} {\bibinfo {title} {Enhanced triplet superconductivity in next-generation ultraclean \text{UTe$_2$}},\ }\bibfield  {journal} {\bibinfo  {journal} {Proceedings of the National Academy of Sciences}\ }\textbf {\bibinfo {volume} {121}},\ \href {https://doi.org/10.1073/pnas.2403067121} {10.1073/pnas.2403067121} (\bibinfo {year} {2024})\BibitemShut {NoStop}%
\bibitem [{\citenamefont {Nakamine}\ \emph {et~al.}(2019)\citenamefont {Nakamine}, \citenamefont {Kitagawa}, \citenamefont {Ishida}, \citenamefont {Tokunaga}, \citenamefont {Sakai}, \citenamefont {Kambe}, \citenamefont {Nakamura}, \citenamefont {Shimizu}, \citenamefont {Homma}, \citenamefont {Li}, \citenamefont {Honda},\ and\ \citenamefont {Aoki}}]{NakamineJPSJ2019}%
  \BibitemOpen
  \bibfield  {author} {\bibinfo {author} {\bibfnamefont {G.}~\bibnamefont {Nakamine}}, \bibinfo {author} {\bibfnamefont {S.}~\bibnamefont {Kitagawa}}, \bibinfo {author} {\bibfnamefont {K.}~\bibnamefont {Ishida}}, \bibinfo {author} {\bibfnamefont {Y.}~\bibnamefont {Tokunaga}}, \bibinfo {author} {\bibfnamefont {H.}~\bibnamefont {Sakai}}, \bibinfo {author} {\bibfnamefont {S.}~\bibnamefont {Kambe}}, \bibinfo {author} {\bibfnamefont {A.}~\bibnamefont {Nakamura}}, \bibinfo {author} {\bibfnamefont {Y.}~\bibnamefont {Shimizu}}, \bibinfo {author} {\bibfnamefont {Y.}~\bibnamefont {Homma}}, \bibinfo {author} {\bibfnamefont {D.}~\bibnamefont {Li}}, \bibinfo {author} {\bibfnamefont {F.}~\bibnamefont {Honda}},\ and\ \bibinfo {author} {\bibfnamefont {D.}~\bibnamefont {Aoki}},\ }\bibfield  {title} {\bibinfo {title} {Superconducting properties of heavy fermion \text{UTe$_2$} revealed by \text{$^{125}$Te}-nuclear magnetic resonance},\ }\href {https://doi.org/10.7566/JPSJ.88.113703} {\bibfield  {journal} {\bibinfo  {journal}
  {J. Phys. Soc. Jpn.}\ }\textbf {\bibinfo {volume} {88}},\ \bibinfo {pages} {113703} (\bibinfo {year} {2019})}\BibitemShut {NoStop}%
\bibitem [{\citenamefont {Nakamine}\ \emph {et~al.}(2021{\natexlab{a}})\citenamefont {Nakamine}, \citenamefont {Kinjo}, \citenamefont {Kitagawa}, \citenamefont {Ishida}, \citenamefont {Tokunaga}, \citenamefont {Sakai}, \citenamefont {Kambe}, \citenamefont {Nakamura}, \citenamefont {Shimizu}, \citenamefont {Homma}, \citenamefont {Li}, \citenamefont {Honda},\ and\ \citenamefont {Aoki}}]{NakaminePRB2021}%
  \BibitemOpen
  \bibfield  {author} {\bibinfo {author} {\bibfnamefont {G.}~\bibnamefont {Nakamine}}, \bibinfo {author} {\bibfnamefont {K.}~\bibnamefont {Kinjo}}, \bibinfo {author} {\bibfnamefont {S.}~\bibnamefont {Kitagawa}}, \bibinfo {author} {\bibfnamefont {K.}~\bibnamefont {Ishida}}, \bibinfo {author} {\bibfnamefont {Y.}~\bibnamefont {Tokunaga}}, \bibinfo {author} {\bibfnamefont {H.}~\bibnamefont {Sakai}}, \bibinfo {author} {\bibfnamefont {S.}~\bibnamefont {Kambe}}, \bibinfo {author} {\bibfnamefont {A.}~\bibnamefont {Nakamura}}, \bibinfo {author} {\bibfnamefont {Y.}~\bibnamefont {Shimizu}}, \bibinfo {author} {\bibfnamefont {Y.}~\bibnamefont {Homma}}, \bibinfo {author} {\bibfnamefont {D.}~\bibnamefont {Li}}, \bibinfo {author} {\bibfnamefont {F.}~\bibnamefont {Honda}},\ and\ \bibinfo {author} {\bibfnamefont {D.}~\bibnamefont {Aoki}},\ }\bibfield  {title} {\bibinfo {title} {Anisotropic response of spin susceptibility in the superconducting state of \text{UTe$_2$} probed with \text{$^{125}$Te} nmr measurement},\ }\href
  {https://doi.org/10.1103/physrevb.103.l100503} {\bibfield  {journal} {\bibinfo  {journal} {Phys. Rev. B}\ }\textbf {\bibinfo {volume} {103}},\ \bibinfo {pages} {L100503} (\bibinfo {year} {2021}{\natexlab{a}})}\BibitemShut {NoStop}%
\bibitem [{\citenamefont {Nakamine}\ \emph {et~al.}(2021{\natexlab{b}})\citenamefont {Nakamine}, \citenamefont {Kinjo}, \citenamefont {Kitagawa}, \citenamefont {Ishida}, \citenamefont {Tokunaga}, \citenamefont {Sakai}, \citenamefont {Kambe}, \citenamefont {Nakamura}, \citenamefont {Shimizu}, \citenamefont {Homma}, \citenamefont {Li}, \citenamefont {Honda},\ and\ \citenamefont {Aoki}}]{NakamineJPSJ2021}%
  \BibitemOpen
  \bibfield  {author} {\bibinfo {author} {\bibfnamefont {G.}~\bibnamefont {Nakamine}}, \bibinfo {author} {\bibfnamefont {K.}~\bibnamefont {Kinjo}}, \bibinfo {author} {\bibfnamefont {S.}~\bibnamefont {Kitagawa}}, \bibinfo {author} {\bibfnamefont {K.}~\bibnamefont {Ishida}}, \bibinfo {author} {\bibfnamefont {Y.}~\bibnamefont {Tokunaga}}, \bibinfo {author} {\bibfnamefont {H.}~\bibnamefont {Sakai}}, \bibinfo {author} {\bibfnamefont {S.}~\bibnamefont {Kambe}}, \bibinfo {author} {\bibfnamefont {A.}~\bibnamefont {Nakamura}}, \bibinfo {author} {\bibfnamefont {Y.}~\bibnamefont {Shimizu}}, \bibinfo {author} {\bibfnamefont {Y.}~\bibnamefont {Homma}}, \bibinfo {author} {\bibfnamefont {D.}~\bibnamefont {Li}}, \bibinfo {author} {\bibfnamefont {F.}~\bibnamefont {Honda}},\ and\ \bibinfo {author} {\bibfnamefont {D.}~\bibnamefont {Aoki}},\ }\bibfield  {title} {\bibinfo {title} {Inhomogeneous superconducting state probed by \text{$^{125}$Te NMR} on \text{UTe$_2$}},\ }\href {https://doi.org/10.7566/JPSJ.90.064709} {\bibfield
  {journal} {\bibinfo  {journal} {J. Phys. Soc. Jpn.}\ }\textbf {\bibinfo {volume} {90}},\ \bibinfo {pages} {064709} (\bibinfo {year} {2021}{\natexlab{b}})},\ \Eprint {https://arxiv.org/abs/https://doi.org/10.7566/JPSJ.90.064709} {https://doi.org/10.7566/JPSJ.90.064709} \BibitemShut {NoStop}%
\bibitem [{\citenamefont {Fujibayashi}\ \emph {et~al.}(2022)\citenamefont {Fujibayashi}, \citenamefont {Nakamine}, \citenamefont {Kinjo}, \citenamefont {Kitagawa}, \citenamefont {Ishida}, \citenamefont {Tokunaga}, \citenamefont {Sakai}, \citenamefont {Kambe}, \citenamefont {Nakamura}, \citenamefont {Shimizu}, \citenamefont {Homma}, \citenamefont {Li}, \citenamefont {Honda},\ and\ \citenamefont {Aoki}}]{FujibayashiJPSJ2022}%
  \BibitemOpen
  \bibfield  {author} {\bibinfo {author} {\bibfnamefont {H.}~\bibnamefont {Fujibayashi}}, \bibinfo {author} {\bibfnamefont {G.}~\bibnamefont {Nakamine}}, \bibinfo {author} {\bibfnamefont {K.}~\bibnamefont {Kinjo}}, \bibinfo {author} {\bibfnamefont {S.}~\bibnamefont {Kitagawa}}, \bibinfo {author} {\bibfnamefont {K.}~\bibnamefont {Ishida}}, \bibinfo {author} {\bibfnamefont {Y.}~\bibnamefont {Tokunaga}}, \bibinfo {author} {\bibfnamefont {H.}~\bibnamefont {Sakai}}, \bibinfo {author} {\bibfnamefont {S.}~\bibnamefont {Kambe}}, \bibinfo {author} {\bibfnamefont {A.}~\bibnamefont {Nakamura}}, \bibinfo {author} {\bibfnamefont {Y.}~\bibnamefont {Shimizu}}, \bibinfo {author} {\bibfnamefont {Y.}~\bibnamefont {Homma}}, \bibinfo {author} {\bibfnamefont {D.}~\bibnamefont {Li}}, \bibinfo {author} {\bibfnamefont {F.}~\bibnamefont {Honda}},\ and\ \bibinfo {author} {\bibfnamefont {D.}~\bibnamefont {Aoki}},\ }\bibfield  {title} {\bibinfo {title} {\text{S}uperconducting \text{O}rder \text{P}arameter in \text{UTe$_2$}
  \text{D}etermined by \text{K}night \text{S}hift \text{M}easurement},\ }\href {https://doi.org/10.7566/JPSJ.91.043705} {\bibfield  {journal} {\bibinfo  {journal} {J. Phys. Soc. Jpn.}\ }\textbf {\bibinfo {volume} {91}},\ \bibinfo {pages} {043705} (\bibinfo {year} {2022})}\BibitemShut {NoStop}%
\bibitem [{\citenamefont {Matsumura}\ \emph {et~al.}(2023)\citenamefont {Matsumura}, \citenamefont {Fujibayashi}, \citenamefont {Kinjo}, \citenamefont {Kitagawa}, \citenamefont {Ishida}, \citenamefont {Tokunaga}, \citenamefont {Sakai}, \citenamefont {Kambe}, \citenamefont {Nakamura}, \citenamefont {Shimizu}, \citenamefont {Homma}, \citenamefont {Li}, \citenamefont {Honda},\ and\ \citenamefont {Aoki}}]{MatsumuraJPSJ2023}%
  \BibitemOpen
  \bibfield  {author} {\bibinfo {author} {\bibfnamefont {H.}~\bibnamefont {Matsumura}}, \bibinfo {author} {\bibfnamefont {H.}~\bibnamefont {Fujibayashi}}, \bibinfo {author} {\bibfnamefont {K.}~\bibnamefont {Kinjo}}, \bibinfo {author} {\bibfnamefont {S.}~\bibnamefont {Kitagawa}}, \bibinfo {author} {\bibfnamefont {K.}~\bibnamefont {Ishida}}, \bibinfo {author} {\bibfnamefont {Y.}~\bibnamefont {Tokunaga}}, \bibinfo {author} {\bibfnamefont {H.}~\bibnamefont {Sakai}}, \bibinfo {author} {\bibfnamefont {S.}~\bibnamefont {Kambe}}, \bibinfo {author} {\bibfnamefont {A.}~\bibnamefont {Nakamura}}, \bibinfo {author} {\bibfnamefont {Y.}~\bibnamefont {Shimizu}}, \bibinfo {author} {\bibfnamefont {Y.}~\bibnamefont {Homma}}, \bibinfo {author} {\bibfnamefont {D.}~\bibnamefont {Li}}, \bibinfo {author} {\bibfnamefont {F.}~\bibnamefont {Honda}},\ and\ \bibinfo {author} {\bibfnamefont {D.}~\bibnamefont {Aoki}},\ }\bibfield  {title} {\bibinfo {title} {\text{Large Reduction in the $a$-axis Knight Shift} on \text{UTe$_2$} with
  \text{$T_c$ = 2.1 K}},\ }\href {https://doi.org/10.7566/jpsj.92.063701} {\bibfield  {journal} {\bibinfo  {journal} {J. Phys. Soc. Jpn.}\ }\textbf {\bibinfo {volume} {92}},\ \bibinfo {pages} {063701} (\bibinfo {year} {2023})}\BibitemShut {NoStop}%
\bibitem [{\citenamefont {Kinjo}\ \emph {et~al.}(2023{\natexlab{b}})\citenamefont {Kinjo}, \citenamefont {Fujibayashi}, \citenamefont {Matsumura}, \citenamefont {Hori}, \citenamefont {Kitagawa}, \citenamefont {Ishida}, \citenamefont {Tokunaga}, \citenamefont {Sakai}, \citenamefont {Kambe}, \citenamefont {Nakamura}, \citenamefont {Shimizu}, \citenamefont {Homma}, \citenamefont {Li}, \citenamefont {Honda},\ and\ \citenamefont {Aoki}}]{KinjoSciAdv2023}%
  \BibitemOpen
  \bibfield  {author} {\bibinfo {author} {\bibfnamefont {K.}~\bibnamefont {Kinjo}}, \bibinfo {author} {\bibfnamefont {H.}~\bibnamefont {Fujibayashi}}, \bibinfo {author} {\bibfnamefont {H.}~\bibnamefont {Matsumura}}, \bibinfo {author} {\bibfnamefont {F.}~\bibnamefont {Hori}}, \bibinfo {author} {\bibfnamefont {S.}~\bibnamefont {Kitagawa}}, \bibinfo {author} {\bibfnamefont {K.}~\bibnamefont {Ishida}}, \bibinfo {author} {\bibfnamefont {Y.}~\bibnamefont {Tokunaga}}, \bibinfo {author} {\bibfnamefont {H.}~\bibnamefont {Sakai}}, \bibinfo {author} {\bibfnamefont {S.}~\bibnamefont {Kambe}}, \bibinfo {author} {\bibfnamefont {A.}~\bibnamefont {Nakamura}}, \bibinfo {author} {\bibfnamefont {Y.}~\bibnamefont {Shimizu}}, \bibinfo {author} {\bibfnamefont {Y.}~\bibnamefont {Homma}}, \bibinfo {author} {\bibfnamefont {D.}~\bibnamefont {Li}}, \bibinfo {author} {\bibfnamefont {F.}~\bibnamefont {Honda}},\ and\ \bibinfo {author} {\bibfnamefont {D.}~\bibnamefont {Aoki}},\ }\bibfield  {title} {\bibinfo {title} {Superconducting spin
  reorientation in spin-triplet multiple superconducting phases of \text{UTe$_2$}},\ }\href {https://doi.org/10.1126/sciadv.adg2736} {\bibfield  {journal} {\bibinfo  {journal} {Sci. Adv.}\ }\textbf {\bibinfo {volume} {9}},\ \bibinfo {pages} {eadg2736} (\bibinfo {year} {2023}{\natexlab{b}})},\ \Eprint {https://arxiv.org/abs/https://www.science.org/doi/pdf/10.1126/sciadv.adg2736} {https://www.science.org/doi/pdf/10.1126/sciadv.adg2736} \BibitemShut {NoStop}%
\bibitem [{\citenamefont {Kittaka}\ \emph {et~al.}(2020)\citenamefont {Kittaka}, \citenamefont {Shimizu}, \citenamefont {Sakakibara}, \citenamefont {Nakamura}, \citenamefont {Li}, \citenamefont {Homma}, \citenamefont {Honda}, \citenamefont {Aoki},\ and\ \citenamefont {Machida}}]{KittakaPRR2020}%
  \BibitemOpen
  \bibfield  {author} {\bibinfo {author} {\bibfnamefont {S.}~\bibnamefont {Kittaka}}, \bibinfo {author} {\bibfnamefont {Y.}~\bibnamefont {Shimizu}}, \bibinfo {author} {\bibfnamefont {T.}~\bibnamefont {Sakakibara}}, \bibinfo {author} {\bibfnamefont {A.}~\bibnamefont {Nakamura}}, \bibinfo {author} {\bibfnamefont {D.}~\bibnamefont {Li}}, \bibinfo {author} {\bibfnamefont {Y.}~\bibnamefont {Homma}}, \bibinfo {author} {\bibfnamefont {F.}~\bibnamefont {Honda}}, \bibinfo {author} {\bibfnamefont {D.}~\bibnamefont {Aoki}},\ and\ \bibinfo {author} {\bibfnamefont {K.}~\bibnamefont {Machida}},\ }\bibfield  {title} {\bibinfo {title} {Orientation of point nodes and nonunitary triplet pairing tuned by the easy-axis magnetization in \text{UTe$_2$}},\ }\bibfield  {journal} {\bibinfo  {journal} {Physical Review Research}\ }\textbf {\bibinfo {volume} {2}},\ \href {https://doi.org/10.1103/physrevresearch.2.032014} {10.1103/physrevresearch.2.032014} (\bibinfo {year} {2020})\BibitemShut {NoStop}%
\bibitem [{\citenamefont {Ishihara}\ \emph {et~al.}(2023{\natexlab{a}})\citenamefont {Ishihara}, \citenamefont {Roppongi}, \citenamefont {Kobayashi}, \citenamefont {Imamura}, \citenamefont {Mizukami}, \citenamefont {Sakai}, \citenamefont {Opletal}, \citenamefont {Tokiwa}, \citenamefont {Haga}, \citenamefont {Hashimoto},\ and\ \citenamefont {Shibauchi}}]{IshiharaNatComm2023}%
  \BibitemOpen
  \bibfield  {author} {\bibinfo {author} {\bibfnamefont {K.}~\bibnamefont {Ishihara}}, \bibinfo {author} {\bibfnamefont {M.}~\bibnamefont {Roppongi}}, \bibinfo {author} {\bibfnamefont {M.}~\bibnamefont {Kobayashi}}, \bibinfo {author} {\bibfnamefont {K.}~\bibnamefont {Imamura}}, \bibinfo {author} {\bibfnamefont {Y.}~\bibnamefont {Mizukami}}, \bibinfo {author} {\bibfnamefont {H.}~\bibnamefont {Sakai}}, \bibinfo {author} {\bibfnamefont {P.}~\bibnamefont {Opletal}}, \bibinfo {author} {\bibfnamefont {Y.}~\bibnamefont {Tokiwa}}, \bibinfo {author} {\bibfnamefont {Y.}~\bibnamefont {Haga}}, \bibinfo {author} {\bibfnamefont {K.}~\bibnamefont {Hashimoto}},\ and\ \bibinfo {author} {\bibfnamefont {T.}~\bibnamefont {Shibauchi}},\ }\bibfield  {title} {\bibinfo {title} {\text{C}hiral superconductivity in \text{UTe$_2$} probed by anisotropic low-energy excitations},\ }\href {https://doi.org/10.1038/s41467-023-38688-y} {\bibfield  {journal} {\bibinfo  {journal} {Nat. Commun.}\ }\textbf {\bibinfo {volume} {14}},\ \bibinfo
  {pages} {2966} (\bibinfo {year} {2023}{\natexlab{a}})}\BibitemShut {NoStop}%
\bibitem [{\citenamefont {Aoki}(2024)}]{AokiJPSJ2024}%
  \BibitemOpen
  \bibfield  {author} {\bibinfo {author} {\bibfnamefont {D.}~\bibnamefont {Aoki}},\ }\bibfield  {title} {\bibinfo {title} {\text{Molten Salt Flux Liquid} \text{Transport Method for} \text{Ultra Clean Single Crystals} \text{UTe$_2$}},\ }\href {https://doi.org/10.7566/jpsj.93.043703} {\bibfield  {journal} {\bibinfo  {journal} {J. Phys. Soc. Jpn.}\ }\textbf {\bibinfo {volume} {93}},\ \bibinfo {pages} {043703} (\bibinfo {year} {2024})}\BibitemShut {NoStop}%
\bibitem [{\citenamefont {Hayes}\ \emph {et~al.}(2025)\citenamefont {Hayes}, \citenamefont {Metz}, \citenamefont {Frank}, \citenamefont {Saha}, \citenamefont {Butch}, \citenamefont {Mishra}, \citenamefont {Hirschfeld},\ and\ \citenamefont {Paglione}}]{HayesPRX2025}%
  \BibitemOpen
  \bibfield  {author} {\bibinfo {author} {\bibfnamefont {I.~M.}\ \bibnamefont {Hayes}}, \bibinfo {author} {\bibfnamefont {T.~E.}\ \bibnamefont {Metz}}, \bibinfo {author} {\bibfnamefont {C.~E.}\ \bibnamefont {Frank}}, \bibinfo {author} {\bibfnamefont {S.~R.}\ \bibnamefont {Saha}}, \bibinfo {author} {\bibfnamefont {N.~P.}\ \bibnamefont {Butch}}, \bibinfo {author} {\bibfnamefont {V.}~\bibnamefont {Mishra}}, \bibinfo {author} {\bibfnamefont {P.~J.}\ \bibnamefont {Hirschfeld}},\ and\ \bibinfo {author} {\bibfnamefont {J.}~\bibnamefont {Paglione}},\ }\bibfield  {title} {\bibinfo {title} {Robust nodal behavior in the thermal conductivity of superconducting \text{UTe$_2$}},\ }\bibfield  {journal} {\bibinfo  {journal} {Physical Review X}\ }\textbf {\bibinfo {volume} {15}},\ \href {https://doi.org/10.1103/physrevx.15.021029} {10.1103/physrevx.15.021029} (\bibinfo {year} {2025})\BibitemShut {NoStop}%
\bibitem [{\citenamefont {Yosida}(1958)}]{YosidaPhyRev1958}%
  \BibitemOpen
  \bibfield  {author} {\bibinfo {author} {\bibfnamefont {K.}~\bibnamefont {Yosida}},\ }\bibfield  {title} {\bibinfo {title} {Paramagnetic susceptibility in superconductors},\ }\href {https://doi.org/10.1103/physrev.110.769} {\bibfield  {journal} {\bibinfo  {journal} {Physical Review}\ }\textbf {\bibinfo {volume} {110}},\ \bibinfo {pages} {769–770} (\bibinfo {year} {1958})}\BibitemShut {NoStop}%
\bibitem [{\citenamefont {MacLaughlin}(1976)}]{DEMacLaughlin1976}%
  \BibitemOpen
  \bibfield  {author} {\bibinfo {author} {\bibfnamefont {D.~E.}\ \bibnamefont {MacLaughlin}},\ }\bibinfo {title} {Magnetic resonance in the superconducting state},\ in\ \href {https://doi.org/10.1016/s0081-1947(08)60541-x} {\emph {\bibinfo {booktitle} {Solid State Physics}}}\ (\bibinfo  {publisher} {Elsevier},\ \bibinfo {year} {1976})\ p.\ \bibinfo {pages} {1–69}\BibitemShut {NoStop}%
\bibitem [{\citenamefont {Matsumura}\ \emph {et~al.}(2025)\citenamefont {Matsumura}, \citenamefont {Takahashi}, \citenamefont {Kinjo}, \citenamefont {Kitagawa}, \citenamefont {Ishida}, \citenamefont {Tokunaga}, \citenamefont {Sakai}, \citenamefont {Kambe}, \citenamefont {Nakamura}, \citenamefont {Shimizu}, \citenamefont {Homma}, \citenamefont {Li}, \citenamefont {Honda}, \citenamefont {Miyake},\ and\ \citenamefont {Aoki}}]{MatsumuraPRB2025}%
  \BibitemOpen
  \bibfield  {author} {\bibinfo {author} {\bibfnamefont {H.}~\bibnamefont {Matsumura}}, \bibinfo {author} {\bibfnamefont {Y.}~\bibnamefont {Takahashi}}, \bibinfo {author} {\bibfnamefont {K.}~\bibnamefont {Kinjo}}, \bibinfo {author} {\bibfnamefont {S.}~\bibnamefont {Kitagawa}}, \bibinfo {author} {\bibfnamefont {K.}~\bibnamefont {Ishida}}, \bibinfo {author} {\bibfnamefont {Y.}~\bibnamefont {Tokunaga}}, \bibinfo {author} {\bibfnamefont {H.}~\bibnamefont {Sakai}}, \bibinfo {author} {\bibfnamefont {S.}~\bibnamefont {Kambe}}, \bibinfo {author} {\bibfnamefont {A.}~\bibnamefont {Nakamura}}, \bibinfo {author} {\bibfnamefont {Y.}~\bibnamefont {Shimizu}}, \bibinfo {author} {\bibfnamefont {Y.}~\bibnamefont {Homma}}, \bibinfo {author} {\bibfnamefont {D.}~\bibnamefont {Li}}, \bibinfo {author} {\bibfnamefont {F.}~\bibnamefont {Honda}}, \bibinfo {author} {\bibfnamefont {A.}~\bibnamefont {Miyake}},\ and\ \bibinfo {author} {\bibfnamefont {D.}~\bibnamefont {Aoki}},\ }\bibfield  {title} {\bibinfo {title} {\text{$b$}-axis and
  \text{$c$}-axis \text{K}night shift measurements in the superconducting state on ultraclean \text{UTe$_2$} with \text{$T_c$ = 2.1 K}},\ }\bibfield  {journal} {\bibinfo  {journal} {Physical Review B}\ }\textbf {\bibinfo {volume} {111}},\ \href {https://doi.org/10.1103/physrevb.111.174526} {10.1103/physrevb.111.174526} (\bibinfo {year} {2025})\BibitemShut {NoStop}%
\bibitem [{\citenamefont {Ishizuka}\ \emph {et~al.}(2019)\citenamefont {Ishizuka}, \citenamefont {Sumita}, \citenamefont {Daido},\ and\ \citenamefont {Yanase}}]{IshizukaPRL2019}%
  \BibitemOpen
  \bibfield  {author} {\bibinfo {author} {\bibfnamefont {J.}~\bibnamefont {Ishizuka}}, \bibinfo {author} {\bibfnamefont {S.}~\bibnamefont {Sumita}}, \bibinfo {author} {\bibfnamefont {A.}~\bibnamefont {Daido}},\ and\ \bibinfo {author} {\bibfnamefont {Y.}~\bibnamefont {Yanase}},\ }\bibfield  {title} {\bibinfo {title} {\text{I}nsulator-\text{M}etal \text{T}ransition and \text{T}opological \text{S}uperconductivity in \text{UTe$_2$} from a \text{F}irst-\text{P}rinciples \text{C}alculation},\ }\href {https://doi.org/10.1103/physrevlett.123.217001} {\bibfield  {journal} {\bibinfo  {journal} {Phys. Rev. Lett.}\ }\textbf {\bibinfo {volume} {123}},\ \bibinfo {pages} {217001} (\bibinfo {year} {2019})}\BibitemShut {NoStop}%
\bibitem [{\citenamefont {Heffner}\ \emph {et~al.}(1989)\citenamefont {Heffner}, \citenamefont {Willis}, \citenamefont {Smith}, \citenamefont {Birrer}, \citenamefont {Baines}, \citenamefont {Gygax}, \citenamefont {Hitti}, \citenamefont {Lippelt}, \citenamefont {Ott}, \citenamefont {Schenck},\ and\ \citenamefont {MacLaughlin}}]{HeffnerPRL1989}%
  \BibitemOpen
  \bibfield  {author} {\bibinfo {author} {\bibfnamefont {R.~H.}\ \bibnamefont {Heffner}}, \bibinfo {author} {\bibfnamefont {J.~O.}\ \bibnamefont {Willis}}, \bibinfo {author} {\bibfnamefont {J.~L.}\ \bibnamefont {Smith}}, \bibinfo {author} {\bibfnamefont {P.}~\bibnamefont {Birrer}}, \bibinfo {author} {\bibfnamefont {C.}~\bibnamefont {Baines}}, \bibinfo {author} {\bibfnamefont {F.~N.}\ \bibnamefont {Gygax}}, \bibinfo {author} {\bibfnamefont {B.}~\bibnamefont {Hitti}}, \bibinfo {author} {\bibfnamefont {E.}~\bibnamefont {Lippelt}}, \bibinfo {author} {\bibfnamefont {H.~R.}\ \bibnamefont {Ott}}, \bibinfo {author} {\bibfnamefont {A.}~\bibnamefont {Schenck}},\ and\ \bibinfo {author} {\bibfnamefont {D.~E.}\ \bibnamefont {MacLaughlin}},\ }\bibfield  {title} {\bibinfo {title} {Muon-spin relaxation studies of weak magnetic correlations in ${\mathrm{u}}_{1\mathrm{\ensuremath{-}}\mathrm{x}}$${\mathrm{th}}_{\mathrm{x}}$${\mathrm{be}}_{13}$},\ }\href {https://doi.org/10.1103/PhysRevB.40.806} {\bibfield  {journal} {\bibinfo
  {journal} {Phys. Rev. B}\ }\textbf {\bibinfo {volume} {40}},\ \bibinfo {pages} {806} (\bibinfo {year} {1989})}\BibitemShut {NoStop}%
\bibitem [{\citenamefont {Matsuno}\ \emph {et~al.}(2017)\citenamefont {Matsuno}, \citenamefont {Morita}, \citenamefont {Kotegawa}, \citenamefont {Tou}, \citenamefont {Haga}, \citenamefont {Yamamoto},\ and\ \citenamefont {Ōnuki}}]{MatsunoJPCP2017}%
  \BibitemOpen
  \bibfield  {author} {\bibinfo {author} {\bibfnamefont {H.}~\bibnamefont {Matsuno}}, \bibinfo {author} {\bibfnamefont {K.}~\bibnamefont {Morita}}, \bibinfo {author} {\bibfnamefont {H.}~\bibnamefont {Kotegawa}}, \bibinfo {author} {\bibfnamefont {H.}~\bibnamefont {Tou}}, \bibinfo {author} {\bibfnamefont {Y.}~\bibnamefont {Haga}}, \bibinfo {author} {\bibfnamefont {E.}~\bibnamefont {Yamamoto}},\ and\ \bibinfo {author} {\bibfnamefont {Y.}~\bibnamefont {Ōnuki}},\ }\bibfield  {title} {\bibinfo {title} {9be-nmr studies on anomalous superconducting phase diagram in \text{UBe$_{13}$}},\ }\href {https://doi.org/10.1088/1742-6596/807/5/052015} {\bibfield  {journal} {\bibinfo  {journal} {Journal of Physics: Conference Series}\ }\textbf {\bibinfo {volume} {807}},\ \bibinfo {pages} {052015} (\bibinfo {year} {2017})}\BibitemShut {NoStop}%
\bibitem [{\citenamefont {Tou}\ \emph {et~al.}(1996)\citenamefont {Tou}, \citenamefont {Kitaoka}, \citenamefont {Asayama}, \citenamefont {Kimura}, \citenamefont {\ifmmode~\bar{O}\else \={O}\fi{}nuki}, \citenamefont {Yamamoto},\ and\ \citenamefont {Maezawa}}]{TouPRL1996}%
  \BibitemOpen
  \bibfield  {author} {\bibinfo {author} {\bibfnamefont {H.}~\bibnamefont {Tou}}, \bibinfo {author} {\bibfnamefont {Y.}~\bibnamefont {Kitaoka}}, \bibinfo {author} {\bibfnamefont {K.}~\bibnamefont {Asayama}}, \bibinfo {author} {\bibfnamefont {N.}~\bibnamefont {Kimura}}, \bibinfo {author} {\bibfnamefont {Y.}~\bibnamefont {\ifmmode~\bar{O}\else \={O}\fi{}nuki}}, \bibinfo {author} {\bibfnamefont {E.}~\bibnamefont {Yamamoto}},\ and\ \bibinfo {author} {\bibfnamefont {K.}~\bibnamefont {Maezawa}},\ }\bibfield  {title} {\bibinfo {title} {Odd-parity superconductivity with parallel spin pairing in ${\mathrm{upt}}_{3}$: Evidence from ${}^{195}\mathrm{Pt}$ knight shift study},\ }\href {https://doi.org/10.1103/PhysRevLett.77.1374} {\bibfield  {journal} {\bibinfo  {journal} {Phys. Rev. Lett.}\ }\textbf {\bibinfo {volume} {77}},\ \bibinfo {pages} {1374} (\bibinfo {year} {1996})}\BibitemShut {NoStop}%
\bibitem [{\citenamefont {Tou}\ \emph {et~al.}(1998)\citenamefont {Tou}, \citenamefont {Kitaoka}, \citenamefont {Ishida}, \citenamefont {Asayama}, \citenamefont {Kimura}, \citenamefont {O\ifmmode\bar\else\textasciimacron\fi{}nuki}, \citenamefont {Yamamoto}, \citenamefont {Haga},\ and\ \citenamefont {Maezawa}}]{TouPRL1998}%
  \BibitemOpen
  \bibfield  {author} {\bibinfo {author} {\bibfnamefont {H.}~\bibnamefont {Tou}}, \bibinfo {author} {\bibfnamefont {Y.}~\bibnamefont {Kitaoka}}, \bibinfo {author} {\bibfnamefont {K.}~\bibnamefont {Ishida}}, \bibinfo {author} {\bibfnamefont {K.}~\bibnamefont {Asayama}}, \bibinfo {author} {\bibfnamefont {N.}~\bibnamefont {Kimura}}, \bibinfo {author} {\bibfnamefont {Y.}~\bibnamefont {O\ifmmode\bar\else\textasciimacron\fi{}nuki}}, \bibinfo {author} {\bibfnamefont {E.}~\bibnamefont {Yamamoto}}, \bibinfo {author} {\bibfnamefont {Y.}~\bibnamefont {Haga}},\ and\ \bibinfo {author} {\bibfnamefont {K.}~\bibnamefont {Maezawa}},\ }\bibfield  {title} {\bibinfo {title} {Nonunitary spin-triplet superconductivity in ${\mathrm{upt}}_{3}$: Evidence from $^{195}pt$ knight shift study},\ }\href {https://doi.org/10.1103/PhysRevLett.80.3129} {\bibfield  {journal} {\bibinfo  {journal} {Phys. Rev. Lett.}\ }\textbf {\bibinfo {volume} {80}},\ \bibinfo {pages} {3129} (\bibinfo {year} {1998})}\BibitemShut {NoStop}%
\bibitem [{\citenamefont {Gannon}\ \emph {et~al.}(2012)\citenamefont {Gannon}, \citenamefont {Halperin}, \citenamefont {Rastovski}, \citenamefont {Eskildsen}, \citenamefont {Dai},\ and\ \citenamefont {Stunault}}]{GannonPRB2012}%
  \BibitemOpen
  \bibfield  {author} {\bibinfo {author} {\bibfnamefont {W.~J.}\ \bibnamefont {Gannon}}, \bibinfo {author} {\bibfnamefont {W.~P.}\ \bibnamefont {Halperin}}, \bibinfo {author} {\bibfnamefont {C.}~\bibnamefont {Rastovski}}, \bibinfo {author} {\bibfnamefont {M.~R.}\ \bibnamefont {Eskildsen}}, \bibinfo {author} {\bibfnamefont {P.}~\bibnamefont {Dai}},\ and\ \bibinfo {author} {\bibfnamefont {A.}~\bibnamefont {Stunault}},\ }\bibfield  {title} {\bibinfo {title} {Magnetization in the superconducting state of upt${}_{3}$ from polarized neutron diffraction},\ }\href {https://doi.org/10.1103/PhysRevB.86.104510} {\bibfield  {journal} {\bibinfo  {journal} {Phys. Rev. B}\ }\textbf {\bibinfo {volume} {86}},\ \bibinfo {pages} {104510} (\bibinfo {year} {2012})}\BibitemShut {NoStop}%
\bibitem [{\citenamefont {Gannon}\ \emph {et~al.}(2017)\citenamefont {Gannon}, \citenamefont {Halperin}, \citenamefont {Eskildsen}, \citenamefont {Dai}, \citenamefont {Hansen}, \citenamefont {Lefmann},\ and\ \citenamefont {Stunault}}]{GannonPRB2017}%
  \BibitemOpen
  \bibfield  {author} {\bibinfo {author} {\bibfnamefont {W.~J.}\ \bibnamefont {Gannon}}, \bibinfo {author} {\bibfnamefont {W.~P.}\ \bibnamefont {Halperin}}, \bibinfo {author} {\bibfnamefont {M.~R.}\ \bibnamefont {Eskildsen}}, \bibinfo {author} {\bibfnamefont {P.}~\bibnamefont {Dai}}, \bibinfo {author} {\bibfnamefont {U.~B.}\ \bibnamefont {Hansen}}, \bibinfo {author} {\bibfnamefont {K.}~\bibnamefont {Lefmann}},\ and\ \bibinfo {author} {\bibfnamefont {A.}~\bibnamefont {Stunault}},\ }\bibfield  {title} {\bibinfo {title} {Spin susceptibility of the topological superconductor ${\mathbf{upt}}_{3}$ from polarized neutron diffraction},\ }\href {https://doi.org/10.1103/PhysRevB.96.041111} {\bibfield  {journal} {\bibinfo  {journal} {Phys. Rev. B}\ }\textbf {\bibinfo {volume} {96}},\ \bibinfo {pages} {041111} (\bibinfo {year} {2017})}\BibitemShut {NoStop}%
\bibitem [{\citenamefont {Aoyama}\ \emph {et~al.}(2019)\citenamefont {Aoyama}, \citenamefont {Kotegawa}, \citenamefont {Kimura}, \citenamefont {Yamamoto}, \citenamefont {Haga}, \citenamefont {Ōnuki},\ and\ \citenamefont {Tou}}]{AoyamaJPSJ2019}%
  \BibitemOpen
  \bibfield  {author} {\bibinfo {author} {\bibfnamefont {T.}~\bibnamefont {Aoyama}}, \bibinfo {author} {\bibfnamefont {H.}~\bibnamefont {Kotegawa}}, \bibinfo {author} {\bibfnamefont {N.}~\bibnamefont {Kimura}}, \bibinfo {author} {\bibfnamefont {E.}~\bibnamefont {Yamamoto}}, \bibinfo {author} {\bibfnamefont {Y.}~\bibnamefont {Haga}}, \bibinfo {author} {\bibfnamefont {Y.}~\bibnamefont {Ōnuki}},\ and\ \bibinfo {author} {\bibfnamefont {H.}~\bibnamefont {Tou}},\ }\bibfield  {title} {\bibinfo {title} {Evidence for weak spin–orbit interaction experienced by cooper pairs in the spin-triplet superconductor \text{UPt$_3$}: \text{$^{195}$Pt-NMR} study},\ }\href {https://doi.org/10.7566/jpsj.88.064706} {\bibfield  {journal} {\bibinfo  {journal} {Journal of the Physical Society of Japan}\ }\textbf {\bibinfo {volume} {88}},\ \bibinfo {pages} {064706} (\bibinfo {year} {2019})}\BibitemShut {NoStop}%
\bibitem [{\citenamefont {Yang}\ \emph {et~al.}(2021)\citenamefont {Yang}, \citenamefont {Luo}, \citenamefont {Yi}, \citenamefont {Shi}, \citenamefont {Zhou},\ and\ \citenamefont {Zheng}}]{YangScAd2021}%
  \BibitemOpen
  \bibfield  {author} {\bibinfo {author} {\bibfnamefont {J.}~\bibnamefont {Yang}}, \bibinfo {author} {\bibfnamefont {J.}~\bibnamefont {Luo}}, \bibinfo {author} {\bibfnamefont {C.}~\bibnamefont {Yi}}, \bibinfo {author} {\bibfnamefont {Y.}~\bibnamefont {Shi}}, \bibinfo {author} {\bibfnamefont {Y.}~\bibnamefont {Zhou}},\ and\ \bibinfo {author} {\bibfnamefont {G.-q.}\ \bibnamefont {Zheng}},\ }\bibfield  {title} {\bibinfo {title} {Spin-triplet superconductivity in \text{K$_2$Cr$_3$As$_3$}},\ }\bibfield  {journal} {\bibinfo  {journal} {Science Advances}\ }\textbf {\bibinfo {volume} {7}},\ \href {https://doi.org/10.1126/sciadv.abl4432} {10.1126/sciadv.abl4432} (\bibinfo {year} {2021})\BibitemShut {NoStop}%
\bibitem [{\citenamefont {Sakai}\ \emph {et~al.}(2022)\citenamefont {Sakai}, \citenamefont {Opletal}, \citenamefont {Tokiwa}, \citenamefont {Yamamoto}, \citenamefont {Tokunaga}, \citenamefont {Kambe},\ and\ \citenamefont {Haga}}]{Sakai2022PRM}%
  \BibitemOpen
  \bibfield  {author} {\bibinfo {author} {\bibfnamefont {H.}~\bibnamefont {Sakai}}, \bibinfo {author} {\bibfnamefont {P.}~\bibnamefont {Opletal}}, \bibinfo {author} {\bibfnamefont {Y.}~\bibnamefont {Tokiwa}}, \bibinfo {author} {\bibfnamefont {E.}~\bibnamefont {Yamamoto}}, \bibinfo {author} {\bibfnamefont {Y.}~\bibnamefont {Tokunaga}}, \bibinfo {author} {\bibfnamefont {S.}~\bibnamefont {Kambe}},\ and\ \bibinfo {author} {\bibfnamefont {Y.}~\bibnamefont {Haga}},\ }\bibfield  {title} {\bibinfo {title} {Single crystal growth of superconducting \text{UTe$_2$} by molten salt flux method},\ }\href {https://doi.org/10.1103/physrevmaterials.6.073401} {\bibfield  {journal} {\bibinfo  {journal} {Phys. Rev. Materials}\ }\textbf {\bibinfo {volume} {6}},\ \bibinfo {pages} {073401} (\bibinfo {year} {2022})}\BibitemShut {NoStop}%
\bibitem [{\citenamefont {Carter}\ \emph {et~al.}(1976)\citenamefont {Carter}, \citenamefont {Kahan},\ and\ \citenamefont {Bennett}}]{GCCarter1976}%
  \BibitemOpen
  \bibfield  {author} {\bibinfo {author} {\bibfnamefont {G.~C.}\ \bibnamefont {Carter}}, \bibinfo {author} {\bibfnamefont {D.~J.}\ \bibnamefont {Kahan}},\ and\ \bibinfo {author} {\bibfnamefont {L.~H.}\ \bibnamefont {Bennett}},\ }\href@noop {} {\emph {\bibinfo {title} {Metallic shifts in NMR}}}\ (\bibinfo  {publisher} {Oxford : New York : Pergamon Press},\ \bibinfo {year} {1976})\BibitemShut {NoStop}%
\bibitem [{\citenamefont {Tokunaga}\ \emph {et~al.}(2019)\citenamefont {Tokunaga}, \citenamefont {Sakai}, \citenamefont {Kambe}, \citenamefont {Hattori}, \citenamefont {Higa}, \citenamefont {Nakamine}, \citenamefont {Kitagawa}, \citenamefont {Ishida}, \citenamefont {Nakamura}, \citenamefont {Shimizu}, \citenamefont {Homma}, \citenamefont {Li}, \citenamefont {Honda},\ and\ \citenamefont {Aoki}}]{TokunagaJPSJ2019}%
  \BibitemOpen
  \bibfield  {author} {\bibinfo {author} {\bibfnamefont {Y.}~\bibnamefont {Tokunaga}}, \bibinfo {author} {\bibfnamefont {H.}~\bibnamefont {Sakai}}, \bibinfo {author} {\bibfnamefont {S.}~\bibnamefont {Kambe}}, \bibinfo {author} {\bibfnamefont {T.}~\bibnamefont {Hattori}}, \bibinfo {author} {\bibfnamefont {N.}~\bibnamefont {Higa}}, \bibinfo {author} {\bibfnamefont {G.}~\bibnamefont {Nakamine}}, \bibinfo {author} {\bibfnamefont {S.}~\bibnamefont {Kitagawa}}, \bibinfo {author} {\bibfnamefont {K.}~\bibnamefont {Ishida}}, \bibinfo {author} {\bibfnamefont {A.}~\bibnamefont {Nakamura}}, \bibinfo {author} {\bibfnamefont {Y.}~\bibnamefont {Shimizu}}, \bibinfo {author} {\bibfnamefont {Y.}~\bibnamefont {Homma}}, \bibinfo {author} {\bibfnamefont {D.}~\bibnamefont {Li}}, \bibinfo {author} {\bibfnamefont {F.}~\bibnamefont {Honda}},\ and\ \bibinfo {author} {\bibfnamefont {D.}~\bibnamefont {Aoki}},\ }\bibfield  {title} {\bibinfo {title} {\text{$^{125}$Te-NMR} \text{S}tudy on a \text{S}ingle \text{C}rystal of \text{H}eavy
  \text{F}ermion superconductor \text{UTe$_2$}},\ }\href {https://doi.org/10.7566/jpsj.88.073701} {\bibfield  {journal} {\bibinfo  {journal} {J. Phys. Soc. Jpn.}\ }\textbf {\bibinfo {volume} {88}},\ \bibinfo {pages} {073701} (\bibinfo {year} {2019})}\BibitemShut {NoStop}%
\bibitem [{\citenamefont {Fujibayashi}\ \emph {et~al.}(2023)\citenamefont {Fujibayashi}, \citenamefont {Kinjo}, \citenamefont {Nakamine}, \citenamefont {Kitagawa}, \citenamefont {Ishida}, \citenamefont {Tokunaga}, \citenamefont {Sakai}, \citenamefont {Kambe}, \citenamefont {Nakamura}, \citenamefont {Shimizu}, \citenamefont {Homma}, \citenamefont {Li}, \citenamefont {Honda},\ and\ \citenamefont {Aoki}}]{FujibayashiJPSJ2023}%
  \BibitemOpen
  \bibfield  {author} {\bibinfo {author} {\bibfnamefont {H.}~\bibnamefont {Fujibayashi}}, \bibinfo {author} {\bibfnamefont {K.}~\bibnamefont {Kinjo}}, \bibinfo {author} {\bibfnamefont {G.}~\bibnamefont {Nakamine}}, \bibinfo {author} {\bibfnamefont {S.}~\bibnamefont {Kitagawa}}, \bibinfo {author} {\bibfnamefont {K.}~\bibnamefont {Ishida}}, \bibinfo {author} {\bibfnamefont {Y.}~\bibnamefont {Tokunaga}}, \bibinfo {author} {\bibfnamefont {H.}~\bibnamefont {Sakai}}, \bibinfo {author} {\bibfnamefont {S.}~\bibnamefont {Kambe}}, \bibinfo {author} {\bibfnamefont {A.}~\bibnamefont {Nakamura}}, \bibinfo {author} {\bibfnamefont {Y.}~\bibnamefont {Shimizu}}, \bibinfo {author} {\bibfnamefont {Y.}~\bibnamefont {Homma}}, \bibinfo {author} {\bibfnamefont {D.}~\bibnamefont {Li}}, \bibinfo {author} {\bibfnamefont {F.}~\bibnamefont {Honda}},\ and\ \bibinfo {author} {\bibfnamefont {D.}~\bibnamefont {Aoki}},\ }\bibfield  {title} {\bibinfo {title} {\text{L}ow-\text{T}emperature \text{M}agnetic \text{F}luctuations
  \text{I}nvestigated by \text{$^{125}$Te-NMR} on the \text{U}ranium-\text{B}ased \text{S}uperconductor \text{UTe$_2$}},\ }\href {https://doi.org/10.7566/JPSJ.92.053702} {\bibfield  {journal} {\bibinfo  {journal} {J. Phys. Soc. Jpn.}\ }\textbf {\bibinfo {volume} {92}},\ \bibinfo {pages} {053702} (\bibinfo {year} {2023})}\BibitemShut {NoStop}%
\bibitem [{\citenamefont {Lee}\ \emph {et~al.}(2025)\citenamefont {Lee}, \citenamefont {Woods}, \citenamefont {Rosa}, \citenamefont {Thomas}, \citenamefont {Bauer}, \citenamefont {Lin},\ and\ \citenamefont {Movshovich}}]{LeePRR2025}%
  \BibitemOpen
  \bibfield  {author} {\bibinfo {author} {\bibfnamefont {S.}~\bibnamefont {Lee}}, \bibinfo {author} {\bibfnamefont {A.~J.}\ \bibnamefont {Woods}}, \bibinfo {author} {\bibfnamefont {P.~F.~S.}\ \bibnamefont {Rosa}}, \bibinfo {author} {\bibfnamefont {S.~M.}\ \bibnamefont {Thomas}}, \bibinfo {author} {\bibfnamefont {E.~D.}\ \bibnamefont {Bauer}}, \bibinfo {author} {\bibfnamefont {S.-Z.}\ \bibnamefont {Lin}},\ and\ \bibinfo {author} {\bibfnamefont {R.}~\bibnamefont {Movshovich}},\ }\bibfield  {title} {\bibinfo {title} {\text{A}nisotropic field-induced changes in the superconducting order parameter of \text{UTe$_2$}},\ }\bibfield  {journal} {\bibinfo  {journal} {Physical Review Research}\ }\textbf {\bibinfo {volume} {7}},\ \href {https://doi.org/10.1103/physrevresearch.7.l022053} {10.1103/physrevresearch.7.l022053} (\bibinfo {year} {2025})\BibitemShut {NoStop}%
\bibitem [{\citenamefont {{de Gennes}}(1989)}]{deGennes}%
  \BibitemOpen
  \bibfield  {author} {\bibinfo {author} {\bibfnamefont {P.~G.}\ \bibnamefont {{de Gennes}}},\ }\href@noop {} {\emph {\bibinfo {title} {Superconductivity of Metals and Alloys}}}\ (\bibinfo  {publisher} {Addison-Welsley},\ \bibinfo {year} {1989})\BibitemShut {NoStop}%
\bibitem [{\citenamefont {Ishihara}\ \emph {et~al.}(2023{\natexlab{b}})\citenamefont {Ishihara}, \citenamefont {Kobayashi}, \citenamefont {Imamura}, \citenamefont {Konczykowski}, \citenamefont {Sakai}, \citenamefont {Opletal}, \citenamefont {Tokiwa}, \citenamefont {Haga}, \citenamefont {Hashimoto},\ and\ \citenamefont {Shibauchi}}]{IshiharaPRR2023}%
  \BibitemOpen
  \bibfield  {author} {\bibinfo {author} {\bibfnamefont {K.}~\bibnamefont {Ishihara}}, \bibinfo {author} {\bibfnamefont {M.}~\bibnamefont {Kobayashi}}, \bibinfo {author} {\bibfnamefont {K.}~\bibnamefont {Imamura}}, \bibinfo {author} {\bibfnamefont {M.}~\bibnamefont {Konczykowski}}, \bibinfo {author} {\bibfnamefont {H.}~\bibnamefont {Sakai}}, \bibinfo {author} {\bibfnamefont {P.}~\bibnamefont {Opletal}}, \bibinfo {author} {\bibfnamefont {Y.}~\bibnamefont {Tokiwa}}, \bibinfo {author} {\bibfnamefont {Y.}~\bibnamefont {Haga}}, \bibinfo {author} {\bibfnamefont {K.}~\bibnamefont {Hashimoto}},\ and\ \bibinfo {author} {\bibfnamefont {T.}~\bibnamefont {Shibauchi}},\ }\bibfield  {title} {\bibinfo {title} {\text{A}nisotropic enhancement of lower critical field in ultraclean crystals of spin-triplet superconductor candidate \text{UTe$_2$}},\ }\href {https://doi.org/10.1103/physrevresearch.5.l022002} {\bibfield  {journal} {\bibinfo  {journal} {Phys. Rev. Research}\ }\textbf {\bibinfo {volume} {5}},\ \bibinfo {pages}
  {L022002} (\bibinfo {year} {2023}{\natexlab{b}})}\BibitemShut {NoStop}%
\bibitem [{\citenamefont {Werthamer}\ \emph {et~al.}(1966)\citenamefont {Werthamer}, \citenamefont {Helfand},\ and\ \citenamefont {Hohenberg}}]{WHH}%
  \BibitemOpen
  \bibfield  {author} {\bibinfo {author} {\bibfnamefont {N.~R.}\ \bibnamefont {Werthamer}}, \bibinfo {author} {\bibfnamefont {E.}~\bibnamefont {Helfand}},\ and\ \bibinfo {author} {\bibfnamefont {P.~C.}\ \bibnamefont {Hohenberg}},\ }\bibfield  {title} {\bibinfo {title} {Temperature and purity dependence of the superconducting critical field, \text{$H_{c2}$} iii. electron spin and spin-orbit effects},\ }\href {https://doi.org/10.1103/physrev.147.295} {\bibfield  {journal} {\bibinfo  {journal} {Physical Review}\ }\textbf {\bibinfo {volume} {147}},\ \bibinfo {pages} {295–302} (\bibinfo {year} {1966})}\BibitemShut {NoStop}%
\bibitem [{\citenamefont {Tokiwa}\ \emph {et~al.}(2025)\citenamefont {Tokiwa}, \citenamefont {Opletal}, \citenamefont {Sakai}, \citenamefont {Kubo}, \citenamefont {Kambe}, \citenamefont {Yamamoto}, \citenamefont {Kimata}, \citenamefont {Awaji}, \citenamefont {Sasaki}, \citenamefont {Aoki}, \citenamefont {Yanase}, \citenamefont {Tokunaga},\ and\ \citenamefont {Haga}}]{TokiwaPRL2025}%
  \BibitemOpen
  \bibfield  {author} {\bibinfo {author} {\bibfnamefont {Y.}~\bibnamefont {Tokiwa}}, \bibinfo {author} {\bibfnamefont {P.}~\bibnamefont {Opletal}}, \bibinfo {author} {\bibfnamefont {H.}~\bibnamefont {Sakai}}, \bibinfo {author} {\bibfnamefont {K.}~\bibnamefont {Kubo}}, \bibinfo {author} {\bibfnamefont {S.}~\bibnamefont {Kambe}}, \bibinfo {author} {\bibfnamefont {E.}~\bibnamefont {Yamamoto}}, \bibinfo {author} {\bibfnamefont {M.}~\bibnamefont {Kimata}}, \bibinfo {author} {\bibfnamefont {S.}~\bibnamefont {Awaji}}, \bibinfo {author} {\bibfnamefont {T.}~\bibnamefont {Sasaki}}, \bibinfo {author} {\bibfnamefont {D.}~\bibnamefont {Aoki}}, \bibinfo {author} {\bibfnamefont {Y.}~\bibnamefont {Yanase}}, \bibinfo {author} {\bibfnamefont {Y.}~\bibnamefont {Tokunaga}},\ and\ \bibinfo {author} {\bibfnamefont {Y.}~\bibnamefont {Haga}},\ }\href {https://doi.org/10.1103/z8yx-yzdh} {\bibinfo {title} {\text{S}elf-\text{R}econstruction of order parameter in spin-triplet superconductor \text{UTe$_2$}}} (\bibinfo {year}
  {2025})\BibitemShut {NoStop}%
\bibitem [{\citenamefont {Haga}\ \emph {et~al.}(2022)\citenamefont {Haga}, \citenamefont {Opletal}, \citenamefont {Tokiwa}, \citenamefont {Yamamoto}, \citenamefont {Tokunaga}, \citenamefont {Kambe},\ and\ \citenamefont {Sakai}}]{Haga2022JPCM}%
  \BibitemOpen
  \bibfield  {author} {\bibinfo {author} {\bibfnamefont {Y.}~\bibnamefont {Haga}}, \bibinfo {author} {\bibfnamefont {P.}~\bibnamefont {Opletal}}, \bibinfo {author} {\bibfnamefont {Y.}~\bibnamefont {Tokiwa}}, \bibinfo {author} {\bibfnamefont {E.}~\bibnamefont {Yamamoto}}, \bibinfo {author} {\bibfnamefont {Y.}~\bibnamefont {Tokunaga}}, \bibinfo {author} {\bibfnamefont {S.}~\bibnamefont {Kambe}},\ and\ \bibinfo {author} {\bibfnamefont {H.}~\bibnamefont {Sakai}},\ }\bibfield  {title} {\bibinfo {title} {\text{E}ffect of uranium deficiency on normal and superconducting properties in unconventional superconductor \text{UTe$_2$}},\ }\href {https://doi.org/10.1088/1361-648x/ac5201} {\bibfield  {journal} {\bibinfo  {journal} {Journal of Physics: Condensed Matter}\ }\textbf {\bibinfo {volume} {34}},\ \bibinfo {pages} {175601} (\bibinfo {year} {2022})}\BibitemShut {NoStop}%
\bibitem [{\citenamefont {Miyake}\ \emph {et~al.}(2019)\citenamefont {Miyake}, \citenamefont {Shimizu}, \citenamefont {Sato}, \citenamefont {Li}, \citenamefont {Nakamura}, \citenamefont {Homma}, \citenamefont {Honda}, \citenamefont {Flouquet}, \citenamefont {Tokunaga},\ and\ \citenamefont {Aoki}}]{MiyakeJPSJ2019}%
  \BibitemOpen
  \bibfield  {author} {\bibinfo {author} {\bibfnamefont {A.}~\bibnamefont {Miyake}}, \bibinfo {author} {\bibfnamefont {Y.}~\bibnamefont {Shimizu}}, \bibinfo {author} {\bibfnamefont {Y.~J.}\ \bibnamefont {Sato}}, \bibinfo {author} {\bibfnamefont {D.}~\bibnamefont {Li}}, \bibinfo {author} {\bibfnamefont {A.}~\bibnamefont {Nakamura}}, \bibinfo {author} {\bibfnamefont {Y.}~\bibnamefont {Homma}}, \bibinfo {author} {\bibfnamefont {F.}~\bibnamefont {Honda}}, \bibinfo {author} {\bibfnamefont {J.}~\bibnamefont {Flouquet}}, \bibinfo {author} {\bibfnamefont {M.}~\bibnamefont {Tokunaga}},\ and\ \bibinfo {author} {\bibfnamefont {D.}~\bibnamefont {Aoki}},\ }\bibfield  {title} {\bibinfo {title} {Metamagnetic transition in heavy fermion superconductor \text{UTe$_2$}},\ }\href {https://doi.org/10.7566/JPSJ.88.063706} {\bibfield  {journal} {\bibinfo  {journal} {Journal of the Physical Society of Japan}\ }\textbf {\bibinfo {volume} {88}},\ \bibinfo {pages} {063706} (\bibinfo {year} {2019})},\ \Eprint
  {https://arxiv.org/abs/https://doi.org/10.7566/JPSJ.88.063706} {https://doi.org/10.7566/JPSJ.88.063706} \BibitemShut {NoStop}%
\bibitem [{\citenamefont {Rosuel}\ \emph {et~al.}(2023)\citenamefont {Rosuel}, \citenamefont {Marcenat}, \citenamefont {Knebel}, \citenamefont {Klein}, \citenamefont {Pourret}, \citenamefont {Marquardt}, \citenamefont {Niu}, \citenamefont {Rousseau}, \citenamefont {Demuer}, \citenamefont {Seyfarth}, \citenamefont {Lapertot}, \citenamefont {Aoki}, \citenamefont {Braithwaite}, \citenamefont {Flouquet},\ and\ \citenamefont {Brison}}]{RosuelPRX2023}%
  \BibitemOpen
  \bibfield  {author} {\bibinfo {author} {\bibfnamefont {A.}~\bibnamefont {Rosuel}}, \bibinfo {author} {\bibfnamefont {C.}~\bibnamefont {Marcenat}}, \bibinfo {author} {\bibfnamefont {G.}~\bibnamefont {Knebel}}, \bibinfo {author} {\bibfnamefont {T.}~\bibnamefont {Klein}}, \bibinfo {author} {\bibfnamefont {A.}~\bibnamefont {Pourret}}, \bibinfo {author} {\bibfnamefont {N.}~\bibnamefont {Marquardt}}, \bibinfo {author} {\bibfnamefont {Q.}~\bibnamefont {Niu}}, \bibinfo {author} {\bibfnamefont {S.}~\bibnamefont {Rousseau}}, \bibinfo {author} {\bibfnamefont {A.}~\bibnamefont {Demuer}}, \bibinfo {author} {\bibfnamefont {G.}~\bibnamefont {Seyfarth}}, \bibinfo {author} {\bibfnamefont {G.}~\bibnamefont {Lapertot}}, \bibinfo {author} {\bibfnamefont {D.}~\bibnamefont {Aoki}}, \bibinfo {author} {\bibfnamefont {D.}~\bibnamefont {Braithwaite}}, \bibinfo {author} {\bibfnamefont {J.}~\bibnamefont {Flouquet}},\ and\ \bibinfo {author} {\bibfnamefont {J.}~\bibnamefont {Brison}},\ }\bibfield  {title} {\bibinfo {title}
  {Field-induced tuning of the pairing state in a superconductor},\ }\bibfield  {journal} {\bibinfo  {journal} {Physical Review X}\ }\textbf {\bibinfo {volume} {13}},\ \href {https://doi.org/10.1103/physrevx.13.011022} {10.1103/physrevx.13.011022} (\bibinfo {year} {2023})\BibitemShut {NoStop}%
\bibitem [{\citenamefont {Wu}\ \emph {et~al.}(2025)\citenamefont {Wu}, \citenamefont {Chen}, \citenamefont {Weinberger}, \citenamefont {Cabala}, \citenamefont {Sechovsk\'y}, \citenamefont {Vali\ifmmode~\check{s}\else \v{s}\fi{}ka}, \citenamefont {Alireza}, \citenamefont {Eaton},\ and\ \citenamefont {Grosche}}]{WuPRL2025}%
  \BibitemOpen
  \bibfield  {author} {\bibinfo {author} {\bibfnamefont {Z.}~\bibnamefont {Wu}}, \bibinfo {author} {\bibfnamefont {J.}~\bibnamefont {Chen}}, \bibinfo {author} {\bibfnamefont {T.~I.}\ \bibnamefont {Weinberger}}, \bibinfo {author} {\bibfnamefont {A.}~\bibnamefont {Cabala}}, \bibinfo {author} {\bibfnamefont {V.}~\bibnamefont {Sechovsk\'y}}, \bibinfo {author} {\bibfnamefont {M.}~\bibnamefont {Vali\ifmmode~\check{s}\else \v{s}\fi{}ka}}, \bibinfo {author} {\bibfnamefont {P.~L.}\ \bibnamefont {Alireza}}, \bibinfo {author} {\bibfnamefont {A.~G.}\ \bibnamefont {Eaton}},\ and\ \bibinfo {author} {\bibfnamefont {F.~M.}\ \bibnamefont {Grosche}},\ }\bibfield  {title} {\bibinfo {title} {Magnetic signatures of pressure-induced multicomponent superconductivity in ${\mathrm{ute}}_{2}$},\ }\href {https://doi.org/10.1103/PhysRevLett.134.236501} {\bibfield  {journal} {\bibinfo  {journal} {Phys. Rev. Lett.}\ }\textbf {\bibinfo {volume} {134}},\ \bibinfo {pages} {236501} (\bibinfo {year} {2025})}\BibitemShut {NoStop}%
\bibitem [{\citenamefont {Tokiwa}\ \emph {et~al.}(2023)\citenamefont {Tokiwa}, \citenamefont {Sakai}, \citenamefont {Kambe}, \citenamefont {Opletal}, \citenamefont {Yamamoto}, \citenamefont {Kimata}, \citenamefont {Awaji}, \citenamefont {Sasaki}, \citenamefont {Yanase}, \citenamefont {Haga},\ and\ \citenamefont {Tokunaga}}]{TokiwaPRB2023A}%
  \BibitemOpen
  \bibfield  {author} {\bibinfo {author} {\bibfnamefont {Y.}~\bibnamefont {Tokiwa}}, \bibinfo {author} {\bibfnamefont {H.}~\bibnamefont {Sakai}}, \bibinfo {author} {\bibfnamefont {S.}~\bibnamefont {Kambe}}, \bibinfo {author} {\bibfnamefont {P.}~\bibnamefont {Opletal}}, \bibinfo {author} {\bibfnamefont {E.}~\bibnamefont {Yamamoto}}, \bibinfo {author} {\bibfnamefont {M.}~\bibnamefont {Kimata}}, \bibinfo {author} {\bibfnamefont {S.}~\bibnamefont {Awaji}}, \bibinfo {author} {\bibfnamefont {T.}~\bibnamefont {Sasaki}}, \bibinfo {author} {\bibfnamefont {Y.}~\bibnamefont {Yanase}}, \bibinfo {author} {\bibfnamefont {Y.}~\bibnamefont {Haga}},\ and\ \bibinfo {author} {\bibfnamefont {Y.}~\bibnamefont {Tokunaga}},\ }\bibfield  {title} {\bibinfo {title} {Anomalous vortex dynamics in the spin-triplet superconductor \text{UTe$_2$}},\ }\bibfield  {journal} {\bibinfo  {journal} {Physical Review B}\ }\textbf {\bibinfo {volume} {108}},\ \href {https://doi.org/10.1103/physrevb.108.144502} {10.1103/physrevb.108.144502} (\bibinfo
  {year} {2023})\BibitemShut {NoStop}%
\bibitem [{\citenamefont {Li}\ \emph {et~al.}(2022)\citenamefont {Li}, \citenamefont {Kang}, \citenamefont {Zhao}, \citenamefont {Lei}, \citenamefont {Zhou}, \citenamefont {Song}, \citenamefont {Li}, \citenamefont {Zheng}, \citenamefont {Nie}, \citenamefont {Wu},\ and\ \citenamefont {Chen}}]{LiPRB2022}%
  \BibitemOpen
  \bibfield  {author} {\bibinfo {author} {\bibfnamefont {J.}~\bibnamefont {Li}}, \bibinfo {author} {\bibfnamefont {B.~L.}\ \bibnamefont {Kang}}, \bibinfo {author} {\bibfnamefont {D.}~\bibnamefont {Zhao}}, \bibinfo {author} {\bibfnamefont {B.}~\bibnamefont {Lei}}, \bibinfo {author} {\bibfnamefont {Y.~B.}\ \bibnamefont {Zhou}}, \bibinfo {author} {\bibfnamefont {D.~W.}\ \bibnamefont {Song}}, \bibinfo {author} {\bibfnamefont {S.~J.}\ \bibnamefont {Li}}, \bibinfo {author} {\bibfnamefont {L.~X.}\ \bibnamefont {Zheng}}, \bibinfo {author} {\bibfnamefont {L.~P.}\ \bibnamefont {Nie}}, \bibinfo {author} {\bibfnamefont {T.}~\bibnamefont {Wu}},\ and\ \bibinfo {author} {\bibfnamefont {X.~H.}\ \bibnamefont {Chen}},\ }\bibfield  {title} {\bibinfo {title} {\text{$^{77}$Se-NMR} evidence for spin-singlet superconductivity with exotic superconducting fluctuations in \text{FeSe}},\ }\bibfield  {journal} {\bibinfo  {journal} {Physical Review B}\ }\textbf {\bibinfo {volume} {105}},\ \href
  {https://doi.org/10.1103/physrevb.105.054514} {10.1103/physrevb.105.054514} (\bibinfo {year} {2022})\BibitemShut {NoStop}%
\bibitem [{\citenamefont {Kinjo}\ \emph {et~al.}(2019)\citenamefont {Kinjo}, \citenamefont {Kitagawa}, \citenamefont {Nakai}, \citenamefont {Ishida}, \citenamefont {Sugawara},\ and\ \citenamefont {Sato}}]{KinjoJPSJ2019}%
  \BibitemOpen
  \bibfield  {author} {\bibinfo {author} {\bibfnamefont {K.}~\bibnamefont {Kinjo}}, \bibinfo {author} {\bibfnamefont {S.}~\bibnamefont {Kitagawa}}, \bibinfo {author} {\bibfnamefont {Y.}~\bibnamefont {Nakai}}, \bibinfo {author} {\bibfnamefont {K.}~\bibnamefont {Ishida}}, \bibinfo {author} {\bibfnamefont {H.}~\bibnamefont {Sugawara}},\ and\ \bibinfo {author} {\bibfnamefont {H.}~\bibnamefont {Sato}},\ }\bibfield  {title} {\bibinfo {title} {\text{M}agnetic \text{F}ield \text{E}ffect on s-wave \text{S}uperconductor \text{LaRu$_4$P$_{12}$} \text{S}tudied by \text{$^{31}$P-NMR}},\ }\href {https://doi.org/10.7566/jpsj.88.065002} {\bibfield  {journal} {\bibinfo  {journal} {Journal of the Physical Society of Japan}\ }\textbf {\bibinfo {volume} {88}},\ \bibinfo {pages} {065002} (\bibinfo {year} {2019})}\BibitemShut {NoStop}%
\bibitem [{\citenamefont {Koutroulakis}\ \emph {et~al.}(2008)\citenamefont {Koutroulakis}, \citenamefont {Mitrović}, \citenamefont {Horvatić}, \citenamefont {Berthier}, \citenamefont {Lapertot},\ and\ \citenamefont {Flouquet}}]{KoutroulakisPRL2008}%
  \BibitemOpen
  \bibfield  {author} {\bibinfo {author} {\bibfnamefont {G.}~\bibnamefont {Koutroulakis}}, \bibinfo {author} {\bibfnamefont {V.~F.}\ \bibnamefont {Mitrović}}, \bibinfo {author} {\bibfnamefont {M.}~\bibnamefont {Horvatić}}, \bibinfo {author} {\bibfnamefont {C.}~\bibnamefont {Berthier}}, \bibinfo {author} {\bibfnamefont {G.}~\bibnamefont {Lapertot}},\ and\ \bibinfo {author} {\bibfnamefont {J.}~\bibnamefont {Flouquet}},\ }\bibfield  {title} {\bibinfo {title} {\text{F}ield \text{D}ependence of the \text{G}round \text{S}tate in the \text{E}xotic \text{S}uperconductor \text{CeCoIn$_5$}: \text{A} \text{N}uclear \text{M}agnetic \text{R}esonance \text{I}nvestigation},\ }\bibfield  {journal} {\bibinfo  {journal} {Physical Review Letters}\ }\textbf {\bibinfo {volume} {101}},\ \href {https://doi.org/10.1103/physrevlett.101.047004} {10.1103/physrevlett.101.047004} (\bibinfo {year} {2008})\BibitemShut {NoStop}%
\bibitem [{\citenamefont {Chronister}\ \emph {et~al.}(2021)\citenamefont {Chronister}, \citenamefont {Pustogow}, \citenamefont {Kikugawa}, \citenamefont {Sokolov}, \citenamefont {Jerzembeck}, \citenamefont {Hicks}, \citenamefont {Mackenzie}, \citenamefont {Bauer},\ and\ \citenamefont {Brown}}]{ChronisterPNAS2021}%
  \BibitemOpen
  \bibfield  {author} {\bibinfo {author} {\bibfnamefont {A.}~\bibnamefont {Chronister}}, \bibinfo {author} {\bibfnamefont {A.}~\bibnamefont {Pustogow}}, \bibinfo {author} {\bibfnamefont {N.}~\bibnamefont {Kikugawa}}, \bibinfo {author} {\bibfnamefont {D.~A.}\ \bibnamefont {Sokolov}}, \bibinfo {author} {\bibfnamefont {F.}~\bibnamefont {Jerzembeck}}, \bibinfo {author} {\bibfnamefont {C.~W.}\ \bibnamefont {Hicks}}, \bibinfo {author} {\bibfnamefont {A.~P.}\ \bibnamefont {Mackenzie}}, \bibinfo {author} {\bibfnamefont {E.~D.}\ \bibnamefont {Bauer}},\ and\ \bibinfo {author} {\bibfnamefont {S.~E.}\ \bibnamefont {Brown}},\ }\bibfield  {title} {\bibinfo {title} {\text{E}vidence for even parity unconventional superconductivity in \text{Sr$_2$RuO$_4$}},\ }\bibfield  {journal} {\bibinfo  {journal} {Proceedings of the National Academy of Sciences}\ }\textbf {\bibinfo {volume} {118}},\ \href {https://doi.org/10.1073/pnas.2025313118} {10.1073/pnas.2025313118} (\bibinfo {year} {2021})\BibitemShut {NoStop}%
\bibitem [{\citenamefont {Aoki}\ \emph {et~al.}(2022{\natexlab{b}})\citenamefont {Aoki}, \citenamefont {Sakai}, \citenamefont {Opletal}, \citenamefont {Tokiwa}, \citenamefont {Ishizuka}, \citenamefont {Yanase}, \citenamefont {Harima}, \citenamefont {Nakamura}, \citenamefont {Li}, \citenamefont {Homma}, \citenamefont {Shimizu}, \citenamefont {Knebel}, \citenamefont {Flouquet},\ and\ \citenamefont {Haga}}]{aokiJPSJ2022}%
  \BibitemOpen
  \bibfield  {author} {\bibinfo {author} {\bibfnamefont {D.}~\bibnamefont {Aoki}}, \bibinfo {author} {\bibfnamefont {H.}~\bibnamefont {Sakai}}, \bibinfo {author} {\bibfnamefont {P.}~\bibnamefont {Opletal}}, \bibinfo {author} {\bibfnamefont {Y.}~\bibnamefont {Tokiwa}}, \bibinfo {author} {\bibfnamefont {J.}~\bibnamefont {Ishizuka}}, \bibinfo {author} {\bibfnamefont {Y.}~\bibnamefont {Yanase}}, \bibinfo {author} {\bibfnamefont {H.}~\bibnamefont {Harima}}, \bibinfo {author} {\bibfnamefont {A.}~\bibnamefont {Nakamura}}, \bibinfo {author} {\bibfnamefont {D.}~\bibnamefont {Li}}, \bibinfo {author} {\bibfnamefont {Y.}~\bibnamefont {Homma}}, \bibinfo {author} {\bibfnamefont {Y.}~\bibnamefont {Shimizu}}, \bibinfo {author} {\bibfnamefont {G.}~\bibnamefont {Knebel}}, \bibinfo {author} {\bibfnamefont {J.}~\bibnamefont {Flouquet}},\ and\ \bibinfo {author} {\bibfnamefont {Y.}~\bibnamefont {Haga}},\ }\bibfield  {title} {\bibinfo {title} {First observation of the de haas–van alphen effect and fermi surfaces in the
  unconventional superconductor \text{UTe$_2$}},\ }\href {https://doi.org/10.7566/JPSJ.91.083704} {\bibfield  {journal} {\bibinfo  {journal} {Journal of the Physical Society of Japan}\ }\textbf {\bibinfo {volume} {91}},\ \bibinfo {pages} {083704} (\bibinfo {year} {2022}{\natexlab{b}})}\BibitemShut {NoStop}%
\bibitem [{\citenamefont {Eaton}\ \emph {et~al.}(2024)\citenamefont {Eaton}, \citenamefont {Weinberger}, \citenamefont {Popiel}, \citenamefont {Wu}, \citenamefont {Hickey}, \citenamefont {Cabala}, \citenamefont {Pospíšil}, \citenamefont {Prokleška}, \citenamefont {Haidamak}, \citenamefont {Bastien}, \citenamefont {Opletal}, \citenamefont {Sakai}, \citenamefont {Haga}, \citenamefont {Nowell}, \citenamefont {Benjamin}, \citenamefont {Sechovský}, \citenamefont {Lonzarich}, \citenamefont {Grosche},\ and\ \citenamefont {Vališka}}]{EatonNatComm2024}%
  \BibitemOpen
  \bibfield  {author} {\bibinfo {author} {\bibfnamefont {A.~G.}\ \bibnamefont {Eaton}}, \bibinfo {author} {\bibfnamefont {T.~I.}\ \bibnamefont {Weinberger}}, \bibinfo {author} {\bibfnamefont {N.~J.~M.}\ \bibnamefont {Popiel}}, \bibinfo {author} {\bibfnamefont {Z.}~\bibnamefont {Wu}}, \bibinfo {author} {\bibfnamefont {A.~J.}\ \bibnamefont {Hickey}}, \bibinfo {author} {\bibfnamefont {A.}~\bibnamefont {Cabala}}, \bibinfo {author} {\bibfnamefont {J.}~\bibnamefont {Pospíšil}}, \bibinfo {author} {\bibfnamefont {J.}~\bibnamefont {Prokleška}}, \bibinfo {author} {\bibfnamefont {T.}~\bibnamefont {Haidamak}}, \bibinfo {author} {\bibfnamefont {G.}~\bibnamefont {Bastien}}, \bibinfo {author} {\bibfnamefont {P.}~\bibnamefont {Opletal}}, \bibinfo {author} {\bibfnamefont {H.}~\bibnamefont {Sakai}}, \bibinfo {author} {\bibfnamefont {Y.}~\bibnamefont {Haga}}, \bibinfo {author} {\bibfnamefont {R.}~\bibnamefont {Nowell}}, \bibinfo {author} {\bibfnamefont {S.~M.}\ \bibnamefont {Benjamin}}, \bibinfo {author} {\bibfnamefont
  {V.}~\bibnamefont {Sechovský}}, \bibinfo {author} {\bibfnamefont {G.~G.}\ \bibnamefont {Lonzarich}}, \bibinfo {author} {\bibfnamefont {F.~M.}\ \bibnamefont {Grosche}},\ and\ \bibinfo {author} {\bibfnamefont {M.}~\bibnamefont {Vališka}},\ }\bibfield  {title} {\bibinfo {title} {Quasi-2d fermi surface in the anomalous superconductor \text{UTe$_2$}},\ }\bibfield  {journal} {\bibinfo  {journal} {Nature Communications}\ }\textbf {\bibinfo {volume} {15}},\ \href {https://doi.org/10.1038/s41467-023-44110-4} {10.1038/s41467-023-44110-4} (\bibinfo {year} {2024})\BibitemShut {NoStop}%
\bibitem [{\citenamefont {Ambika}\ \emph {et~al.}(2022)\citenamefont {Ambika}, \citenamefont {Ding}, \citenamefont {Rana}, \citenamefont {Frank}, \citenamefont {Green}, \citenamefont {Ran}, \citenamefont {Butch},\ and\ \citenamefont {Furukawa}}]{AmbikaPRB2022}%
  \BibitemOpen
  \bibfield  {author} {\bibinfo {author} {\bibfnamefont {D.~V.}\ \bibnamefont {Ambika}}, \bibinfo {author} {\bibfnamefont {Q.-P.}\ \bibnamefont {Ding}}, \bibinfo {author} {\bibfnamefont {K.}~\bibnamefont {Rana}}, \bibinfo {author} {\bibfnamefont {C.~E.}\ \bibnamefont {Frank}}, \bibinfo {author} {\bibfnamefont {E.~L.}\ \bibnamefont {Green}}, \bibinfo {author} {\bibfnamefont {S.}~\bibnamefont {Ran}}, \bibinfo {author} {\bibfnamefont {N.~P.}\ \bibnamefont {Butch}},\ and\ \bibinfo {author} {\bibfnamefont {Y.}~\bibnamefont {Furukawa}},\ }\bibfield  {title} {\bibinfo {title} {\text{P}ossible coexistence of antiferromagnetic and ferromagnetic spin fluctuations in the spin-triplet superconductor \text{UTe$_2$} revealed by \text{Te} \text{NMR} under pressure},\ }\href {https://doi.org/10.1103/physrevb.105.l220403} {\bibfield  {journal} {\bibinfo  {journal} {Phys. Rev. B}\ }\textbf {\bibinfo {volume} {105}},\ \bibinfo {pages} {L220403} (\bibinfo {year} {2022})}\BibitemShut {NoStop}%
\bibitem [{\citenamefont {Suetsugu}\ \emph {et~al.}(2024)\citenamefont {Suetsugu}, \citenamefont {Shimomura}, \citenamefont {Kamimura}, \citenamefont {Asaba}, \citenamefont {Asaeda}, \citenamefont {Kosuge}, \citenamefont {Sekino}, \citenamefont {Ikemori}, \citenamefont {Kasahara}, \citenamefont {Kohsaka}, \citenamefont {Lee}, \citenamefont {Yanase}, \citenamefont {Sakai}, \citenamefont {Opletal}, \citenamefont {Tokiwa}, \citenamefont {Haga},\ and\ \citenamefont {Matsuda}}]{SuetsuguSciAdv2024}%
  \BibitemOpen
  \bibfield  {author} {\bibinfo {author} {\bibfnamefont {S.}~\bibnamefont {Suetsugu}}, \bibinfo {author} {\bibfnamefont {M.}~\bibnamefont {Shimomura}}, \bibinfo {author} {\bibfnamefont {M.}~\bibnamefont {Kamimura}}, \bibinfo {author} {\bibfnamefont {T.}~\bibnamefont {Asaba}}, \bibinfo {author} {\bibfnamefont {H.}~\bibnamefont {Asaeda}}, \bibinfo {author} {\bibfnamefont {Y.}~\bibnamefont {Kosuge}}, \bibinfo {author} {\bibfnamefont {Y.}~\bibnamefont {Sekino}}, \bibinfo {author} {\bibfnamefont {S.}~\bibnamefont {Ikemori}}, \bibinfo {author} {\bibfnamefont {Y.}~\bibnamefont {Kasahara}}, \bibinfo {author} {\bibfnamefont {Y.}~\bibnamefont {Kohsaka}}, \bibinfo {author} {\bibfnamefont {M.}~\bibnamefont {Lee}}, \bibinfo {author} {\bibfnamefont {Y.}~\bibnamefont {Yanase}}, \bibinfo {author} {\bibfnamefont {H.}~\bibnamefont {Sakai}}, \bibinfo {author} {\bibfnamefont {P.}~\bibnamefont {Opletal}}, \bibinfo {author} {\bibfnamefont {Y.}~\bibnamefont {Tokiwa}}, \bibinfo {author} {\bibfnamefont {Y.}~\bibnamefont {Haga}},\
  and\ \bibinfo {author} {\bibfnamefont {Y.}~\bibnamefont {Matsuda}},\ }\bibfield  {title} {\bibinfo {title} {Fully gapped pairing state in spin-triplet superconductor ute 2},\ }\href {https://doi.org/10.1126/sciadv.adk3772} {\bibfield  {journal} {\bibinfo  {journal} {Sci. Adv.}\ }\textbf {\bibinfo {volume} {10}},\ \bibinfo {pages} {eadk3772} (\bibinfo {year} {2024})}\BibitemShut {NoStop}%
\bibitem [{\citenamefont {Tei}\ \emph {et~al.}(2024)\citenamefont {Tei}, \citenamefont {Mizushima},\ and\ \citenamefont {Fujimoto}}]{TeiPRB2024}%
  \BibitemOpen
  \bibfield  {author} {\bibinfo {author} {\bibfnamefont {J.}~\bibnamefont {Tei}}, \bibinfo {author} {\bibfnamefont {T.}~\bibnamefont {Mizushima}},\ and\ \bibinfo {author} {\bibfnamefont {S.}~\bibnamefont {Fujimoto}},\ }\bibfield  {title} {\bibinfo {title} {Pairing symmetries of multiple superconducting phases in \text{UTe$_2$} : Competition between ferromagnetic and antiferromagnetic fluctuations},\ }\href@noop {} {\bibfield  {journal} {\bibinfo  {journal} {Phys. Rev. B}\ }\textbf {\bibinfo {volume} {109}} (\bibinfo {year} {2024})}\BibitemShut {NoStop}%
\bibitem [{\citenamefont {Hakuno}\ \emph {et~al.}(2024)\citenamefont {Hakuno}, \citenamefont {Nogaki},\ and\ \citenamefont {Yanase}}]{HakunoPRB2024}%
  \BibitemOpen
  \bibfield  {author} {\bibinfo {author} {\bibfnamefont {R.}~\bibnamefont {Hakuno}}, \bibinfo {author} {\bibfnamefont {K.}~\bibnamefont {Nogaki}},\ and\ \bibinfo {author} {\bibfnamefont {Y.}~\bibnamefont {Yanase}},\ }\bibfield  {title} {\bibinfo {title} {Magnetism and superconductivity in mixed-dimensional periodic anderson model for \text{UTe$_2$}},\ }\bibfield  {journal} {\bibinfo  {journal} {Phys. Rev. B}\ }\textbf {\bibinfo {volume} {109}},\ \href {https://doi.org/10.1103/physrevb.109.104509} {10.1103/physrevb.109.104509} (\bibinfo {year} {2024})\BibitemShut {NoStop}%
\bibitem [{\citenamefont {Tokunaga}\ \emph {et~al.}(2023)\citenamefont {Tokunaga}, \citenamefont {Sakai}, \citenamefont {Kambe}, \citenamefont {Opletal}, \citenamefont {Tokiwa}, \citenamefont {Haga}, \citenamefont {Kitagawa}, \citenamefont {Ishida}, \citenamefont {Aoki}, \citenamefont {Knebel}, \citenamefont {Lapertot}, \citenamefont {Kr\"amer},\ and\ \citenamefont {Horvati\ifmmode~\acute{c}\else \'{c}\fi{}}}]{TokunagaPRL2023}%
  \BibitemOpen
  \bibfield  {author} {\bibinfo {author} {\bibfnamefont {Y.}~\bibnamefont {Tokunaga}}, \bibinfo {author} {\bibfnamefont {H.}~\bibnamefont {Sakai}}, \bibinfo {author} {\bibfnamefont {S.}~\bibnamefont {Kambe}}, \bibinfo {author} {\bibfnamefont {P.}~\bibnamefont {Opletal}}, \bibinfo {author} {\bibfnamefont {Y.}~\bibnamefont {Tokiwa}}, \bibinfo {author} {\bibfnamefont {Y.}~\bibnamefont {Haga}}, \bibinfo {author} {\bibfnamefont {S.}~\bibnamefont {Kitagawa}}, \bibinfo {author} {\bibfnamefont {K.}~\bibnamefont {Ishida}}, \bibinfo {author} {\bibfnamefont {D.}~\bibnamefont {Aoki}}, \bibinfo {author} {\bibfnamefont {G.}~\bibnamefont {Knebel}}, \bibinfo {author} {\bibfnamefont {G.}~\bibnamefont {Lapertot}}, \bibinfo {author} {\bibfnamefont {S.}~\bibnamefont {Kr\"amer}},\ and\ \bibinfo {author} {\bibfnamefont {M.}~\bibnamefont {Horvati\ifmmode~\acute{c}\else \'{c}\fi{}}},\ }\bibfield  {title} {\bibinfo {title} {Longitudinal spin fluctuations driving field-reinforced superconductivity in \text{UTe$_2$}},\ }\href
  {https://doi.org/10.1103/PhysRevLett.131.226503} {\bibfield  {journal} {\bibinfo  {journal} {Phys. Rev. Lett.}\ }\textbf {\bibinfo {volume} {131}},\ \bibinfo {pages} {226503} (\bibinfo {year} {2023})}\BibitemShut {NoStop}%
\bibitem [{\citenamefont {Tokiwa}\ \emph {et~al.}(2024)\citenamefont {Tokiwa}, \citenamefont {Opletal}, \citenamefont {Sakai}, \citenamefont {Kambe}, \citenamefont {Yamamoto}, \citenamefont {Kimata}, \citenamefont {Awaji}, \citenamefont {Sasaki}, \citenamefont {Aoki}, \citenamefont {Haga},\ and\ \citenamefont {Tokunaga}}]{TokiwaPRB2024}%
  \BibitemOpen
  \bibfield  {author} {\bibinfo {author} {\bibfnamefont {Y.}~\bibnamefont {Tokiwa}}, \bibinfo {author} {\bibfnamefont {P.}~\bibnamefont {Opletal}}, \bibinfo {author} {\bibfnamefont {H.}~\bibnamefont {Sakai}}, \bibinfo {author} {\bibfnamefont {S.}~\bibnamefont {Kambe}}, \bibinfo {author} {\bibfnamefont {E.}~\bibnamefont {Yamamoto}}, \bibinfo {author} {\bibfnamefont {M.}~\bibnamefont {Kimata}}, \bibinfo {author} {\bibfnamefont {S.}~\bibnamefont {Awaji}}, \bibinfo {author} {\bibfnamefont {T.}~\bibnamefont {Sasaki}}, \bibinfo {author} {\bibfnamefont {D.}~\bibnamefont {Aoki}}, \bibinfo {author} {\bibfnamefont {Y.}~\bibnamefont {Haga}},\ and\ \bibinfo {author} {\bibfnamefont {Y.}~\bibnamefont {Tokunaga}},\ }\bibfield  {title} {\bibinfo {title} {Reinforcement of superconductivity by quantum critical fluctuations of metamagnetism in \text{UTe$_2$}},\ }\bibfield  {journal} {\bibinfo  {journal} {Physical Review B}\ }\textbf {\bibinfo {volume} {109}},\ \href {https://doi.org/10.1103/physrevb.109.l140502}
  {10.1103/physrevb.109.l140502} (\bibinfo {year} {2024})\BibitemShut {NoStop}%
\bibitem [{\citenamefont {Ishizuka}\ and\ \citenamefont {Yanase}(2021)}]{IshizukaPRB2021}%
  \BibitemOpen
  \bibfield  {author} {\bibinfo {author} {\bibfnamefont {J.}~\bibnamefont {Ishizuka}}\ and\ \bibinfo {author} {\bibfnamefont {Y.}~\bibnamefont {Yanase}},\ }\bibfield  {title} {\bibinfo {title} {Periodic anderson model for magnetism and superconductivity in {UTe}2},\ }\bibfield  {journal} {\bibinfo  {journal} {Phys. Rev. B}\ }\textbf {\bibinfo {volume} {103}},\ \href {https://doi.org/10.1103/physrevb.103.094504} {10.1103/physrevb.103.094504} (\bibinfo {year} {2021})\BibitemShut {NoStop}%
\bibitem [{\citenamefont {Anderson}\ and\ \citenamefont {Brinkman}(1973)}]{AndersonPRL1973}%
  \BibitemOpen
  \bibfield  {author} {\bibinfo {author} {\bibfnamefont {P.~W.}\ \bibnamefont {Anderson}}\ and\ \bibinfo {author} {\bibfnamefont {W.~F.}\ \bibnamefont {Brinkman}},\ }\bibfield  {title} {\bibinfo {title} {\text{A}nisotropic superfluidity in $^{3}\mathrm{He}$: A possible interpretation of its stability as a spin-fluctuation effect},\ }\href {https://doi.org/10.1103/PhysRevLett.30.1108} {\bibfield  {journal} {\bibinfo  {journal} {Phys. Rev. Lett.}\ }\textbf {\bibinfo {volume} {30}},\ \bibinfo {pages} {1108} (\bibinfo {year} {1973})}\BibitemShut {NoStop}%
\end{thebibliography}

%

\end{document}